# PUBLICATIONS

**OF THE**

# UNITED STATES

# NAVAL OBSERVATORY

**SECOND SERIES**

---

VOLUME XXVII – PARTS I AND II

## $W1_{J00}$
## RESULTS OF THE OBSERVATIONS
**MADE WITH THE**
## SIX-INCH TRANSIT CIRCLE
## 1977-1982

**AND**

## $W2_{J00}$
## RESULTS OF THE USNO POLE-TO-POLE OBSERVATIONS
**MADE WITH THE**
## SIX-INCH AND SEVEN-INCH TRANSIT CIRCLES
## 1985-1996

**PREFACE**

The astronomical results contained in this publication represent a continuation of previous work of the United States Naval Observatory Six-inch and Seven-inch transit circles. The results are also the last from United States Naval Observatory transit circles, since those instruments are no longer in service.

**TABLE OF CONTENTS**

# Part I

# $W1_{J00}$

## RESULTS OF THE OBSERVATIONS

**MADE WITH THE**

## SIX-INCH TRANSIT CIRCLE

## 1977-1982

**OBSERVATIONS OF THE SUN, PLANETS, AND MINOR PLANETS**
**CATALOG OF 7,267 STARS ON THE**
**INTERNATIONAL CELESTIAL REFERENCE FRAME (ICRF)**
**AND ABSOLUTE FRAME REFERENCED TO EQUINOX OF J2000.0**

**By**

**T.J. RAFFERTY, E.R. HOLDENRIED, and S.E. URBAN**

# Introduction

**Observing Program** – These are the results of observations made with the Six-inch Transit Circle in Washington, D.C., between September 1977 and August 1982. This catalog was the first Washington transit circle catalog to be referred to the Equinox of J2000.0, and will be denoted as the $W1_{J00}$. The observing program was structured to be absolute, in the sense that the positions were not explicitly relying on any previous observations. The absolute positions were defined with respect to an internally consistent frame that was unique to the particular instrument. The uniqueness of this instrumental frame was a function, not only of the instrument itself, but also of the methods used to form the catalog. Comparisons of the $W1_{J00}$ absolute catalog with the Hipparcos Star Catalog (European Space Agency 1997) showed significant differences which we believe were caused by the limitations of the instrument and perhaps also by the very complex reduction methods themselves. Therefore, given here are both the absolute positions reduced in the traditional way as well as the position differentially adjusted to the system of the Hipparcos Star Catalog.

The majority of the observations in this program fall into two categories, stars north of declination -30° from the Fifth Fundamental Catalog (FK5) (Fricke *et al.* 1988), and stars from the *Catalog Of 3539 Zodiacal Stars For The Equinox 1950.0* compiled by James Robertson (1940). The zodiacal stars are located in a zone approximately 16 degrees wide centered on the ecliptic and are brighter than or equal to visual magnitude 9.5. Robertson describes the composition of the catalog as follows: "It includes all the stars from Gill's *Catalog Of 2798 Zodiacal Stars for the Epoch 1900*, with 2 omissions (Nos 103 and 437 of Gill's catalog) and 11 additions; also 361 in Hedrick's Catalog and 27 Backlund-Hough stars which are not in Gill's list; and finally all additional stars as bright as 7.0 magnitude (344 stars) are included, giving a total of 3,539 stars all within 8 degrees of the ecliptic. The program has added 37 "filler" stars and 128 components of multiple stars that were presented in the original list as single stars. Table 1 summarizes the types of stars observed and the number of observations for each class of object. The figures given in the column "Number in Class" are the number of individual stars in that class which were not included in another. Some stars could fit in more than one class, for example 12 of the Zodiacal stars could also be classified as stars observed by photographic zenith tubes (PZT); but, for the purposes of that column, were counted only as Zodiacals. The Blaauw stars were selected from a list prepared by Adrian Blaauw (1955) of Leiden of O and B type stars and Cepheids which were of particular interest for studies of galactic dynamics. The AGK3R stars (Corbin 1978) were a selection from that catalog which had poor observational histories or other problems. The 203 clock stars are a list of fundamental stars that have been observed for this purpose by previous Six-inch transit circle catalogs going back to 1936.

The PZT stars were drawn from a list compiled by Haruo Yasuda of Tokyo in response to a proposal by Commission 8 (Positional Astronomy) of the International Astronomical Union (IAU), made at the XIV General Assembly, which comprised all the stars observed by northern PZTs then in use. Martha Carpenter of the University of Virginia requested that a list of stars be observed that could be used in studies of the Hyades cluster, hence the Carpenter stars. Also from the University of Virginia, Philip Ianna supplied a list for proper motion studies at high galactic latitudes that bears his name. The Radio Source Reference stars were chosen from a list of stars to be used in a process of linking radio sources to the optical reference frame. Papers by Wade (1970), and Sandage and Kristian (1970) give more details. In the miscellaneous category were included stars that are themselves radio sources such as R Aquilae, and some that exhibit very large proper motions such as 61 Cygni B (61 Cygni A, which also has a high proper motion, was observed as an FK5 star). Generally the order of class priority (from highest to lowest) was Clock or Azimuth, FK5, Zodiacal, PZT, and Blaauw, with all the others having about equal priority.

Table 1: List of star classes and solar system objects included and their number of observations.

| Observing Program (1977-1982) | | | | | | |
|---|---|---|---|---|---|---|
| Star Class | Number of | | Number in | | Solar System | Number of |
| | Observations | | Class | | Objects | Observations |
| | | | | | | |
| Zodiacal | 29,648 | | 3,470 | | Ceres (1) | 226 |
| Clock | 14,813 | | 203 | | Pallas (2) | 173 |
| FK5 | 11,072 | | 929 | | Juno (3) | 103 |
| Refraction (lower culmination) | 2,772 | | 114 | | Vesta (4) | 201 |
| Azimuth (upper culmination) | 2,269 | | 18 | | Hebe (6) | 110 |
| Azimuth (lower culmination) | 2,246 | | 18 | | Iris (7) | 141 |
| Day | 2,971 | | 59 | | Flora (8) | 74 |
| AGK3R | 7,401 | | 1,601 | | Metis (9) | 63 |
| PZT | 2,572 | | 302 | | Eunomia (15) | 89 |
| Blaauw | 2,424 | | 284 | | Sun | 861 |
| Ianna | 1,246 | | 147 | | Mercury | 328 |
| Radio Source Reference | 1,322 | | 195 | | Venus | 621 |
| Carpenter | 841 | | 113 | | Mars | 252 |
| Miscellaneous | 205 | | 23 | | Jupiter | 327 |
| | | | | | Saturn | 304 |
| | | | | | Uranus | 288 |
| | | | | | Neptune | 222 |
| | | | | | | |
| Total | 81,802 | | 7,476 | | Total | 4,383 |

Every effort was made to obtain at least eight observations of each star distributed equally between Clamps and Circles (the terms Clamp and Circle will be defined later in the section on Instrumentation). However, some stars were added to the program too late to obtain that ideal distribution. For the late arrivals it was decided that a minimum of three observations in right ascension and declination would be required. This was realized for all stars except two Radio Source Reference stars which were kept with only two observations in declination. Both stars had three or more observations in right ascension. The FK5 stars that served as Clock, Azimuth, or Refraction stars accrued many more than the minimum number of observations. Of the 7,285 stars observed, only 18 are not included in the final results due to too few or discordant observations. Also some solar system observations were found to be discordant and rejected.

Although 458 observations were made of the Moon primarily for determining $\Delta T$, they have not been included here because Lunar Laser Ranging techniques better provide these data.

Table 2 shows the ranges of the magnitude of the stars, the standard deviations of a single observation, the standard error of the mean positons, number of observation per star, and the sky coverage.

Table 2

Summary of the Star Positions in $W1_{J00}$ Catalog

| | Range | Average |
|---|---|---|
| Magnitudes | -1.6 to 10.4 mag | 7.18 |
| RA standard deviation of a single obs'n | 03 to 920 mas | 268 mas |
| Dec standard deviation of single obs'n | 02 to 990 mas | 291 mas |
| RA standard error of the mean | 15 to 460 mas | 98 mas |
| Dec standard error of the mean | 10 to 400 mas | 107 mas |
| RA Number of observations/Star | 3 to 187 | 10 |
| Dec Number of observations/Star | 2 to 179 | 10 |
| Declination Coverage | -38° 52′ 40″ to +89° 02′ 16″ | |

**Personnel and Acknowledgements** - The program was carried out under the direction of J.A. Hughes, director of the Transit Circle Division. Many Division members served as observers and are listed in Table 3.

Table 3: Observers (1977-1982)

| code | name | tenure |
|---|---|---|
| RB | R.L. Branham | September 1977 to August 1982 |
| TC | T.E. Corbin | September 1977 to August 1982 |
| SD | S.J. Dick | August 1979 to August 1982 |
| SG | F.S. Gauss | September 1977 to July 1982 |
| EH | E.R. Holdenried | September 1977 to August 1982 |
| EJ | E.S. Jackson | September 1977 to August 1982 |
| BK | B.L. Klock | September 1977 to August 1982 |
| RM | R.J. Miller | April 1978 to August 1982 |
| DR | M.D. Robinson | September 1977 to August 1982 |
| RR | R.W. Rhynsburger | September 1977 to November 1980 |
| TR | T.J. Rafferty | September 1977 to July 1982 |
| CS | C.A. Smith | September 1977 to August 1982 |
| RS | R.C. Stone | October 1981 to August 1982 |
| ZY | Z.G. Yao | May 1982 to August 1982 |

The preliminary daily reductions and editing of the data were carried out by a team composed of various observers and included at one time or another Dick, Holdenried, Miller, Rafferty and Stone; Rhynsburger reviewed the editing until he retired in late 1980. The final reductions were carried out by Holdenried and Rafferty in consultation with Smith and David Scott. Yao provided the apparent places for the solar system objects along with some auxiliary quantities such as horizontal parallax and phase angle.

Instrumentation support was provided by Klock, Gauss, Harold E. Durgin, Rafferty, James H. Davis, Edward D. McClain, Phillip K. Miller, Joseph J. Bone, Jr., and Herbert T. Gaskins, as well as the personnel of the USNO Instrument Shop: Harold M. Durham, John W. Pohlman Jr,. Raymond O. Ragsdale, George Phillips, Guiseppe Reolon, Joseph D. Gauvreau, Edward C. Matthews, John W. Bowles, James B. Beasley, and Stephen J. Boretos.

Clayton Smith died before the completion of the final reductions. His broad experience and expert advice were heavily relied upon and were sorely missed particularly in the later stages of the discussions. Any flaws in the resulting catalog should in no way reflect upon him except to underscore how useful it would have been to have him available to review the finished product. It is our wish to present this catalog as a memorial to him.

## Instrument, Accessories, and Procedures

The Six-inch Transit Circle, built by the Warner and Swasey Company, has operated from the U.S. Naval Observatory site in Washington, D.C, since 1897. The instrument, following the Repsold design, is constructed of steel. The telescope was supported by pivots which rest on bronze buttons embedded in wyes fixed to two steel cages, each of which is attached to the top of two brick piers. The two brick piers sit on two arms of a large concrete cross shaped structure that lies below ground level. The other two arms of this structure support the two piers of the collimating telescopes. Before the start of this observing program, all piers were insulated with several inches of a sprayed-on polystyrene material. For more information on the Six-inch Transit Circle see Watts (1950) and Hemenway (1966), and for more information on transit circles in general see Podobed (1962).

**Pivots** - A hardened steel cylinder 2.24 inches in diameter forms each pivot and rests on two bronze buttons spaced 90° apart within the wyes. Most of the weight of the moving parts of the instrument (about 400 pounds) was supported on counterpoise wheels and was balanced by means of overhead levers with iron weights. The weight borne by each pivot was about 17 pounds.

**New Objective** - The original J.A. Brashear Six-inch objective was replaced by a new lens from the Farrand Optical Corporation of New York before the start of the $W1_{J00}$ observing program. The new lens, with an equivalent focal length of 72 inches (182.9 cm), was designed to be temperature-compensating, which eliminated the seasonal use of shims to correct the focus as was necessary with the old objective.

**New Graduated Circle** - The altitude of the telescope with respect to the cages was determined by photoelectrically scanning, through microscopes, the image of a graduated circle attached to the rotation axis. The graduated circle was fabricated from a glass annulus 68.5 cm in diameter mounted upon a steel wheel. The annulus was graduated by division lines spaced at intervals of $0\overset{\circ}{.}05$. In previous programs, the divisions were engraved on an inlay of silver or gold. The employment of a glass circle installed at the start of this observing program was necessitated by the change to the photoelectric scanning system (the previous system was photographic) and the loss of craftsman with the ability to engrave a circle. The glass proved to be a difficult medium to work with. Problems were encountered depositing the division lines on its surface and mounting the thin annulus to the steel wheel. The Teledyne-Gurley Corporation, after a number of failed attempts, delivered an apparently good glass circle in 1974. That same year, after taking a set of diameter corrections, the observing program was begun. Another set of diameter corrections taken in 1977 revealed that this glass circle had suffered significant changes large enough to render the observations taken during that period unacceptable. Another glass circle, produced by the Heidenhain Corporation of Germany, was installed in mid-1977. The observing program was effectively restarted at that time. The Heidenhain glass circle was very successful and was used for all observations described in this catalog. Details about the Teledyne-Gurley and Heidenhain glass circles can be found in a paper by Rafferty and Klock (1982).

**New Circle Scanners** - The photographic cameras used during the $W5_{50}$ program (Hughes and Scott 1982) to record the positions of the graduated circle were replaced with photoelectric scanners for the $W1_{J00}$ observing program. A description of the scanners can be found in a paper by Rafferty and Klock (1986). New microscope tubes and mounts were built by the U.S. Naval Observatory's Instrument Shop to improve the system's stability. The microscopes were mounted in pairs 180° apart (that is, on opposite ends of a diameter) to minimize the eccentricity error caused by the non-coincidence of the physical center of rotation of the circle and the geometric center. Six microscopes on three diameters were mounted on each cage during the observing program.

**Data Acquisition and Control System** - The IBM 1800 Data Acquisition and Control System, installed in 1969 during the $W5_{50}$ observing program (Hughes and Scott 1982), was used during the $W1_{J00}$ until 1980 when it was replaced by a Hewlett Packard HP1000. These data acquisition and control systems collected all the environmental data, readings from the micrometer screw resolvers, circle scanners, and clock timing data. They also computed apparent places prior to the observation and thereby determined the speed of the micrometer motor drive and the altitude to which the telescope was set (manually), and performed daily preliminary reductions. An important improvement over previous (passive) systems was the ability to provide instant feedback to the observer. This took the form of messages on a terminal, tones and lights, and a voice synthesizer. This data acquisition and control system, developed by Gauss, greatly improved the observing efficiency.

**Micrometer** - The same visual, two-axis, micrometer used during the previous programs was used for this observing program. Resolvers, one for each micrometer screw, were used to determine the rotation of the screws and were recorded by resolver-to-digital converters. Determinations of progressive and periodic screw errors were made in 1973 and 1982. The eyepiece, a 10mm focal length Plössl, was fitted with a dove prism to allow the observer to reverse the apparent field of view to minimize personal bias. Tungsten wires, 0.0002 inches in diameter, were mounted on the moveable slides forming a box 8 arc seconds in right ascension and 6 arc seconds in declination within which observations were centered. The right ascension wires were moved by a motor drive, with the speed automatically set as a function of declination and finely adjusted by the observer using hand block buttons. The reading of the right ascension screw was made at intervals of 4 seconds while the observer kept the object centered. The declination screw was set manually and the readings recorded when the observer pressed a hand block button. Field illumination allowed the wires to appear black against a slightly brighter background.

**Magnitude Screens** - Two moveable screens were mounted in the center of the telescope tube that could be inserted into the optical path. They could be used either separately or together to reduce the apparent magnitude of the brighter stars. One screen reduced the star's apparent brightness by about 1.5 magnitudes, the other by about 3 magnitudes, and the two together about 5 magnitudes. The screens were not used during the observations of the Sun, Moon, and planets.

**Instrument Reversal** - The Six-inch was reversed (rotated 180° in azimuth) interchanging the east and west pivots approximately every 30 days. The orientation of the instrument was referenced to the location of the clamping device, which fixed the altitude of the instrument after it was pointed to that of a star, and thus referred to either as Clamp West or Clamp East. By changing the direction of the screws, circle readings, and the orientation of the lens with respect to the sky, the reversing of the instrument helped minimize systematic errors. Every effort was made to make an equal number of observations of each star on each Clamp. Corrections were determined in both right ascension and declination that were correlated with the Clamp and are described later.

**Circle Rotation** - In July 1980, the wheel of the graduated circle, while remaining in the plane of the meridian, was rotated with respect to the tube of the transit circle. Observations taken before the circle rotation are distinguished by the nomenclature Circle One and those after the rotation as Circle Two. The circle rotation was done to mitigate systematic errors that might be correlated with time or the graduated circle. An effort was made to take an equal number of observations of each star on each of the Circles. Corrections were determined in both right ascension and declination that were Circle dependent and are described later.

**Observing Tours** - Observations were grouped into "tours". On most nights, two observing tours were taken dividing the night in half. Determinations of corrections to the transit circle, such as the collimation, level, azimuth with respect to the marks, nadir, and flexure (to be described later), were made at the start and end of each nighttime observing tour and at intervals of three hours. Selected groups of stars were observed in each nighttime observing tour to determine corrections to the clock, azimuth, and refraction (also to be described later). Daytime observing tours usually started three hours before the meridian transit of the Sun and ended three hours after. Determinations of the instrument's collimation, level, and azimuth with respect to the marks, nadir, and flexure were made at the start and end of each daytime observing tour and at intervals of two hours. The daytime observing tours included observations of the Sun, Mercury, Venus, and bright stars from a group around the equator. More details concerning the daytime observations are given later.

**Environmental Data** - The air temperature was measured to 0.1 degrees Celsius using Hy-cal platinum resistance probes mounted above the transit circle. The air pressure was read to 0.1mm of mercury with Exactel Servomanometers located inside the transit house. The barometric readings were temperature compensated. The dew point was read to 5 degrees Fahrenheit using Honeywell probes treated with lithium chloride activation solution and dried. The dew point probes were also located inside the transit house and were retreated every six months. All the environmental units were sampled by the control system every 30 seconds.

**Clock System** - The recording of the right ascension measurements was synchronized with the sidereal clock. Rubidium standards were used at the start of the observing program and were replaced with cesium standards in 1980. A once-per-second pulse from a clock was used to trigger an interrupt-driven routine in the data acquisition computer that maintained the time in a common area accessible to all programs. The routines that recorded the right ascension data could always be guaranteed of obtaining the correct time, as the interrupt was at the highest possible level. Electronically, the buttons from the observer's hand block were "anded" with the clock pulse, so that both had to occur before a signal was recorded. Although the time was only signaled once per second, the signal was accurate to approximately 30 microseconds.

## Right Ascensions

**Corrections for Pivot Irregularities** - The Six-inch is supported at the ends of its axis of rotation by two steel cylinders, called pivots, that rest on brass buttons attached to cages which are, in turn, affixed to two piers, aligned East and West and so designated. This design permits a maximum of mechanical rigidity as well as defining the rotation of the telescope simply by the interaction of the pivots with the buttons. Over the years, this design also has shown the benefit that all the wear has occurred in the brass buttons. Thus periodic replacement of these buttons, when wear is detected, has ensured that only the pivots determine changes in the plane of rotation. If the pivots were exactly cylindrical, this plane would not change orientation as the telescope rotated. However, in spite of careful machining, the pivots are, at some level of measurement, irregularly shaped ellipsoids. In the past the effects of these irregularities have been measured and applied to the observations to correct them to a fixed plane of rotation.

The irregularities of the pivots can be resolved into two orthogonal components, one component in the azimuthal direction and the other 90° from that in the direction of the level. The pivots on the Six-inch are referred to as the Clamp Pivot or the Inductosyn Pivot, depending whether the pivot is next to the clamping mechanism or next to a circular resolver called an Inductosyn. The process to determine the pivot irregularities made use of an axial microscope, which could be mounted on either the cage on the East Pier or the cage on the West Pier, to measure the positions in the field of view of tiny dots of mercury evaporated onto the flat vertical face of each pivot. The positions of these dots were monitored as the telescope rotated through 360°. Although there was no reason to expect the measurements to depend on the cage from which they are taken, it was often the case that sets were taken from both cages. Thus, the corrections applied to the observations are a mean of sets of measures taken in 1963, 1982, 1984 and 1989.

Tables 4 and 5 give the combined (East and West) values of the pivot irregularities in level and azimuth for both pivots, Inductosyn Pivot and Clamp Pivot, as a function of zenith distance.

The actual correction to the time of transit $dT$ caused by pivot irregularities was calculated from the following:

$$dT = \{((\zeta_{90} - \zeta_{270}) / 2) + (\lambda_S - \lambda_{180} + ((\zeta_{90} - \zeta_{270}) / 2)) \cos z + \zeta_S \sin z\} \sec \delta$$

where: $\zeta$ = azimuthal component of pivot irregularities

$\lambda$ = level component of pivot irregularities
$\delta$ = declination of observed object (equator distance
for objects observed at lower culmination)
z = zenith distance of observed object

The subscripts denote that the pivot irregularities are a function of the pointing of the telescope. Thus $\zeta_S$ denotes the azimuthal component of the pivot irregularity when the telescope is pointed in the direction of the observed object, S; $\lambda_{180}$ denotes the level component when the telescope is pointed at a zenith distance of 180°. The pivot irregularities are also a function of how the pivots are oriented with respect to the East and West cages that is if the Inductosyn pivot is supported by the East cage, the preceding tables are inverted with respect to zenith distance.

Table 4: Azimuth Pivot Irregularities (Inductosyn pivot supported by the West Cage)
units = seconds of time
σ = standard deviation of mean

| Azimuth Correction | | | | |
|---|---|---|---|---|
| zenith distance | Inductosyn Pivot | | Clamp Pivot | |
| 200° | 0.000240 | σ = 0.000680 | 0.000590 | σ = 0.000353 |
| 190 | 0.001640 | 0.000559 | 0.001860 | 0.000565 |
| 180 | 0.002400 | 0.001621 | 0.007000 | 0.000956 |
| 170 | 0.002360 | 0.000789 | 0.002440 | 0.000230 |
| 160 | 0.004090 | 0.000355 | 0.001210 | 0.000456 |
| 150 | 0.003280 | 0.000401 | 0.000140 | 0.000528 |
| 140 | 0.002450 | 0.000380 | 0.000690 | 0.000285 |
| 130 | 0.002330 | 0.000425 | 0.002500 | 0.000400 |
| 120 | 0.001430 | 0.000560 | 0.003430 | 0.000282 |
| 110 | 0.000610 | 0.000516 | 0.003250 | 0.000261 |
| 100 | 0.001400 | 0.000355 | 0.000860 | 0.000391 |
| 90 | 0.002510 | 0.000884 | 0.004690 | 0.000387 |
| 80 | 0.002960 | 0.000722 | 0.001740 | 0.000491 |
| 70 | 0.002460 | 0.000732 | 0.001120 | 0.000463 |
| 60 | 0.003470 | 0.000436 | 0.000390 | 0.000235 |
| 50 | 0.002620 | 0.000561 | 0.001070 | 0.000292 |
| 40 | 0.002450 | 0.000742 | 0.002450 | 0.000365 |
| 30 | 0.001050 | 0.000332 | 0.002620 | 0.000434 |
| 20 | 0.000040 | 0.000447 | 0.002090 | 0.000345 |
| 10 | 0.002140 | 0.000436 | 0.000440 | 0.000656 |
| 0 | 0.002090 | 0.000487 | 0.008360 | 0.000545 |
| 10 | 0.002470 | 0.000617 | 0.002580 | 0.000467 |
| 20 | 0.002660 | 0.000750 | 0.001000 | 0.000416 |
| 30 | 0.002560 | 0.000448 | 0.001590 | 0.000367 |
| 40 | 0.002540 | 0.000646 | 0.002050 | 0.000263 |
| 50 | 0.001680 | 0.000741 | 0.002270 | 0.000440 |
| 60 | 0.002040 | 0.000499 | 0.001330 | 0.000347 |
| 70 | 0.000170 | 0.000509 | 0.000800 | 0.000338 |
| 80 | 0.001330 | 0.000584 | 0.001280 | 0.000175 |
| 90 | 0.002650 | 0.000722 | 0.007780 | 0.000850 |
| 100 | 0.001790 | 0.000596 | 0.001700 | 0.000392 |
| 110 | 0.001940 | 0.000800 | 0.002900 | 0.000445 |
| 120 | 0.002750 | 0.000992 | 0.003260 | 0.000221 |
| 130 | 0.002910 | 0.000596 | 0.003160 | 0.000430 |
| 140 | 0.002530 | 0.000688 | 0.002370 | 0.000446 |
| 150 | 0.002040 | 0.000614 | 0.002660 | 0.000406 |
| 160 | 0.000240 | 0.000680 | 0.000590 | 0.000353 |
| 170 | 0.001640 | 0.000559 | 0.001860 | 0.000565 |
| 180 | 0.002400 | 0.001621 | 0.007000 | 0.000956 |
| 190 | 0.002360 | 0.000789 | 0.002440 | 0.000230 |
| 200 | 0.004090 | 0.000355 | 0.001210 | 0.000456 |
| 210 | 0.003280 | 0.000401 | 0.000140 | 0.000528 |

Table 5: Level Pivot Irregularities (Inductosyn pivot supported by the West Cage)
units = seconds of time
σ = standard deviation of mean

| Level Correction | | | | |
|---|---|---|---|---|
| zenith distance | Inductosyn Pivot | | Clamp Pivot | |
| 200° | 0.001510 | σ = 0.000798 | 0.000910 | σ = 0.000387 |
| 190 | 0.003140 | 0.000819 | 0.001590 | 0.000512 |
| 180 | 0.003400 | 0.001393 | 0.007060 | 0.000764 |
| 170 | 0.002970 | 0.000768 | 0.002320 | 0.000225 |
| 160 | 0.004030 | 0.000428 | 0.001220 | 0.000388 |
| 150 | 0.003030 | 0.000357 | 0.000060 | 0.000468 |
| 140 | 0.001690 | 0.000490 | 0.000460 | 0.000275 |
| 130 | 0.001130 | 0.000638 | 0.002280 | 0.000397 |
| 120 | 0.000480 | 0.000587 | 0.003370 | 0.000269 |
| 110 | 0.000460 | 0.000723 | 0.003080 | 0.000251 |
| 100 | 0.002640 | 0.000747 | 0.000530 | 0.000395 |
| 90 | 0.003010 | 0.000838 | 0.004760 | 0.000335 |
| 80 | 0.003120 | 0.000713 | 0.001750 | 0.000422 |
| 70 | 0.002750 | 0.000593 | 0.000790 | 0.000443 |
| 60 | 0.003260 | 0.000385 | 0.000580 | 0.000253 |
| 50 | 0.002210 | 0.000501 | 0.001350 | 0.000310 |
| 40 | 0.001520 | 0.000713 | 0.002560 | 0.000325 |
| 30 | 0.000260 | 0.000606 | 0.002600 | 0.000371 |
| 20 | 0.001390 | 0.000642 | 0.001700 | 0.000434 |
| 10 | 0.003000 | 0.000576 | 0.000730 | 0.000581 |
| 0 | 0.002390 | 0.000595 | 0.008440 | 0.000511 |
| 10 | 0.002890 | 0.000779 | 0.002400 | 0.000417 |
| 20 | 0.002810 | 0.000787 | 0.001420 | 0.000445 |
| 30 | 0.002150 | 0.000427 | 0.001820 | 0.000331 |
| 40 | 0.001650 | 0.000643 | 0.001980 | 0.000236 |
| 50 | 0.001190 | 0.000604 | 0.002270 | 0.000370 |
| 60 | 0.001060 | 0.000585 | 0.001380 | 0.000324 |
| 70 | 0.000460 | 0.000559 | 0.000710 | 0.000307 |
| 80 | 0.001890 | 0.000567 | 0.001480 | 0.000259 |
| 90 | 0.003410 | 0.000713 | 0.008190 | 0.000835 |
| 100 | 0.002670 | 0.000750 | 0.001540 | 0.000354 |
| 110 | 0.002520 | 0.000820 | 0.002840 | 0.000400 |
| 120 | 0.002380 | 0.000860 | 0.003050 | 0.000285 |
| 130 | 0.002470 | 0.000529 | 0.003040 | 0.000421 |
| 140 | 0.002000 | 0.000606 | 0.002400 | 0.000401 |
| 150 | 0.001370 | 0.000607 | 0.002940 | 0.000389 |
| 160 | 0.001510 | 0.000798 | 0.000910 | 0.000387 |
| 170 | 0.003140 | 0.000819 | 0.001590 | 0.000512 |
| 180 | 0.003400 | 0.001393 | 0.007060 | 0.000764 |
| 190 | 0.002970 | 0.000768 | 0.002320 | 0.000225 |
| 200 | 0.004030 | 0.000428 | 0.001220 | 0.000388 |
| 210 | 0.003030 | 0.000357 | 0.000060 | 0.000468 |

To interpolate to zenith distances between table entries a cubic spline was utilized. Figure 1 shows the effect of the corrections for pivot irregularities, i.e. *dT*, on the time of transit versus zenith distance. Because the *dT* relation has a secant δ factor to render it in seconds of time, *dT* grows very large with increasing declination and likewise zenith distance and is undefined at the pole, a zenith distance of approximately 51°. Thus in Figure 1, the corrections plotted are restricted to values of $\pm 0^{s}.050$.

Figure 1: Corrections *dT* to transit times, caused by pivot irregularities.

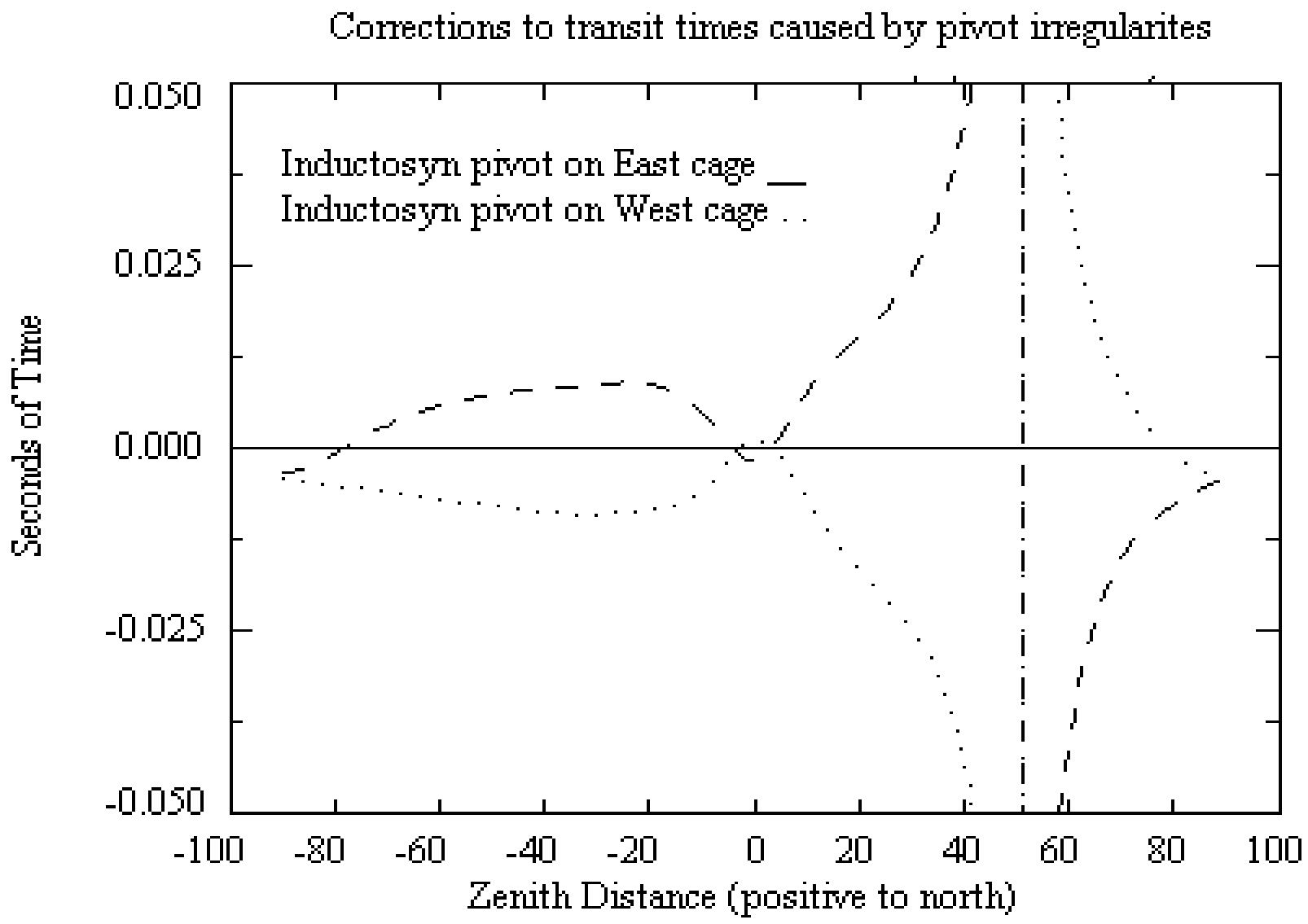

**Micrometer Screw Corrections** - The deviations of the threads of the micrometer screw from nominal were measured in 1973, before the program began, and in 1982, after the program ended. In all cases the periodic deviations, or "errors", were found to be insignificant. However a significant progressive error was found that changed between the two determinations. Because it was not possible to tell how the progressive error evolved with time, it was decided to apply a mean. In 1973, twenty-one sets of measures were taken, in 1982 twenty-six. However, in 1982 a set consisted of measures taken in both the direct (screw readings increasing) and reverse direction. In 1973 a set consisted of measures taken in mostly one direction, only three being taken in the reverse direction. Table 5 contains a weighted mean of all of the sets.

The screw errors applied for settings between those of the entries in the Table 6 were calculated from an 8th order polynomial fit (least squares regression) to the mean "measures". The 4th column in Table 6 gives the differences between the values computed by this polynomial and the actual measure. The polynomial was the following:

$$\Delta\Sigma = A + Bv + Cv^2 + Dv^3 + Ev^4 + Fv^5 + Gv^6 + Hv^7 + Iv^8$$

where: $\Delta\Sigma$ = correction to be applied to screw reading
$v$ = actual screw reading

*A* = +0.543421
*B* = -0.665053
*C* = +0.600815
*D* = -0.130127
*E* = +0.151529
*F* = -0.100799
*G* = +0.381665
*H* = -0.769225
*I* = +0.640310

Table 6: Mean progressive errors of the right ascension micrometer screw
σ = standard deviation of mean
rev = revolution of the micrometer screw

| screw reading (rev) | screw error (rev) | σ (rev) | fit residuals (rev) | screw error (sec of time) |
|---|---|---|---|---|
| 2.0 | 0.0000 | 0.0000 | -0.0003 | 0.0000 |
| 3.0 | 0.0006 | 0.0002 | 0.0004 | 0.0022 |
| 4.0 | 0.0022 | 0.0005 | 0.0005 | 0.0084 |
| 5.0 | 0.0027 | 0.0006 | 0.0004 | 0.0104 |
| 6.0 | 0.0026 | 0.0006 | 0.0004 | 0.0099 |
| 7.0 | 0.0032 | 0.0007 | 0.0003 | 0.0122 |
| 8.0 | 0.0040 | 0.0008 | 0.0001 | 0.0154 |
| 9.0 | 0.0047 | 0.0009 | 0.0000 | 0.0178 |
| 10.0 | 0.0049 | 0.0009 | 0.0003 | 0.0187 |
| 11.0 | 0.0060 | 0.0009 | 0.0002 | 0.0227 |
| 12.0 | 0.0067 | 0.0011 | 0.0004 | 0.0256 |
| 13.0 | 0.0068 | 0.0011 | 0.0000 | 0.0259 |
| 14.0 | 0.0065 | 0.0011 | 0.0007 | 0.0247 |
| 15.0 | 0.0076 | 0.0012 | 0.0001 | 0.0290 |
| 16.0 | 0.0082 | 0.0012 | 0.0004 | 0.0311 |
| 17.0 | 0.0077 | 0.0011 | 0.0002 | 0.0293 |
| 18.0 | 0.0078 | 0.0010 | 0.0000 | 0.0297 |
| 19.0 | 0.0079 | 0.0010 | 0.0002 | 0.0300 |
| 20.0 | 0.0067 | 0.0009 | 0.0006 | 0.0257 |
| 21.0 | 0.0067 | 0.0010 | 0.0003 | 0.0255 |
| 22.0 | 0.0068 | 0.0010 | 0.0004 | 0.0259 |
| 23.0 | 0.0062 | 0.0008 | 0.0005 | 0.0236 |
| 24.0 | 0.0049 | 0.0008 | 0.0000 | 0.0185 |
| 25.0 | 0.0040 | 0.0006 | 0.0002 | 0.0153 |
| 26.0 | 0.0025 | 0.0005 | 0.0002 | 0.0096 |
| 27.0 | 0.0010 | 0.0003 | 0.0006 | 0.0038 |
| 28.0 | 0.0012 | 0.0005 | 0.0007 | 0.0045 |
| 29.0 | 0.0000 | 0.0000 | 0.0002 | 0.0000 |

The right ascension micrometer screw has a total travel of about 30 revolutions, beginning at zero; however, mechanical stops restrict the usable travel to between 2 and 29 revolutions to protect the screw and the micrometer slide. The preceding polynomial is defined only for revolutions between 2 and 29. When applying the screw error, the correction at the nominal center of the field was subtracted from the error, which has the effect of making the error zero at this point.

The scale of the right ascension screw was found using observations of stars located in a zone from $-10°$ to +30° declination. From 40153 observations, the scale was determined to be $3\overset{s}{.}8121 \pm .00004$ per revolution at the equator.

**Other Instrumental Constants** - The collimation axis of the telescope, the level and the azimuth with respect to the marks, were determined at intervals of between 2 and 3 hours during a tour. The collimation axis was found by using two horizontal collimating telescopes located in the meridian on either side (to the north and south) of the main telescope. When the transit circle was pointed at the zenith, ports in the tube allowed an unobstructed line of sight between the collimators. The north collimator was equipped with a slide that moved in the focal plane and to which was attached a pair of closely placed, parallel vertical wires. The south collimator had a single, fixed vertical wire mounted in its focal plane. The collimators were aligned by looking though the north collimator at the image of a wire in the south collimator and, by means of the moveable slide, centering that image between the image of the wire pair affixed to the slide. The transit circle was then used to measure the positions of the wires in each collimator. The axis of collimation was ½ the sum of these measured positions. For each tour an average collimation was adopted.

To monitor the level of the telescope, the images of the micrometer wires as reflected in a mercury basin positioned beneath the instrument at the nadir point were measured with respect to the collimation axis. This was performed every 2-3 hours during a tour generally at the same time that a collimation was taken. A rate was computed and used to interpolate between levels to the time of transit of an observed object. Theoretically, tides should affect the level. However a study failed to reveal any correlation and thus no additional correction for tides was applied.

The azimuth of the telescope was determined relative to two fixed light sources located several hundred feet to the north and south of the instrument along the meridian. These light sources, called marks or mires, were viewed through special long focus lenses; the marks being at the focal point of the lenses while the lenses themselves were within the confines of the transit pavilion. Thus the images of the marks were in parallel ray space and could be viewed directly by the telescope. Observations of the marks were also performed every 2-3 hours at about the same time those of the level and collimation. Like the level, a rate was computed and used to interpolate to the time of transit.

**Clock Corrections** - A special set of 203 FK stars between declinations +30° and $-30°$ that are uniformly distributed in right ascension and declination were selected as clock stars. For each night tour a clock correction was developed which was the average of the observed minus calculated right ascensions, $(O - C)_\alpha$'s, of the clock stars observed during the tour. It was the practice to observe at least 5 clock stars per tour with the average being 7-8. Because the observed right ascensions of all objects in a tour are reduced differentially with respect to the clock stars, the instrumental frame inevitably will be linked closely to the zero point and systematics defined by these stars. In the case of the present catalog, that zero point is the FK5. However, it is the aim of an absolute catalog such as this to provide positions that are based on an independent instrumental (observed) system. Therefore the object of the clock star analysis was to, as much as possible, disengage the observed right ascension system from the systematics of the FK5 so that in the end the clock stars could be said to represent the instrumental frame and that such a system would be independent of the FK5 except possibly in the orientation of the equinox. To this end the following procedure was developed following leads well delineated by C.A. Smith and D.K. Scott in their reductions of previous Washington catalogs, such as the $W5_{50}$ (Hughes and Scott 1982) and the $WL_{50}$ (Hughes *et al.* 1992).

In the analysis described in this section all reference to $(O - C)_\alpha$'s refers to that of the clock stars. The strategy was to analyze these $(O - C)_\alpha$'s in such a way as to remove their dependence on the FK5 and also the observer. This was accomplished by producing quantities which could be applied to the *C*'s to correct them for systematic differences between the instrumental frame and the FK5 and for observer bias. The process is broken into three steps; determination of observer bias, an analysis of systematic signatures dependent on sidereal time and the development of corrections to individual clock star. After observer biases were determined and applied to the clock corrections the new clock corrections were analyzed for a sidereal time dependent function. This function was then applied and improved clock corrections calculated along with new $(O - C)_\alpha$'s for the individual clock stars. The newly calculated $(O - C)_\alpha$'s were then also analyzed for a time (i.e. right ascension) dependent function. This function was then used to develop corrections unique to the individual clock stars. With both the periodic and individual corrections applied, new clock corrections were derived. These new clock corrections were assumed to be definitive of the instrumental frame.

It was convincingly demonstrated by C.A. Smith in his discussion of the $WL_{50}$ (Hughes *et al.* 1992) that observer bias does contaminate the clock corrections. It was therefore thought prudent that such a bias be looked for here, even though a so called “impersonal” micrometer was employed and great care was taken to ensure that observations were balanced with respect to the “reversal” of the eyepiece. This investigation was done by differencing clock corrections of successive tours, where single tour is allowed to appear in two different pairs, and then collecting the differences by observer. Because successive tours were observed by all combinations of two observers, the following analysis was adopted. All the differences where observer *X* was the observer of record for one, either the first or the second, of paired adjacent tours were collected into a group. This group was subdivided into two classes, those differences formed from a pair when *X'*s tour was the first in the pair, and those formed when *X'*s tour was the last of the pair. Tours were paired only if they were separated by less than 24 hours. The mean of those differences are presented in Table 7.

The differences seemed of marginal significance so the same treatment was tried on a subset of the data, pairs formed only if the tours were on the same night. The results were virtually identical to the above. The fact that the average difference where observer X was the observer of the first tour of a pair (column labeled "1st of pair") is nearly equal to the negative of the average difference where observer X was the observer of the second tour (column labeled "2nd of pair") seems to underline the reality of these biases.

Table 7: Observer Biases from Preliminary Comparison of Clock Corrections of Adjacent Tours (separated by position of observer within pair)
units = seconds of time
σ = standard deviation of unit weight

| | 1st of pair | | | 2nd of pair | | | | |
|---|---|---|---|---|---|---|---|---|
| Ob-server | mean diff. | n | σ | mean diff. | n | σ | Student t | df |
| RB | -0.0117 | 120 | 0.0252 | 0.0081 | 119 | 0.0290 | -1.026 | 237 |
| TC | 0.0008 | 172 | 0.0205 | 0.0002 | 173 | 0.0236 | 0.434 | 343 |
| SD | 0.0163 | 103 | 0.0267 | -0.0111 | 101 | 0.0258 | 1.403 | 202 |
| SG | -0.0020 | 149 | 0.0240 | 0.0026 | 148 | 0.0253 | 0.184 | 295 |
| EH | 0.0091 | 167 | 0.0234 | -0.0078 | 166 | 0.0230 | 0.525 | 331 |
| EJ | -0.0052 | 197 | 0.0228 | 0.0034 | 209 | 0.0227 | -0.807 | 404 |
| BK | -0.0056 | 15 | 0.0270 | 0.0061 | 15 | 0.0230 | 0.051 | 28 |
| RM | -0.0024 | 134 | 0.0221 | 0.0003 | 137 | 0.0212 | -0.811 | 269 |
| TR | -0.0047 | 153 | 0.0275 | 0.0055 | 134 | 0.0239 | 0.238 | 285 |
| RR | 0.0052 | 92 | 0.0210 | -0.0044 | 90 | 0.0208 | 0.253 | 180 |
| DR | -0.0083 | 117 | 0.0243 | 0.0120 | 119 | 0.0266 | 1.106 | 234 |
| CS | 0.0077 | 160 | 0.0247 | -0.0086 | 166 | 0.0226 | -0.340 | 324 |
| RS | 0.0016 | 30 | 0.0243 | -0.0019 | 32 | 0.0298 | -0.049 | 60 |
| ZY | -0.0156 | 5 | 0.0181 | 0.0206 | 5 | 0.0194 | 0.422 | 8 |
| | | | | | | | | |
| Ave. | 0.0001 | | | 0.0001 | | | | |
| Total | 0.1614 | | | 0.1614 | | | | |

The observer corrections were formed by taking the mean of the 1st of pair differences with the negative of the 2nd of pair for each observer. These corrections were then applied to the clock corrections of the individual tours and the process repeated for four iterations. The results are given in Table 8.

The labels 1st - 4th refer to the iteration number. The columns labeled mean and σ are the mean differences and the standard deviations (unit weight) of the differences of each iteration. At each iteration one half of the mean differences from the previous iteration was added to an accumulated difference to form the observer corrections applied during that iteration. For example, to form the observer corrections applied during the 3rd iteration, one half of the mean difference from the 2nd iteration was applied to the result of the 1st iteration. In the case of observer ZY the final correction was formed as follows:

$$(1st_iteration) + (2nd_iteration)/2 + (3rd)/2 + (4th)/2 = final_correction$$

$$0\overset{s}{.}0181 + 0\overset{s}{.}0025/2 + 0\overset{s}{.}0014/2 + 0\overset{s}{.}0000/2 = 0\overset{s}{.}0162$$

Table 8: Final Observer Biases from Comparisons of Clock Corrections of Adjacent Tours
units = seconds of time
σ = standard deviation of unit weight

| Ob-server | 1st mean difference | σ | 2nd mean difference | σ | 3rd mean difference | σ | 4th mean difference | σ | Final cor-rection | n |
|---|---|---|---|---|---|---|---|---|---|---|
| RB | -0.0099 | 0.0272 | 0.0003 | 0.0264 | 0.0003 | 0.0263 | 0.0003 | 0.0263 | -0.0099 | 239 |
| TC | 0.0003 | 0.0221 | 0.0005 | 0.0210 | 0.0005 | 0.0210 | 0.0005 | 0.0210 | 0.0003 | 345 |
| SD | 0.0137 | 0.0263 | -0.0006 | 0.0254 | -0.0006 | 0.0254 | -0.0006 | 0.0254 | 0.0137 | 204 |
| SG | -0.0023 | 0.0246 | -0.0005 | 0.0236 | -0.0006 | 0.0236 | -0.0006 | 0.0236 | -0.0023 | 297 |
| EH | 0.0085 | 0.0232 | -0.0017 | 0.0231 | -0.0008 | 0.0231 | -0.0008 | 0.0231 | 0.0077 | 333 |
| EJ | -0.0042 | 0.0227 | 0.0007 | 0.0218 | 0.0007 | 0.0218 | 0.0007 | 0.0218 | -0.0042 | 406 |
| BK | -0.0058 | 0.0246 | 0.0008 | 0.0220 | 0.0008 | 0.0221 | 0.0008 | 0.0221 | -0.0058 | 30 |
| RM | -0.0013 | 0.0216 | -0.0001 | 0.0206 | -0.0002 | 0.0206 | -0.0002 | 0.0206 | -0.0013 | 241 |
| TR | -0.0051 | 0.0258 | 0.0005 | 0.0253 | 0.0004 | 0.0253 | 0.0004 | 0.0253 | -0.0051 | 287 |
| RR | 0.0048 | 0.0209 | -0.0003 | 0.0211 | -0.0003 | 0.0211 | -0.0003 | 0.0211 | 0.0048 | 182 |
| DR | -0.0102 | 0.0255 | 0.0014 | 0.0247 | 0.0006 | 0.0247 | 0.0006 | 0.0247 | -0.0095 | 236 |
| CS | 0.0081 | 0.0236 | -0.0004 | 0.0219 | -0.0004 | 0.0219 | -0.0004 | 0.0219 | 0.0081 | 326 |
| RS | 0.0018 | 0.0270 | 0.0001 | 0.0250 | 0.0001 | 0.0250 | 0.0001 | 0.0250 | 0.0018 | 62 |
| ZY | -0.0181 | 0.0179 | 0.0025 | 0.0196 | 0.0014 | 0.0195 | 0.0007 | 0.0195 | -0.0162 | 10 |

In this case the mean difference for the 4th iteration was set to $0\overset{s}{.}0000$ because any difference whose absolute value was less than $0\overset{s}{.}0010$ was regarded as zero.

After each clock correction was corrected for observer bias using the final observer corrections described above and applied to each tour, the analysis could then proceed in the search for a periodic signature in right ascension. Least squares regression analysis was performed on a function that is the difference of two truncated Fourier series of the form

$$C_1\sin\theta + C_2\cos\theta + C_3\cos2\theta + C_4\cos3\theta$$

The difference function was constructed from paired tours as

$$\Delta CC_1CC_2 = C_1\,(\sin(a_1) - \sin(a_2)) + C_2\,(\cos(a_1) - \cos(a_2)) + C_3\,(\cos(2a_1) - \cos(2a_2)) + C_4\,(\cos(3a_1) - \cos(3a_2))$$

where: $\Delta CC_1CC_2$ = clock correction of $tour_1$ minus $tour_2$
$a_1$ = sidereal time associated with $tour_1$ (the first tour of pair)
$a_2$ = sidereal time associated with $tour_2$ (the second tour of pair)

The data set used was essentially the same as that employed in the preceding analysis of observer bias, that is differences in the clock corrections of paired tours separated by less than 24 hours. The regression analysis produced the constants:

| | estimate | *s.d.*. | student *t* | *df* |
|---|---|---|---|---|
| $C_1 =$ | $+^{s}.001738$ | ±.00056 | 3.107 | 1609 |
| $C_2 =$ | $+^{s}.001138$ | .00059 | 1.943 | |
| $C_3 =$ | $-^{s}.001414$ | .00054 | −2.615 | |
| $C_4 =$ | $-^{s}.001464$ | .00061 | −2.387 | |

When the data set formed by restricting the pairs of tours to be on the same night was processed in a similar fashion, and as in the case of the observer corrections, the results were identical. However the standard deviations were higher so the constants used were from the first solution, i.e. paired tours not necessarily on the same night but within 24 hours. Then using the constants from above in the relation

$$\Delta CC = C_1 \sin(a) + C_2 \cos(a_1) + C_3 \cos(2a) + C_4 \cos(3a)$$

where $\Delta CC$ = correction to the clock correction CC
$a$ = sidereal time associated with the tour

a new clock correction was calculated for each tour.

Tours on the same night were then combined into a single tour with observer corrections and periodic corrections applied to each clock star. A clock correction was calculated for this combined tour because as Smith says in his discussion of the $WL_{50}$ (Hughes *et al.* 1992): "The shorter the time base-line over which a clock corrections is computed, the more difficult it becomes to decouple the instrumental right ascension system from small amplitude, long period systematic errors in the right ascension system of the initially assumed clock star positions.". The quantity $((O - C)_\alpha - CC_{combine})$, where $CC_{combine}$ is the clock correction of the combined tour, was formed for each clock star observation in the combined tour, these quantities were collected by star and averaged. This average was applied as correction to the calculated position of each individual clock star. The process was repeated for a total of three iterations with the results given in Table 9.

Table 9: Corrections to the position of the clock stars.

| | | | | | Iteration | | | | | | | | | | Iteration | | | |
|---|---|---|---|---|---|---|---|---|---|---|---|---|---|---|---|---|---|---|
| FK5 | RA | | Dec | | $1^{st}$ | $2^{nd}$ | $3^{rd}$ | | | FK5 | RA | | Dec | | $1^{st}$ | $2^{nd}$ | $3^{rd}$ | |
| No. | *h* | *m* | ° | ′ | s | s | s | n | | No. | *h* | *m* | ° | ′ | s | s | s | n |
| 7 | 00 | 13 | 15 | 11 | 0.00047 | 0.00011 | 0.00001 | 52 | | 588 | 15 | 50 | 04 | 28 | -0.00688 | -0.00014 | 0.00000 | 70 |
| 9 | 00 | 19 | -08 | 49 | 0.00506 | 0.00032 | 0.00003 | 44 | | 594 | 16 | 00 | -22 | 37 | 0.00594 | 0.00020 | 0.00002 | 75 |
| 13 | 00 | 30 | -03 | 57 | 0.00329 | -0.00005 | -0.00002 | 59 | | 597 | 16 | 05 | -19 | 48 | 0.00666 | 0.00019 | 0.00001 | 90 |
| 19 | 00 | 38 | 29 | 18 | -0.00205 | 0.00015 | 0.00004 | 37 | | 603 | 16 | 14 | -03 | 41 | -0.00289 | -0.00004 | -0.00001 | 77 |
| 27 | 00 | 47 | 24 | 16 | -0.00615 | -0.00006 | 0.00004 | 42 | | 609 | 16 | 21 | 19 | 09 | -0.00143 | -0.00004 | 0.00000 | 60 |
| 28 | 00 | 48 | 07 | 35 | 0.00533 | 0.00026 | 0.00003 | 21 | | 622 | 16 | 37 | -10 | 34 | 0.00278 | 0.00016 | 0.00001 | 75 |
| 36 | 01 | 02 | 07 | 53 | -0.00141 | -0.00002 | 0.00001 | 44 | | 624 | 16 | 41 | -17 | 44 | 0.00201 | 0.00010 | 0.00000 | 52 |
| 45 | 01 | 19 | 27 | 15 | 0.00076 | 0.00018 | 0.00003 | 63 | | 629 | 16 | 52 | 14 | 58 | 0.00432 | 0.00025 | 0.00002 | 59 |
| 47 | 01 | 24 | -08 | 10 | 0.00181 | -0.00006 | -0.00001 | 28 | | 633 | 16 | 57 | 09 | 22 | -0.00246 | -0.00019 | -0.00003 | 78 |
| 50 | 01 | 31 | 15 | 20 | -0.00467 | -0.00005 | 0.00001 | 33 | | 635 | 17 | 05 | 12 | 44 | -0.00245 | -0.00014 | -0.00001 | 65 |
| 56 | 01 | 41 | 05 | 29 | -0.00728 | -0.00021 | 0.00002 | 31 | | 641 | 17 | 15 | 24 | 50 | 0.00060 | 0.00001 | -0.00001 | 52 |
| 60 | 01 | 45 | 09 | 09 | -0.00211 | 0.00008 | 0.00003 | 45 | | 644 | 17 | 22 | -24 | 59 | 0.00428 | 0.00024 | 0.00000 | 74 |
| 66 | 01 | 54 | 20 | 48 | 0.00681 | 0.00050 | 0.00006 | 28 | | 665 | 17 | 43 | 04 | 33 | -0.00158 | -0.00007 | 0.00000 | 75 |
| 71 | 02 | 00 | -21 | 04 | 0.00793 | 0.00055 | 0.00006 | 31 | | 668 | 17 | 47 | 02 | 42 | -0.00241 | -0.00028 | -0.00002 | 59 |
| 80 | 02 | 16 | -06 | 25 | 0.00423 | 0.00019 | 0.00000 | 28 | | 673 | 17 | 59 | -09 | 46 | -0.00408 | -0.00017 | -0.00001 | 41 |
| 83 | 02 | 22 | -23 | 48 | 0.01188 | 0.00035 | 0.00004 | 35 | | 680 | 18 | 07 | 09 | 33 | -0.00700 | -0.00031 | 0.00000 | 63 |
| 85 | 02 | 28 | 08 | 27 | 0.00361 | 0.00030 | 0.00005 | 29 | | 682 | 18 | 13 | -21 | 03 | 0.00494 | 0.00014 | 0.00002 | 56 |
| 91 | 02 | 39 | 00 | 19 | -0.00230 | 0.00016 | 0.00004 | 42 | | 688 | 18 | 21 | -02 | 53 | 0.00080 | 0.00009 | 0.00002 | 56 |
| 98 | 02 | 44 | 10 | 06 | -0.00407 | 0.00008 | 0.00004 | 42 | | 692 | 18 | 27 | -25 | 25 | 0.01374 | 0.00075 | 0.00006 | 11 |
| 102 | 02 | 51 | -21 | 00 | -0.00509 | 0.00021 | 0.00003 | 12 | | 703 | 18 | 45 | 20 | 32 | -0.00029 | 0.00001 | -0.00003 | 35 |
| 104 | 02 | 56 | -08 | 53 | -0.00004 | 0.00003 | 0.00001 | 69 | | 712 | 18 | 59 | 15 | 04 | -0.00321 | -0.00026 | -0.00004 | 82 |
| 107 | 03 | 02 | 04 | 05 | -0.00032 | -0.00005 | 0.00000 | 66 | | 716 | 19 | 05 | 13 | 51 | -0.00138 | -0.00020 | -0.00004 | 57 |
| 114 | 03 | 11 | 19 | 43 | -0.00125 | -0.00001 | 0.00000 | 57 | | 720 | 19 | 09 | -21 | 01 | 0.00393 | 0.00003 | -0.00003 | 64 |
| 121 | 03 | 24 | 09 | 01 | -0.00255 | -0.00011 | -0.00001 | 55 | | 725 | 19 | 17 | 11 | 35 | -0.00107 | -0.00023 | -0.00005 | 55 |
| 127 | 03 | 32 | -09 | 27 | -0.00090 | -0.00008 | -0.00002 | 63 | | 730 | 19 | 25 | 03 | 06 | -0.00039 | -0.00016 | -0.00006 | 46 |
| 135 | 03 | 43 | -09 | 46 | -0.00332 | -0.00001 | 0.00000 | 23 | | 732 | 19 | 30 | 27 | 57 | -0.01205 | -0.00062 | -0.00007 | 37 |
| 139 | 03 | 47 | 24 | 06 | -0.01420 | -0.00075 | -0.00005 | 13 | | 736 | 19 | 36 | -24 | 53 | 0.01282 | 0.00045 | 0.00000 | 57 |
| 149 | 03 | 58 | -13 | 30 | 0.00394 | 0.00043 | 0.00002 | 32 | | 741 | 19 | 46 | 10 | 36 | -0.00410 | -0.00042 | -0.00006 | 65 |
| 151 | 04 | 03 | 05 | 59 | -0.00556 | -0.00037 | -0.00005 | 31 | | 746 | 19 | 52 | 01 | 00 | -0.00142 | -0.00026 | -0.00006 | 62 |
| 154 | 04 | 11 | -06 | 50 | 0.00067 | -0.00007 | -0.00003 | 50 | | 749 | 19 | 55 | 06 | 24 | -0.00239 | -0.00009 | -0.00003 | 89 |
| 159 | 04 | 19 | 15 | 37 | -0.00504 | -0.00060 | -0.00011 | 19 | | 753 | 20 | 02 | -27 | 42 | 0.00631 | 0.00039 | 0.00001 | 25 |
| 164 | 04 | 28 | 19 | 10 | -0.00782 | -0.00060 | -0.00008 | 26 | | 756 | 20 | 11 | 00 | 49 | -0.00085 | -0.00040 | -0.00006 | 61 |
| 169 | 04 | 36 | -03 | 21 | 0.00320 | -0.00006 | -0.00006 | 45 | | 762 | 20 | 21 | -14 | 46 | 0.00729 | 0.00027 | 0.00000 | 83 |
| 174 | 04 | 42 | 22 | 57 | -0.00194 | -0.00042 | -0.00006 | 45 | | 768 | 20 | 33 | 11 | 18 | -0.00984 | -0.00037 | -0.00004 | 65 |
| 176 | 04 | 45 | -03 | 15 | -0.00125 | -0.00021 | -0.00007 | 44 | | 774 | 20 | 39 | 15 | 54 | -0.00717 | -0.00062 | -0.00007 | 47 |
| 179 | 04 | 51 | 05 | 36 | 0.00042 | -0.00027 | -0.00008 | 47 | | 786 | 20 | 54 | 28 | 03 | -0.00458 | -0.00016 | -0.00002 | 67 |
| 184 | 05 | 03 | 21 | 35 | -0.00316 | -0.00029 | -0.00006 | 38 | | 794 | 21 | 09 | -11 | 22 | 0.00301 | 0.00015 | 0.00000 | 63 |
| 186 | 05 | 05 | -22 | 22 | 0.00417 | 0.00003 | -0.00004 | 26 | | 800 | 21 | 15 | 05 | 14 | -0.00205 | 0.00004 | 0.00002 | 64 |
| 195 | 05 | 17 | -06 | 50 | -0.00247 | -0.00026 | -0.00006 | 49 | | 804 | 21 | 22 | 19 | 48 | -0.00075 | 0.00014 | 0.00001 | 54 |
| 208 | 05 | 34 | 09 | 29 | -0.00348 | -0.00045 | -0.00009 | 40 | | 806 | 21 | 26 | -22 | 24 | 0.00287 | 0.00022 | 0.00004 | 37 |
| 226 | 05 | 56 | -14 | 10 | -0.00029 | -0.00033 | -0.00006 | 77 | | 808 | 21 | 31 | -05 | 34 | -0.00423 | -0.00001 | 0.00002 | 39 |
| 232 | 06 | 07 | 14 | 46 | -0.00463 | -0.00088 | -0.00013 | 27 | | 812 | 21 | 40 | -16 | 39 | 0.00819 | 0.00027 | 0.00001 | 66 |
| 244 | 06 | 23 | 04 | 35 | -0.00051 | -0.00027 | -0.00007 | 58 | | 815 | 21 | 44 | 09 | 52 | -0.00827 | -0.00025 | -0.00003 | 42 |
| 246 | 06 | 27 | -04 | 45 | -0.00108 | -0.00018 | -0.00003 | 20 | | 823 | 21 | 53 | 25 | 55 | 0.00438 | 0.00021 | 0.00002 | 56 |
| 256 | 06 | 45 | 12 | 53 | -0.00432 | -0.00038 | -0.00006 | 42 | | 826 | 22 | 01 | 13 | 07 | -0.00268 | 0.00034 | 0.00006 | 40 |
| 266 | 06 | 54 | -12 | 02 | 0.00027 | -0.00016 | -0.00003 | 61 | | 827 | 22 | 05 | 00 | 19 | 0.00270 | 0.00026 | 0.00005 | 64 |
| 269 | 07 | 04 | 20 | 34 | 0.00011 | -0.00002 | -0.00001 | 57 | | 834 | 22 | 10 | 06 | 11 | 0.00070 | 0.00021 | 0.00004 | 46 |
| 279 | 07 | 20 | 21 | 58 | -0.00067 | -0.00005 | -0.00002 | 63 | | 840 | 22 | 16 | -07 | 46 | 0.01172 | 0.00036 | 0.00001 | 16 |
| 285 | 07 | 27 | 08 | 17 | -0.00341 | -0.00032 | -0.00005 | 66 | | 842 | 22 | 21 | -01 | 23 | 0.00297 | 0.00024 | 0.00002 | 79 |
| 288 | 07 | 34 | -22 | 17 | 0.00711 | 0.00036 | 0.00002 | 25 | | 850 | 22 | 35 | 00 | 07 | -0.00211 | -0.00014 | -0.00001 | 49 |
| 293 | 07 | 41 | -09 | 33 | -0.00414 | -0.00003 | -0.00001 | 65 | | 857 | 22 | 43 | 30 | 13 | 0.00275 | -0.00022 | -0.00005 | 23 |
| 305 | 08 | 03 | 27 | 47 | -0.00396 | 0.00015 | 0.00005 | 29 | | 862 | 22 | 49 | 24 | 36 | -0.00899 | -0.00031 | -0.00002 | 65 |
| 308 | 08 | 07 | -24 | 18 | 0.00697 | 0.00020 | 0.00000 | 45 | | 866 | 22 | 54 | -15 | 49 | 0.00233 | 0.00026 | 0.00004 | 27 |
| 311 | 08 | 13 | -15 | 47 | 0.00708 | 0.00015 | 0.00002 | 52 | | 870 | 23 | 03 | 28 | 04 | -0.00063 | 0.00003 | 0.00001 | 59 |
| | | | | | | Iteration | | | | | | | | | | Iteration | | |
| FK5 | RA | | Dec | | $1^{st}$ | $2^{nd}$ | $3^{rd}$ | | | FK5 | RA | | Dec | | $1^{st}$ | $2^{nd}$ | $3^{rd}$ | |
| No. | *h* | *m* | ° | ′ | s | s | s | n | | No. | *h* | *m* | ° | ′ | s | s | s | n |
| 316 | 08 | 25 | -03 | 54 | 0.00184 | 0.00005 | 0.00001 | 71 | | 873 | 23 | 09 | -21 | 10 | -0.00037 | -0.00013 | -0.00001 | 38 |
| 321 | 08 | 32 | 20 | 26 | 0.00541 | 0.00036 | 0.00005 | 52 | | 878 | 23 | 17 | 03 | 16 | -0.00317 | -0.00030 | -0.00002 | 42 |
| 326 | 08 | 44 | 18 | 09 | -0.00401 | -0.00025 | -0.00003 | 39 | | 880 | 23 | 20 | 23 | 44 | -0.00697 | -0.00036 | -0.00002 | 39 |
| 334 | 08 | 55 | 05 | 56 | -0.00156 | 0.00009 | 0.00003 | 100 | | 884 | 23 | 26 | 01 | 15 | 0.00260 | 0.00017 | 0.00003 | 51 |
| 347 | 09 | 14 | 02 | 18 | 0.00866 | 0.00055 | 0.00008 | 54 | | 888 | 23 | 35 | -07 | 27 | -0.00179 | -0.00009 | 0.00000 | 32 |
| 350 | 09 | 18 | 17 | 42 | 0.01265 | 0.00057 | 0.00005 | 65 | | 894 | 23 | 42 | -14 | 32 | 0.00221 | 0.00017 | 0.00003 | 32 |
| 365 | 09 | 41 | 09 | 53 | -0.00144 | 0.00017 | 0.00005 | 71 | | 898 | 23 | 52 | 19 | 07 | 0.00392 | 0.00030 | 0.00004 | 35 |
| 367 | 09 | 45 | 23 | 46 | 0.00079 | 0.00014 | 0.00004 | 66 | | 902 | 23 | 59 | 06 | 51 | -0.00516 | -0.00023 | -0.00001 | 49 |
| 370 | 09 | 51 | -04 | 14 | 0.00112 | 0.00011 | 0.00003 | 55 | | 905 | 00 | 03 | -17 | 20 | 0.00887 | 0.00065 | 0.00005 | 13 |
| 378 | 10 | 00 | 08 | 02 | -0.00050 | 0.00005 | 0.00003 | 81 | | 1022 | 00 | 53 | -01 | 08 | -0.00021 | 0.00008 | 0.00002 | 28 |
| 379 | 10 | 07 | 16 | 45 | -0.00023 | 0.00002 | 0.00002 | 87 | | 1032 | 01 | 11 | 21 | 02 | 0.00835 | 0.00041 | 0.00005 | 33 |
| 384 | 10 | 16 | 23 | 25 | 0.00343 | 0.00030 | 0.00005 | 24 | | 1055 | 02 | 04 | -29 | 17 | 0.00884 | 0.00083 | 0.00009 | 36 |
| 389 | 10 | 26 | -16 | 50 | 0.00506 | 0.00033 | 0.00004 | 71 | | 1056 | 02 | 10 | 19 | 30 | 0.00420 | 0.00005 | 0.00000 | 46 |
| 396 | 10 | 32 | 09 | 18 | 0.00560 | 0.00015 | 0.00001 | 77 | | 1091 | 03 | 15 | -08 | 49 | 0.00388 | 0.00004 | -0.00001 | 31 |
| 404 | 10 | 41 | -01 | 44 | -0.00051 | -0.00007 | 0.00000 | 62 | | 1106 | 03 | 53 | 17 | 19 | 0.00154 | -0.00014 | -0.00004 | 25 |
| 409 | 10 | 49 | 10 | 32 | -0.00615 | -0.00013 | 0.00000 | 80 | | 1144 | 05 | 12 | -16 | 12 | 0.00291 | 0.00019 | -0.00003 | 50 |
| 418 | 11 | 05 | 07 | 20 | -0.00221 | 0.00002 | 0.00001 | 65 | | 1148 | 05 | 27 | 17 | 57 | 0.00021 | -0.00013 | -0.00005 | 44 |
| 422 | 11 | 14 | 20 | 31 | 0.00493 | 0.00019 | 0.00001 | 52 | | 1153 | 05 | 38 | -27 | 12 | -0.01077 | -0.00072 | -0.00013 | 6 |
| 426 | 11 | 19 | -14 | 46 | 0.00335 | 0.00005 | 0.00000 | 40 | | 1155 | 05 | 48 | -04 | 05 | 0.00075 | -0.00021 | -0.00005 | 60 |
| 437 | 11 | 36 | 00 | 49 | 0.00097 | -0.00003 | 0.00000 | 61 | | 1163 | 06 | 04 | 23 | 15 | -0.00250 | -0.00026 | -0.00005 | 76 |
| 445 | 11 | 50 | 01 | 45 | -0.00231 | -0.00016 | -0.00001 | 83 | | 1169 | 06 | 16 | 12 | 16 | -0.00644 | -0.00077 | -0.00013 | 46 |
| 450 | 12 | 05 | 08 | 43 | -0.00629 | -0.00041 | -0.00006 | 34 | | 1175 | 06 | 33 | -01 | 13 | 0.00589 | -0.00010 | -0.00006 | 54 |
| 453 | 12 | 10 | -22 | 37 | 0.00464 | 0.00019 | 0.00001 | 61 | | 1181 | 07 | 00 | -08 | 24 | 0.00481 | 0.00013 | -0.00004 | 44 |
| 457 | 12 | 15 | -17 | 32 | 0.00572 | 0.00022 | 0.00001 | 36 | | 1187 | 07 | 11 | 00 | 29 | -0.00322 | -0.00014 | -0.00003 | 44 |
| 460 | 12 | 19 | 00 | 40 | -0.00479 | -0.00016 | -0.00001 | 65 | | 1201 | 07 | 46 | 10 | 46 | -0.00744 | -0.00017 | 0.00000 | 45 |
| 466 | 12 | 29 | 20 | 53 | 0.00237 | -0.00027 | -0.00005 | 32 | | 1207 | 07 | 53 | 26 | 45 | 0.00312 | 0.00023 | 0.00001 | 70 |
| 471 | 12 | 34 | -23 | 23 | 0.00770 | 0.00031 | 0.00003 | 25 | | 1238 | 09 | 07 | 10 | 40 | -0.00620 | -0.00006 | 0.00004 | 60 |
| 475 | 12 | 39 | -07 | 59 | -0.00657 | -0.00025 | 0.00000 | 48 | | 1245 | 09 | 25 | -05 | 07 | 0.00252 | 0.00021 | 0.00003 | 52 |
| 484 | 12 | 55 | 03 | 23 | -0.00353 | -0.00028 | -0.00003 | 61 | | 1246 | 09 | 31 | 11 | 18 | -0.00863 | -0.00016 | 0.00003 | 64 |
| 488 | 13 | 02 | 10 | 57 | -0.00975 | -0.00023 | -0.00001 | 75 | | 1284 | 11 | 00 | 03 | 37 | -0.00079 | 0.00004 | 0.00001 | 70 |
| 490 | 13 | 09 | -05 | 32 | -0.00128 | -0.00023 | -0.00002 | 62 | | 1297 | 11 | 27 | 02 | 51 | -0.00174 | -0.00001 | 0.00001 | 82 |
| 495 | 13 | 18 | -23 | 10 | 0.00482 | 0.00022 | 0.00003 | 40 | | 1301 | 11 | 44 | -18 | 21 | -0.00322 | -0.00005 | 0.00000 | 57 |
| 501 | 13 | 34 | 00 | 35 | -0.00152 | 0.00015 | 0.00002 | 64 | | 1311 | 12 | 00 | 06 | 36 | 0.00209 | 0.00004 | -0.00001 | 71 |
| 507 | 13 | 47 | 17 | 27 | -0.00048 | -0.00012 | -0.00001 | 66 | | 1348 | 13 | 26 | -12 | 42 | 0.00013 | -0.00007 | -0.00001 | 60 |
| 513 | 13 | 54 | 18 | 23 | 0.00130 | -0.00008 | -0.00001 | 61 | | 1355 | 13 | 41 | -08 | 42 | 0.00389 | 0.00000 | -0.00001 | 70 |
| 516 | 14 | 01 | 01 | 32 | -0.00232 | 0.00004 | 0.00001 | 63 | | 1372 | 14 | 19 | 13 | 00 | 0.00118 | 0.00016 | 0.00002 | 52 |
| 519 | 14 | 06 | -26 | 40 | 0.00871 | 0.00054 | 0.00005 | 54 | | 1381 | 14 | 37 | -12 | 18 | -0.00290 | 0.00023 | 0.00002 | 56 |
| 523 | 14 | 12 | -10 | 16 | 0.00143 | 0.00010 | 0.00002 | 32 | | 1387 | 14 | 50 | -15 | 59 | -0.00295 | 0.00000 | 0.00004 | 37 |
| 533 | 14 | 28 | -02 | 13 | -0.00011 | -0.00018 | -0.00001 | 50 | | 1430 | 16 | 29 | -14 | 33 | -0.00034 | 0.00002 | -0.00001 | 34 |
| 545 | 14 | 43 | -05 | 39 | 0.00466 | 0.00025 | 0.00003 | 54 | | 1459 | 17 | 26 | 04 | 08 | 0.00347 | 0.00017 | 0.00002 | 21 |
| 547 | 14 | 46 | 01 | 53 | -0.00114 | 0.00005 | 0.00000 | 46 | | 1461 | 17 | 34 | -11 | 14 | -0.00327 | -0.00007 | 0.00000 | 71 |
| 551 | 14 | 56 | 14 | 26 | 0.00164 | 0.00022 | 0.00002 | 58 | | 1484 | 18 | 36 | 09 | 07 | 0.00036 | 0.00004 | -0.00001 | 55 |
| 557 | 15 | 04 | 26 | 56 | -0.00160 | -0.00018 | -0.00001 | 79 | | 1495 | 18 | 55 | -16 | 22 | 0.00430 | 0.00026 | 0.00000 | 40 |
| 559 | 15 | 12 | -19 | 47 | 0.00983 | 0.00053 | 0.00004 | 48 | | 1533 | 20 | 29 | -02 | 53 | -0.00100 | -0.00038 | -0.00006 | 24 |
| 564 | 15 | 17 | -09 | 22 | 0.00190 | 0.00044 | 0.00004 | 60 | | 1543 | 20 | 47 | -05 | 01 | -0.00163 | -0.00016 | -0.00005 | 47 |
| 570 | 15 | 25 | 15 | 25 | -0.00497 | -0.00016 | 0.00000 | 75 | | 1552 | 21 | 05 | -17 | 13 | 0.00372 | 0.00019 | 0.00001 | 48 |
| 577 | 15 | 35 | -14 | 47 | 0.00798 | 0.00054 | 0.00005 | 61 | | 1591 | 22 | 30 | -10 | 40 | -0.00422 | 0.00011 | 0.00006 | 10 |
| 582 | 15 | 44 | 06 | 25 | -0.00958 | -0.00051 | -0.00003 | 77 | | | | | | | | | | |

The columns labeled 1st, 2nd and 3rd are the results of, respectively, the 1st, 2nd and 3rd iterations. The final correction adopted for each clock star was the sum over all iterations of the mean residuals. Thus the final correction for star number 0007 was:

$$1st_iteration + 2nd_iteration + 3rd_iteration = final_correction$$

$$0^{s}.00047 + 0^{s}.00011 + 0^{s}.00001 = 0^{s}.00059$$

The condition that the total sum of the final corrections of the all the clock stars be zero was imposed also. The final clock corrections adopted for each tour used the clock stars' $(O - C)_\alpha$'s with the individual corrections tabulated above and the periodic corrections and corrections for observer bias applied. However these corrections were only used in the formation of the clock corrections and not applied when forming the final observed mean catalog positions of the individual clock stars.

**Clamp and Circle Differences** - When the $((O - C)_\alpha - CC)$'s were grouped in zones of zenith distance 5° wide and averaged by Clamp, the resulting quantities were not equal but instead indicated a systematic effect that was Clamp dependent. One half of the difference of these quantities is called the Clamp difference and was applied to the observations to remove this dependence. By similarly grouping the data and averaging by Circle, after having applied the Clamp differences, a dependence with Circle was seen. These quantities are tabulated in Table 10.

If a star had an equal number of observations on each Clamp and on each Circle these systematic errors were satisfactorily removed. Hence a concerted effort was made to ensure that, for most stars, this desirable distribution of observations was realized. The final mean observed catalog position includes the application of the Clamp and Circle differences.

Table 10: Clamp and Circle differences
Units in milliseconds of time × cos (δ)

| Zenith Distance (degrees) | Clamp Difference | | Circle Difference |
|---|---|---|---|
| | Circle One 1/2(E-W) | Circle Two 1/2(E-W) | 1/2(C1-C2) |
| 78.5 to 73.5 | 1.2 | 6.7 | 3.9 |
| 73.5 to 68.5 | 3.8 | 5.0 | 4.2 |
| 68.5 to 63.5 | 0.0 | 1.3 | 3.8 |
| 63.5 to 58.5 | 0.4 | 4.2 | 1.0 |
| 58.5 to 53.5 | -1.0 | 4.0 | 0.0 |
| 53.5 to 48.5 | -0.6 | -0.5 | -0.2 |
| 48.5 to 43.5 | 0.1 | -1.9 | -1.5 |
| 43.5 to 38.5 | 1.3 | 0.0 | 0.1 |
| 38.5 to 33.5 | 0.2 | 3.3 | 1.2 |
| 33.5 to 28.5 | 1.5 | -0.9 | 3.6 |
| 28.5 to 23.5 | 2.1 | -2.7 | 3.4 |
| 23.5 to 18.5 | -0.5 | -1.4 | 1.8 |
| 18.5 to 13.5 | -4.2 | -3.3 | 0.4 |
| 13.5 to 8.5 | -5.0 | -3.9 | -0.3 |
| 8.5 to 3.5 | -4.9 | -5.8 | 0.0 |
| 3.5 to -1.5 | -7.9 | -4.8 | 0.3 |
| -1.5 to -6.5 | -8.9 | -5.2 | -0.1 |
| -6.5 to -11.5 | -6.2 | -6.0 | 0.3 |
| -11.4 to -16.5 | -4.2 | -2.8 | -0.1 |
| -16.5 to -21.5 | -2.1 | -2.7 | 0.2 |
| -21.5 to -26.5 | -0.8 | -0.4 | -0.1 |
| -26.5 to -31.5 | -0.3 | 0.8 | -0.5 |
| -31.5 to -36.5 | 0.7 | 1.9 | -0.2 |
| -36.5 to -41.5 | 1.0 | 2.9 | 0.3 |
| -41.5 to -46.5 | 1.8 | 2.0 | 0.4 |
| -46.5 to -51.5 | 2.2 | 1.8 | 0.3 |
| -51.5 to -56.5 | 0.6 | -1.4 | -0.4 |
| -56.5 to -61.5 | -1.1 | -0.4 | 0.0 |
| -61.5 to -66.5 | -1.0 | -3.6 | -0.3 |
| -66.5 to -71.5 | -1.3 | -1.5 | -1.4 |

**Reductions (General)** - As an object was tracked through the field of view, the readings of the position of the micrometer screw were recorded at 4 second intervals along with the sidereal time. These data were reduced to produce the time of meridian transit of the object by the following algorithm (Chauvenet 1960):

$$Time_of_transit = ST + \tau$$

where: $ST$ = mean (average) sidereal time of the micrometer screw readings.

$\tau$ = hour angle of star east of meridian

and:

$$\sin(\tau - m) = (\tan n)(\tan \delta) + (\sin c)(\sec n)(\sec \delta)$$

$\delta$ = apparent declination of object

$c$ = collimation constant

$c = [(coll - R)(MicroScale)] - [(aberration)(\cos \phi)]$

*coll* = collimation point (average for a tour)

$R$ = mean (average) micrometer screw reading

*MicroScale* = micrometer scale = 3.8121 revolutions/sec

*aberration* = constant of diurnal aberration = 0.0213sec

$\phi$ = latitude of instrument

$m,n$ = Bessel's instrumental constants which are related by small angle approximations to $b$, the level of telescope interpolated to the time of the observation (i.e *ST*), and $a$, the azimuth of telescope also interpolated to *ST*, by the relations:

$$m = b \cos \phi + a \sin \phi$$

$$n = b \sin \phi + a \cos \phi$$

The sidereal time was kept by a clock system constituting of two rubidium, or late in the program, two cesium oscillators (see section on clocks). The mean time of the observation, *ST*, was corrected for the effects of the variation of longitude and the equation of the equinoxes. The variation in longitude was derived from 5 day values of the $x$ and $y$ positions of the pole as disseminated by the Bureau International de l'Heure (BIH) (McCarthy 1984). A cubic spline was used to interpolate the variation to the time of the beginning of a tour.

$$\Delta\alpha = 0\overset{s}{.}0213\ (\rho/r_{eq})(\cos \phi)(\sec \delta)$$

The correction in right ascension for diurnal aberration, above, is given by Explanatory Supplement to the Astronomical Almanac (1992, pg. 132, equation 3.254-3):

where: $\rho$ = geocentric distance at latitude of the observer

$r_{eq}$ = equatorial radius

$\phi$ = latitude of the observer

For the purposes of this calculation $\rho/r_{eq}$ was set equal to one.

The computed apparent place was subtracted from the corrected time of transit to produce an observed minus computed value, $(O - C)_\alpha$. The average of the $(O - C)_\alpha$ 's of the clock stars was subtracted from each time of transit. This average $(O - C)_\alpha$, the clock correction, was determined uniquely for each tour (see Clock Corrections). Most of the detailed analysis of the data that was carried out in the preparation of this catalog involved this latter quantity, the $(O - C)_\alpha$ minus the clock correction, $((O - C)_\alpha - CC)$. A final step in the process was the adding of the mean $((O - C)_\alpha - CC)$ of a star to the mean place calculated for the mean epoch of observation on the equinox of J2000 (proper motion, and orbital motion applied) to produce a new mean position which is regarded as the catalog observed position.

**Fundamental Azimuth Solution** - The azimuth of the telescope was determined periodically during a tour, day or night, relative to the marks. The celestial azimuth of these marks was determined by observing transits at the upper and lower culminations of a list of 18 circumpolar FK5 stars, all within $10^{\circ}\!.5$ of the pole. These stars could be observed at both transits within a short period of time, usually within 12 hours, from September through March. Therefore, combining $(O - C)_{\alpha}$ 's of upper and lower culminations during these months resulted in quantities that are independent of their apparent positions computed from catalog tabulations, and in this sense are fundamental.

The azimuth of the instrument with respect to north was given by the observation of a pair of stars, one of which was an azimuth star, and was calculated from the following:

$$a = [T' - T - A(k' - k)] / (A' - A)$$

where: $a$ = azimuth of the instrument with respect to north
$T = \alpha - ST - c \sec\delta - b \sec\delta \cos z$
$ST$ = mean (average) sidereal time of the micrometer screw readings.
$\alpha$ = apparent right ascension
$\delta$ = apparent declination
$z$ = zenith distance $= \phi - \delta$ (above pole)
$= \delta + \phi + 180°$ (below pole)
$\phi$ = latitude of instrument
$A = \sin z \sec\delta$
$k$ = azimuth of the instrument with respect to the marks

Primed quantities relate to the azimuth star, unprimed to the other star.

It was assumed that during the time between the observations of the two stars that the rate of change of the azimuth of the marks and the clock rate was zero. The usual practice, to ensure that these assumptions held true and to maximize the denominator, was to pair the observation of an azimuth star with that of a clock star, the transit times of which were not greatly separated but which were separated in zenith distance by approximately 90°. However because the Six-inch has historically had very stable marks and with use of the rubidium clocks, it was felt that a slightly modified approach could be used which would give more accurate results and also be more amenable to automated reduction. The new method replaced those quantities that refer to a single clock star with quantities averaged over all clock stars in the tour. The azimuth of the marks, $a_m$, can be inferred from $a$ because $a = k + a_m$. As mentioned above, during the fundamental azimuth season the same azimuth star can be observed at upper and lower culmination within a short period of time and the combination of such observations results in an $a$ that is independent of the computed apparent position (to the extent that the effect of proper motion is small) and is said to be fundamental as are the inferred $a_m$'s $a_m$'s. In this program, the $a_m$'s determined from upper and lower culmination observations of the same azimuth star were combined over a period of a Clamp, about 30 days. The months during which these fundamental azimuth observations were made were called fundamental azimuth seasons.

The $a_m$'s determined during the non-fundamental azimuth season were simply combined into a grand mean averaged over all azimuth stars over a Clamp. The $a_m$'s from both the fundamental and non-fundamental seasons were then collected, arranged in time order, and a cubic spline fit through the weighted points. The $a_m$'s from the non-fundamental season were weighted by the total number of azimuth star observations going into the mean. During the fundamental season each azimuth star produces its own $a_m$ which was assigned a weight of

$$weight = (2v_u v_l)/(v_u + v_l)$$

where: $v_u$ = number of upper culmination observations
$v_l$ = number of lower culmination observations

The cubic spline was used to interpolate $a_m$ to the time of a mark. The interpolated $a_m$ was combined with the mark's $k$ to give the azimuth $a$ of the instrument at the time of the mark. This azimuth was then linearly interpolated to the time of an observation and applied in the reduction of that observation.

Utilizing observations taken during fundamental azimuth season, an iterative procedure was employed to improve the adopted initial catalog positions of the azimuth stars. After the azimuth of the telescope was determined by the preceding procedure, $(O\text{-}C)_\alpha$'s were calculated for the clock and azimuth stars and the mean $(O\text{-}C)_\alpha$'s of the clock stars, $\overline{CC}$, in the tour was subtracted from the $(O\text{-}C)_\alpha$'s of the azimuth stars in that tour to form $((O\text{-}C)_\alpha - \overline{CC})$. The $((O\text{-}C)_\alpha - \overline{CC})$'s of the azimuth stars were collected by star and separated by culmination. A difference of the upper and lower culmination mean $((O\text{-}C)_\alpha - \overline{CC})$'s was formed for each star. The adopted position of each azimuth star was then corrected by half of this difference and the $a_m$'s re-determined following the previously described procedures. The process of correcting the adopted positions of the azimuth stars was repeated a total of two times. The corrections applied are given in the following table. These corrections (shown in Table 11) were not applied in the determination of the final mean observed positions of the azimuth stars.

The third correction of $-0\overset{s}{.}0016$, which was the mean of the all the upper minus lower $((O - C)_\alpha - \overline{CC})$'s, was applied to all the azimuth stars before the final pass to determine the azimuth of the marks, $a_m$. During this last pass the $a_m$'s for all periods, non-fundamental as well as fundamental, were determined. After corrections to the clock stars were estimated (see section on Clock Corrections) the azimuth reductions as described above were revisited and the resulting $a_m$ values, the final values adopted, are plotted in Figures 2 and 3.

Table 11: Corrections to Azimuth Star Positions Derived from Fundamental Azimuth Solutions
(UC = upper culmination and LC = lower culmination)

| FK5 | RA | | Dec | | first iteration | second iteration | UC | LC |
|---|---|---|---|---|---|---|---|---|
| no. | h | m | ° | ′ | s | s | n | n |
| 248 | 6 | 46 | 79 | 34 | 0.035 | 0.020 | 73 | 37 |
| 906 | 1 | 8 | 86 | 15 | 0.111 | 0.058 | 85 | 66 |
| 907 | 2 | 31 | 89 | 15 | 0.277 | 0.111 | 113 | 82 |
| 909 | 7 | 40 | 87 | 1 | 0.054 | 0.077 | 93 | 66 |
| 910 | 9 | 37 | 81 | 19 | 0.005 | 0.004 | 91 | 87 |
| 911 | 10 | 31 | 82 | 33 | 0.022 | 0.012 | 79 | 85 |
| 912 | 16 | 45 | 82 | 2 | 0.114 | 0.011 | 57 | 88 |
| 913 | 17 | 32 | 86 | 35 | 0.149 | 0.025 | 77 | 95 |
| 915 | 20 | 42 | 82 | 31 | 0.049 | 0.051 | 70 | 79 |
| 1636 | 3 | 32 | 84 | 54 | 0.027 | 0.021 | 93 | 62 |
| 1637 | 5 | 31 | 85 | 56 | 0.137 | 0.017 | 79 | 33 |
| 1640 | 9 | 15 | 84 | 10 | 0.103 | 0.005 | 79 | 61 |
| 1642 | 12 | 4 | 85 | 35 | 0.108 | 0.091 | 71 | 86 |
| 1644 | 14 | 50 | 82 | 30 | 0.043 | 0.008 | 62 | 98 |
| 1646 | 18 | 24 | 83 | 10 | 0.077 | 0.048 | 31 | 64 |
| 1647 | 19 | 59 | 84 | 40 | 0.065 | 0.049 | 56 | 71 |
| 1648 | 22 | 13 | 86 | 6 | 0.010 | 0.041 | 90 | 74 |
| 1649 | 22 | 54 | 84 | 20 | 0.086 | 0.021 | 84 | 80 |

The break in the plot at 1977.80 was caused by a shift in the north mark lens. The lens was set back to its original position shortly after the shift was discovered, leaving approximately a two week period with an abnormal $a_m$.

Figures 2 and 3: Azimuth corrections

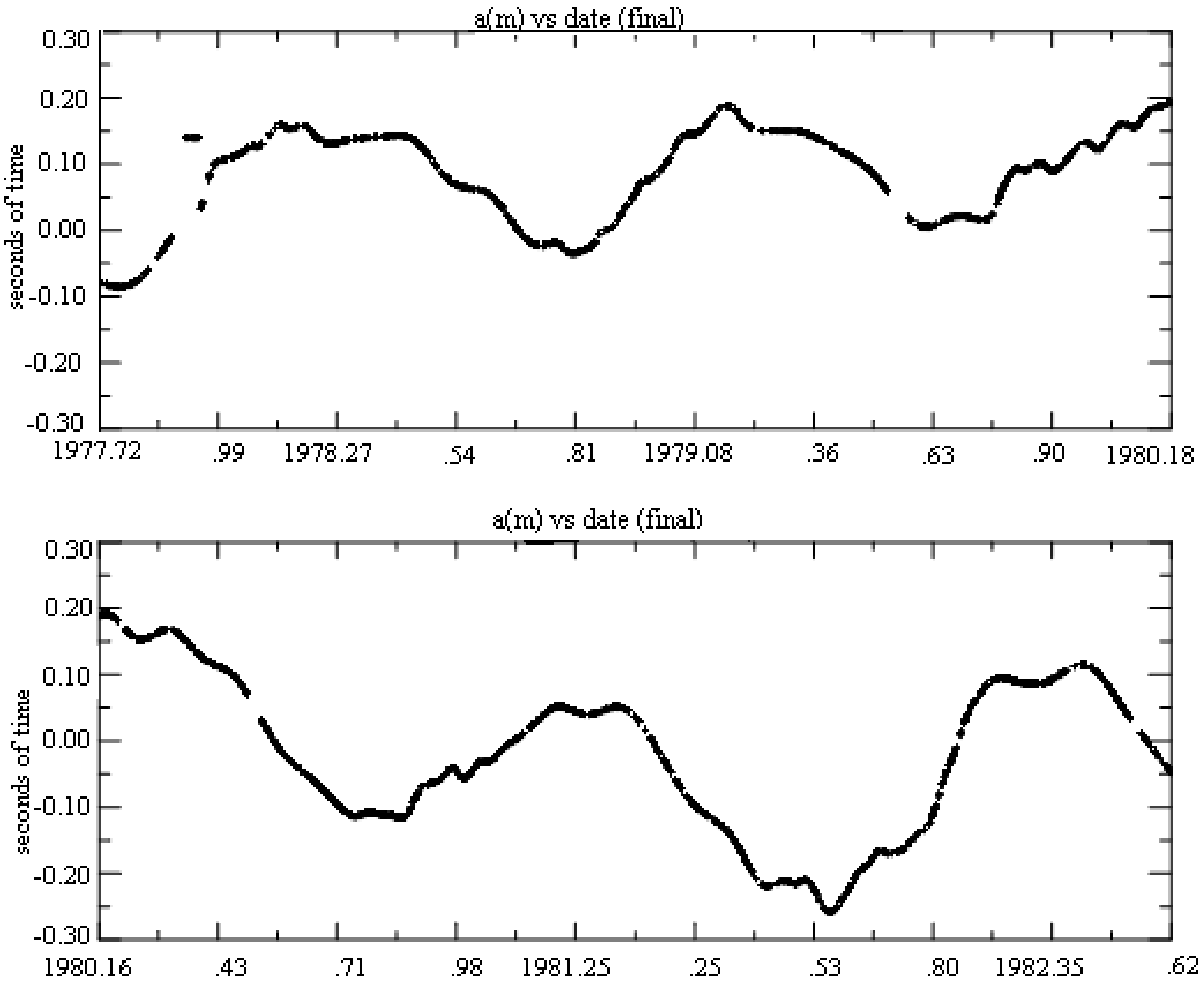

# Declinations

The declinations of objects observed with the transit circle are determined from micrometer measures of the position of an object in the field of view, the pointing of the telescope with respect to the nadir, and the assumed latitude for the telescope's location. Various corrections are applied in this determination of the observed declination positions and are described in the following sections.

**Micrometer Screw Corrections** - The scale of the declination screw was determined by comparing the micrometer measures of the same star observed with the telescope set at slightly different zenith distances. The differences of the setting of the instrument were measured by the circle. By selecting stars covering a wide range of zenith distances and observations from both clamps, the effect of systematic errors in the diameter corrections of the circle could be mitigated. The scale of the declination screw was found to be 0°.00793627/*rev* ± 0.00000103. The declination screw was the same one used during the $W5_{50}$ program (Hughes and Scot, 1982), and the slightly different screw scale used during this program can be attributed to the new objective lens.

Periodic and progressive screw errors were determined at the beginning and end of the program. No significant periodic error (above 0″.01) was found. The progressive screw error was found to vary by 0″.04 over the range of the screw and corrections to remove this error were applied to all micrometer measurements. The mean absolute value of the screw error was 0.″008, with a maximum of 0.″041. Observations of each star on both Clamps resulted in measures taken at different settings of the screw which reduced any systematic error in the screw corrections. The screw corrections were normalized to zero at the center of the screw, where the measures of the nadir and flexure were made.

Zenith distances were obtained by adding the measure made with the declination micrometer screw to the readings of the graduated circle made by photoelectrically scanning the circle through six microscopes. Nadir measurements, obtained by auto-collimation over a mercury basin at the same time as the level measurements, consisted of eight or more settings of the micrometer screw on the reflections of the micrometer wires from the mercury, and were accompanied by photoelectric scans of the graduated circle. A study was done to look for tidal effects on the vertical, but none were found.

**Inclination Corrections** - An inclination correction was applied to the micrometer measures to account for the fact that the micrometer wires were not exactly perpendicular to the right ascension micrometer wires, which are adjusted to be parallel to the rotation plane of the instrument. The inclination correction was determined from star observations. The mean, absolute, value of the correction for the inclination was 0.″036, with a maximum correction of 1.″470. The inclination correction increases as a function of the mean declination screw reading measured from the center of the field.

**Reduction to the Meridian Corrections** - A correction to account for the curvature of the path an object follows in the focal plane of the transit circle as it traverses the field of view was necessary to reference all the declination measurements to the meridian. This correction, referred to as the reduction to the meridian, was given by the equation:

$$RM = (ClpSign)\ (0.000192)\ [\tan(EqD)]\ (R - R_c)^2$$

where: $ClpSign$ = + for Clamp West and – for Clamp East

$EqD$ = equatorial distance, which is equal to the declination, $\delta$, for the upper culmination observations and to ($\delta$ - 180°)for the lower culmination

$R$ = right ascension screw reading of the declination measure

$R_c$ = right ascension screw reading of the collimation point, which defines the instrumental meridian

The mean, absolute, value of the correction for the reduction to the meridian was 0.″083, with a maximum correction of 12.″808 for a star at a declination of 85.°9 and observed 9.7 minutes after crossing the meridian.

**Circle Diameter Corrections** - The circle used was a glass annulus inscribed with lines etched 0.°05 apart and mounted on a steel wheel which is attached to the rotation axis of the transit circle. Slight deviations in the spacing of the lines made it necessary to determine a quantity known as diameter corrections. These corrections, which were derived using a method developed by Erik Høg (1961), were measured numerous times during the program. The corrections were found to have changed after the wheel, on which the circle is mounted, was rotated in 1980. Therefore, two different sets of diameter corrections were applied. The mean, absolute, value of a correction was 0.″358 with a maximum of 1.″962. The diameter corrections vary with the setting of the transit circle, and reversing the instrument monthly and rotating the circle at the middle of the observing program reduced systematic effects in the diameter corrections.

**Personal Equation** - A dove prism mounted in the eyepiece allowed the field to be rotated by 180°. Only observations which were balanced between the two orientations of the prism were kept; so no correction was necessary for a personal bias of the observer.

**Assumed Latitude and Variation of Latitude Corrections** - The observed zenith distances were reduced to preliminary declinations with an assumed latitude of the instrument of +38° 55′ 14.″257 (Hughes *et. al.* 1975). All declination observations were corrected for variation of latitude, using the pole coordinates (*x*, *y*) from the Bureau International de l'Heure (BIH). This is a change from past Six-inch Transit Circle catalogs which used the results from a co-located photographic zenith tube (PZT). An earlier study (Rafferty 1980) indicated that this practice risked introducing PZT instrumental errors into the transit circle catalog. The study showed that a global solution for the pole, such as that provided by the BIH, was less likely to be affected by the bias of a single instrument and hence more free of systematic error. The variation of the latitude was calculated using the equation:

$$\Delta\phi = x \cos\lambda - y \sin\lambda$$

Where: $\Delta\phi$ = variation of latitude referred to the Conventional International Origin
$\lambda$ = astronomical longitude measured west from the Greenwich meridian
(for this catalog $\lambda = 77^{\circ}\, 04'\, 19''\!.0W$)
$x,y$ = BIH coordinates of the pole

The BIH provides the pole coordinates at five-day intervals. They were interpolated to the time of observation by a cubic spline and added to the assumed latitude. A summary of the variation of latitudes used is given in Table 12. The mean, absolute, value of the variation applied to star observations was 0.″276, with a maximum of 0.″478.

Table 12 - Variation of Latitude
units in second of arc

| Fraction of year | 1977 | 1978 | 1979 | 1980 | 1981 | 1982 |
|---|---|---|---|---|---|---|
| 0.00 | | 0.020 | 0.110 | 0.275 | 0.365 | 0.346 |
| 0.05 | | 0.016 | 0.083 | 0.241 | 0.353 | 0.382 |
| 0.10 | | 0.037 | 0.064 | 0.210 | 0.340 | 0.408 |
| 0.15 | | 0.070 | 0.061 | 0.190 | 0.326 | 0.431 |
| 0.20 | | 0.118 | 0.075 | 0.184 | 0.314 | 0.442 |
| 0.25 | | 0.184 | 0.106 | 0.188 | 0.301 | 0.442 |
| 0.30 | | 0.255 | 0.145 | 0.199 | 0.281 | 0.431 |
| 0.35 | | 0.322 | 0.192 | 0.217 | 0.268 | 0.409 |
| 0.40 | | 0.376 | 0.233 | 0.235 | 0.252 | 0.371 |
| 0.45 | | 0.420 | 0.277 | 0.250 | 0.237 | 0.322 |
| 0.50 | | 0.459 | 0.318 | 0.264 | 0.217 | 0.277 |
| 0.55 | | 0.477 | 0.355 | 0.279 | 0.198 | 0.225 |
| 0.60 | | 0.480 | 0.386 | 0.290 | 0.185 | 0.175 |
| 0.65 | | 0.465 | 0.411 | 0.302 | 0.180 | |
| 0.70 | 0.354 | 0.436 | 0.421 | 0.313 | 0.181 | |
| 0.75 | 0.289 | 0.399 | 0.418 | 0.331 | 0.188 | |
| 0.80 | 0.215 | 0.347 | 0.407 | 0.346 | 0.207 | |
| 0.85 | 0.150 | 0.282 | 0.384 | 0.356 | 0.233 | |
| 0.90 | 0.096 | 0.213 | 0.348 | 0.366 | 0.263 | |
| 0.95 | 0.051 | 0.153 | 0.310 | 0.370 | 0.303 | |

**Refraction Corrections** – The refraction corrections were calculated using the same theoretical development as described in the fourth edition of the *Pulkovo Refraction Tables* by Gylden (1956) (also see the introduction to the $WL_{50}$, (Hughes *et al.* 1992). The refraction correction was computed for each observation from the equation:

$$R = (g_s / g_p)\psi \tan z[(1 + mt_o)/(1 + mt)]^{\lambda}[(b/b_o)(1 + \beta_1 t)/(1 + \beta t)]^{A}(1 + q\pi_o)/(1 + q\pi)$$

where:

$g_s / g_p = 1 - 0.001286 - 0.002634\cos 2\phi + 0.000006\cos^2\phi - 0.0000002h$
$h$ = elevation above sea level in meters (88.091m for USNO Six-inch Transit Circle)
$\phi$ = latitude of instrument (+38.°920626945 degrees for USNO Six-inch Transit Circle)
$g_s / g_p$ = 0.998145238 (gravity correction for the USNO Six-inch Transit Circle)
$\psi = \psi_o(1-\Delta\mu)$
$\psi_o$ = 57.5834 (Constant of Refraction)
$(1 - \Delta\mu) = 1 - 0.138618c - 0.224518c^2 - 0.236421c^3 - 0.188840c^4 - 0.118622c^5$

$\lambda = 1.000427 + 0.171936c + 0.250075c^2 + 0.210882c^3 + 0.110539c^4$
$A = 1.000427 + 0.016704c + 0.025601c^2 + 0.023603c^3$
$c = \tan^2(\zeta/2)$
$\tan\zeta = 0.182574 \tan z$
$z$ = observed zenith distance
$m$ = 0.003670, the coefficient of expansion of air
$t_o = 9.31\ C$
$t$ = air temperature in degrees C
$b$ = air pressure in mm
$b_0$ = 751.51 mm
$\beta_1$ = 0.0000184, coefficient of expansion of glass
$\beta$ = 0.0001818, coefficient of expansion of mercury
$\tau$ = temperature of mercury in the barometer in degrees C
$q$ = +0.0002084, factor by which water vapor modifies the index of refraction of dry air
$\pi_o$ = 5.5mm
$\pi$ = partial pressure of water vaper in air given by the equation:
$\pi = 0.95594411 + 0.050121975dp + 0.001417845dp^2 + 0.000173725dp^3 - 0.00000012dp^4$
$dp$ = dew point in degrees F

The correction of refraction was a function of the environmental data and the tangent of the zenith distance, with the mean, absolute, correction applied to the observations was 50″.073, with a maximum value of 318″.846.

**Clamp and Circle Differences** - Differences between an individual star's $(O - C)_\delta$'s made on Clamps East and West were collected by zones of five-degrees declination and a mean was produced for each zone. These are termed Clamp differences. After applying the corrections for the mean Clamp differences, similar differences were developed between observations made before and after the rotation of the circle in 1980. Only FK5 stars were used to determine these values to ensure a large number of observations per star. The resulting quantities are listed in Table 13. The corrections were added to observations made on Clamp West and Circle 2, and subtracted from observations made on Clamp East and Circle 1. The mean, absolute value of the correction for the Clamp differences was 0″.043 with a maximum of 0″.119. The mean, absolute value of the correction for the Circle difference was 0″.032 with a maximum of 0″.093. Both corrections are a function of the pointing of the telescope. To diminish systematic effects in these corrections, every effort was made to equalize the number of observations for each star by clamp and by circle.

**Tour Offsets** - In the reductions for the $WL_{50}$ (Hughes *et al.* 1992), an offset was applied to each observing tour based on the $(O - C)_\delta$'s of the upper culmination observations of FK5 stars. Such tour offsets based on FK5 stars were tested for this program and found to be unnecessary.

Table 13: Corrections for the Clamp and Circle Difference, Declination
units in seconds of arc
Declination in units of degrees
sp = sub-pole

| Declination | Clamp Difference | | Circle Difference |
|---|---|---|---|
| | Circle One | Circle Two | |
| +61 to +65 sp | 0.000 | -0.054 | 0.076 |
| +65 to +70 sp | 0.000 | -0.029 | 0.094 |
| +70 to +75 sp | -0.032 | -0.047 | 0.068 |
| +80 to +85 sp | -0.007 | -0.022 | 0.047 |
| +85 to +90 sp | -0.011 | -0.043 | 0.050 |
| +90 to +85 | 0.022 | -0.054 | 0.036 |
| +85 to +80 | -0.022 | -0.065 | 0.054 |
| +80 to +75 | 0.018 | -0.061 | 0.022 |
| +75 to +70 | -0.050 | -0.058 | 0.018 |
| +70 to +65 | -0.047 | -0.094 | 0.025 |
| +65 to +60 | -0.032 | 0.014 | -0.011 |
| +60 to +55 | -0.029 | -0.025 | 0.000 |
| +55 to +50 | -0.061 | -0.036 | 0.014 |
| +50 to +45 | -0.061 | -0.029 | 0.050 |
| +45 to +40 | -0.065 | -0.011 | 0.022 |
| +40 to +35 | -0.061 | -0.029 | 0.007 |
| +35 to +30 | -0.047 | -0.047 | 0.022 |
| +30 to +25 | -0.068 | 0.000 | 0.022 |
| +25 to +20 | -0.058 | -0.043 | 0.022 |
| +20 to +15 | -0.058 | -0.014 | 0.029 |
| +15 to +10 | -0.047 | 0.000 | 0.040 |
| +10 to +05 | -0.036 | -0.040 | 0.054 |
| +05 to 00 | -0.050 | -0.076 | 0.040 |
| 00 to -05 | -0.036 | -0.061 | 0.029 |
| -05 to -10 | -0.007 | -0.076 | 0.047 |
| -10 to -15 | -0.036 | -0.094 | 0.032 |
| -15 to -20 | -0.026 | -0.042 | 0.029 |
| -20 to -25 | -0.036 | -0.083 | 0.036 |
| -25 to -30 | -0.004 | -0.040 | 0.000 |
| south of -30 | -0.032 | -0.119 | -0.018 |

**Flexure Corrections** - It was common practice to determine the flexure of a transit circle by use of a pair of horizontal collimator telescopes located in the meridian, one to the north and the other to the south of the main transit telescope. After the cross wires in the collimators were aligned, measures were taken on them with the micrometer of the transit circle. The measures in the horizontal direction produced the collimation point which defines the instrumental meridian while those in the vertical direction determined the flexure. Typically, the resulting flexure showed a very large scatter; in this program the nighttime flexures had a mean of 0.″6624 ± 0.2520. Though the flexure determined from the collimators has been known to be poorly determined, the value was necessary when using the circumpolar stars to solve for the correction to the constant of refraction and the correction to the assumed latitude. A further adjustment of this flexure was found to be necessary using solar system observations (to be described later). Various theories have been purposed to explain why the flexure determined from the collimators were so poorly determined, probably the most reasonable postulates that temperature gradients in the tube of the telescope distort the measures (Høg and Miller 1986). During this program the application of flexure as a function of the rate of change of temperature was tested; but very little, if any, improvement could be detected.

## Solar System Observations

All the corrections made to the star observations are also applied to observations of the solar system objects as well as corrections, if necessary, for their appearance, orbital motion, horizontal parallax correction, night-minus-day effects, and Delta-T. The corrections to the solar system objects follow the form used by David K. Scott for the $W3_{50}$ (Adams *et al*. 1964), $W4_{50}$ (Adams and Scott 1968), and $W5_{50}$ (Hughes and Scott 1982).

The corrected observed right ascension for the solar system objects takes the form:

$$\alpha = \alpha_o + \theta_\alpha + v_\alpha + \Delta T_\alpha + \zeta_\alpha$$

where: $\alpha$ = corrected observed right ascension
$\alpha_o$ = observed right ascension
$\theta_\alpha$ = correction in right ascension for orbital motion
$v_\alpha$ = correction in right ascension for visual appearance
$\Delta T_\alpha$ = Delta-T correction in right ascension
$\zeta_\alpha$ = night-minus-day correction in right ascension

The corrected observed declination for the solar system objects takes the form:

$$\delta = \delta_o + \theta_\delta + v_\delta + \Delta T_\delta + \zeta_\delta + \pi$$

where: $\delta$ = corrected observed declination
$\delta_o$ = observed declination
$\theta_\delta$ = correction in declination for orbital motion
$v_\delta$ = correction in declination for visual appearance
$\Delta T_\delta$ = Delta-T correction in declination
$\zeta_\delta$ = night-minus-day correction in declination
$\pi$ = horizontal parallax correction

The observed right ascension ($\alpha_o$) and declination ($\delta_o$) were reduced in the same manner, and include the same corrections, as the stellar observations.

**Orbital Motion Corrections** - Corrections for the orbital motion were applied when the mean measured position does not coincide with the meridian. In right ascension, the correction $\theta_\alpha$ was based on the difference between the sidereal time of the mean micrometer screw reading and the observed time of transit multiplied by the daily orbital motion in right ascension.

$$\theta_\alpha = (T_m - T_t)B$$

where:

$T_m$ = mean sidereal time of observation
$T_t$ = observed time of transit
$B = dy_{mtd} /(dy_{mtd} - \rho_\alpha)$
$dy_{mtd}$ = sidereal seconds per Mean Time Day (86636.555368)
$\rho_\alpha$ = motion in right ascension per Mean Time Day

For the declination observations, the correction $\theta_\delta$ was based on the difference between the mean measured position and the collimation point of the transit circle multiplied by the daily orbital motion in declination

$$\theta_\delta = \pm[(R_m - R_c)Sc_\alpha]/[(\rho_\delta \cos \delta)dy_{mtd}]$$

where:

$\pm$ = + for Clamp East, - for Clamp West
$Sc_\alpha$ = scale of micrometer screw (right ascension)
$R_m$ = mean right ascension screw setting of declination measures
$R_c$ = right ascension screw setting of collimation point
$\rho_\delta$ = motion in declination per Mean Time Day
$dy_{mtd}$ = sidereal seconds per Mean Time Day (86636.555368)

**Visual Appearance Corrections** - A correction for the appearance was sometimes necessary due to either the way measurements were taken of the object or its physical appearance. Measurements of the four limbs of the Sun (north, south, east and west) were averaged to find the center. No correction was applied for the appearance of the Sun's limbs. The minor planets look like point sources and needed no correction for their appearance. The center of the images of Uranus and Neptune were observed with no correction made for defect of illumination. Measurements of the limbs of Saturn were made when possible. Settings were made on the rings when the limbs were not clearly visible, either due to the tilt of the rings or poor atmospheric seeing. No correction for defect of illumination was applied to Saturn. The center of the image of Mercury was observed and required corrections for defect of illumination in both right ascension and declination. The corrections were derived by Newcomb (Eichelberger 1904) and take the form:

$$v_\alpha = [\ S\ q(5 + \cos i)]\cos Q$$

and

$$v_\delta = (-[q(5 + \cos i)]\sin Q)/12$$

where:

$v_\alpha$ = visual appearance correction in right ascension
$v_\delta$ = visual appearance correction in declination
$q$ = defect of illumination
$i$ = angle at Mercury between Earth and Sun
$Q$ = position angle of defect of illumination
$S = (B \times 3600) / (15 \times \cos \delta_{Merc})$
$B = dy_{mtd} / (dy_{mtd} - \rho_\alpha)$
$dy_{mtd}$ = sidereal seconds per Mean Time Day (86636.555368)
$\rho_\alpha$ = motion in right ascension per Mean Time Day
$\delta_{Merc}$ = declination of Mercury

For Venus settings were made on the two illuminated limbs, one in right ascension and the other in declination, necessitating a correction for the semi-diameter. The corrections took the form;

$$v_\alpha = L_\alpha S \pi f$$

and

$$v_\delta = L_\delta S \pi f$$

where:

$v_\alpha$ = visual appearance correction in right ascension
$v_\delta$ = visual appearance correction in declination
$L_\alpha$ = +1 for the following limb, -1 for the preceding limb
$L_\delta$ = +1 for the southern limb, -1 for the northern limb
$\pi$ = horizontal parallax correction
$f$ = 0.95568 for Venus
$S = (B \times 3600) / (15 \times \cos \delta_{Ven})$
$B = dy_{mtd} / (dy_{mtd} - \rho_\alpha)$
$dy_{mtd}$ = sidereal seconds per Mean Time Day (86636.555368)
$\rho_\alpha$ = motion in right ascension per Mean Time Day
$\delta_{Ven}$ = declination of Venus

The $(O - C)$'s for Mercury and Venus were investigated for systematic differences depending on the phase. Additional phase corrections were found to be necessary and are given in Tables 14 and 15. The corrections were determined by grouping the observations by the position angle of the terminator (the angle between the instrumental meridian and a line joining the cusps) for declination and by the differences of the mean heliocentric longitudes between the planet and the Earth for right ascension. The mean differences of the $(O - C)$'s were taken between symmetrical groups following or preceding the Sun and north or south of the Sun. This correction was added to $v_\alpha$ and $v_\delta$ for Mercury and Venus. Adjusting the corrections for scale based on the distance between the Earth and the planet, as was done in the $W5_{50}$ (Hughes and Scott 1982), was found not to be necessary.

Table 14: Additional Phase Correction for Mercury

where: $\Delta L$ = difference in heliocentric longitudes of Mercury and the Earth

$v'_{\alpha}$ = additional phase correction in right ascension in units of seconds of time

$P$ = position angle of the terminator

$v'_{\delta}$ = additional phase correction in declination in units of second of arc

| $\Delta L$ | $v'_{\alpha}$ | n | | $P$ | $v'_{\delta}$ | n |
|---|---|---|---|---|---|---|
| 00° to 20° | 0.007 | 16 | | 0° to 120° | -0.214 | 80 |
| 20 to 40 | 0.013 | 17 | | 120 to 180 | -0.068 | 86 |
| 40 to 60 | 0.019 | 26 | | 180 to 320 | 0.068 | 89 |
| 60 to 80 | 0.024 | 33 | | 320 to 360 | 0.214 | 73 |
| 80 to 100 | 0.030 | 30 | | | | |
| 100 to 120 | 0.036 | 15 | | | | |
| 120 to 140 | 0.042 | 13 | | | | |
| 140 to 160 | 0.048 | 2 | | | | |
| 160 to 200 | -0.053 | 1 | | | | |
| 200 to 220 | -0.048 | 12 | | | | |
| 220 to 240 | -0.042 | 19 | | | | |
| 240 to 260 | -0.036 | 32 | | | | |
| 260 to 280 | -0.030 | 28 | | | | |
| 280 to 300 | -0.024 | 25 | | | | |
| 300 to 320 | -0.019 | 30 | | | | |
| 320 to 340 | -0.013 | 15 | | | | |
| 340 to 360 | -0.007 | 16 | | | | |

Observations of Mars and Jupiter consisted of settings on all four limbs. Settings on the limb formed by the terminator require a correction for defect of illumination. The visual corrections take the form;

$$v_{\alpha} = S\, q[0.5\sin^2 Q - [(1 - \cos i)\sin^2 2Q]/16]$$

and

$$v_{\delta} = q[0.5\sin^2 Q - [(1 - \cos i)\sin^2 2Q]/16]$$

where:

$v_{\alpha}$ = visual appearance correction in right ascension

$v_{\delta}$ = visual appearance correction in declination

$q$ = defect of illumination

$i$ = angle at Mars or Jupiter between Earth and Sun

$Q$ = position angle of defect of illumination

$S = (B \times 3600) / (15 \times \cos \delta_{Pl})$

$B = dy_{mtd} / (dy_{mtd} - \rho_{\alpha})$

$dy_{mtd}$ = sidereal seconds per Mean Time Day (86636.555368)

$\rho_\alpha$ = motion in right ascension per Mean Time Day
$\delta_{Pl}$ = declination of planet (Mars or Jupiter)

Table 15. Additional Phase Correction for Venus

where: $\Delta L$ = difference in heliocentric longitudes of Venus and the Earth
$v'_\alpha$ = additional phase correction in right ascension in units of seconds of time
$P$ = position angle of the terminator
$v_{\delta}$- = additional phase correction in declination in units of second of arc

| $\Delta L$ | $v'_\alpha$ | n | | $P$ | $v'_\delta$ | n |
|---|---|---|---|---|---|---|
| 0° to 20° | 0.019 | 17 | | 0° to 120° | 0.131 | 386 |
| 20 to 40 | 0.026 | 34 | | 120 to 180 | -0.909 | 11 |
| 40 to 60 | 0.033 | 37 | | 180 to 240 | 0.909 | 6 |
| 60 to 80 | 0.039 | 38 | | 240 to 360 | -0.131 | 319 |
| 80 to 100 | 0.046 | 38 | | | | |
| 100 to 120 | 0.053 | 35 | | | | |
| 120 to 140 | 0.060 | 26 | | | | |
| 140 to 160 | 0.067 | 36 | | | | |
| 160 to 180 | 0.073 | 36 | | | | |
| 180 to 200 | -0.073 | 33 | | | | |
| 200 to 220 | -0.067 | 39 | | | | |
| 220 to 240 | -0.060 | 49 | | | | |
| 240 to 260 | -0.053 | 39 | | | | |
| 260 to 280 | -0.046 | 43 | | | | |
| 280 to 320 | -0.033 | 29 | | | | |
| 320 to 340 | -0.026 | 33 | | | | |
| 340 to 360 | -0.019 | 12 | | | | |

**Horizontal Parallax Correction** - The horizontal parallax correction in declination was based on the horizontal parallax at the time of the observation, the Earth's radius vector for the U.S. Naval Observatory, and the difference between the geocentric latitude and observed declination. The correction takes the form;

$$\pi = r_{usno}\, \pi_h \sin(\phi_{geo} - \delta_o)$$

where:

$\pi$ = horizontal parallax correction
$r_{usno}$ = Earth's radius vector (for USNO = .998691)
$\pi_h$ = horizontal parallax
$\phi_{geo}$ = geocentric latitude (for USNO = 38.732576 degrees)
$\delta_o$ = observed declination

**Delta-T ($\Delta T$) -** Unlike the $(O-C)$ positions for the stars, the $(O-C)$'s for the solar system objects must involve Delta-T ($\Delta T$), which is the difference between dynamical time (now referred to as Terrestrial Time or TT) and Universal Time (UT1). The necessary corrections are to the ephemeris places of the solar system objects for the motion of the object during the interval between its transit over the ephemeris meridian and the local (in this case the Six-inch) meridian. The correction is found by multiplying the daily motion of the object by $\Delta T$ and adding it to the calculated position determined from the ephemeris, which in the case of the $W1_{J00}$ was DE200 (Standish 1990). The values of $\Delta T$ used were interpolated from the yearly values in the Astronomical Almanac (Nautical Almanac Office, 2000), which are given in the table below.

| Year | $\Delta T$ (sec of time) |
|---|---|
| 1978.0 | 48.53 |
| 1977.0 | 47.52 |
| 1980.0 | 50.54 |
| 1981.0 | 51.38 |
| 1982.0 | 52.17 |
| 1983.0 | 52.96 |

**Night-minus-Day Corrections -** A comparison of observations of the same stars taken during the day versus night showed a difference necessitating a correction to the day observations. The night-minus-day correction (ξ) was found by subtracting the mean $(O-C)$ for the individual day observation from the mean $(O-C)$ for the same star observed at night. The differences were tested as a function of various factors, such as temperature, declination, and hour angle from the Sun. The night-minus-day corrections given in Tables 16, 17, and 18 were added to the observed positions of the Sun, Mercury, and Venus.

Table 16 – Night-minus-Day correction (ξ) as a function of Local Apparent Solar Time (in hours)
$\xi_\delta$ = Dec correction in seconds of arc and $\xi_\alpha$ = RA correction in seconds of time

| Apparent Time | $\xi_\alpha$ | $\xi_\delta$ |
|---|---|---|
| 8.5 | 0.012 | -0.086 |
| 9.0 | 0.015 | -0.075 |
| 9.5 | 0.018 | -0.064 |
| 10.0 | 0.022 | -0.052 |
| 10.5 | 0.024 | -0.041 |
| 11.0 | 0.025 | -0.030 |
| 11.5 | 0.021 | -0.018 |
| 12.0 | 0.014 | -0.007 |
| 12.5 | 0.003 | 0.004 |
| 13.0 | -0.008 | 0.016 |
| 13.5 | -0.017 | 0.027 |
| 14.0 | -0.023 | 0.038 |
| 14.5 | -0.024 | 0.049 |
| 15.0 | -0.020 | 0.061 |
| 15.5 | -0.012 | 0.072 |

Table 17 – Night-minus-Day correction (ξ) as a function of temperature (in degrees Celsius)
$\xi_\delta$ = Dec correction in seconds of arc and $\xi_\alpha$ = RA correction in seconds of time

| Temperature | $\xi_\alpha$ | $\xi_\delta$ |
|---|---|---|
| -12 | 0.184 | 0.119 |
| -10 | 0.118 | 0.110 |
| -8 | 0.056 | 0.102 |
| -6 | 0.009 | 0.093 |
| -4 | -0.021 | 0.084 |
| -2 | -0.034 | 0.075 |
| 0 | -0.034 | 0.066 |
| 2 | -0.026 | 0.057 |
| 4 | -0.018 | 0.048 |
| 6 | -0.017 | 0.039 |
| 8 | -0.020 | 0.030 |
| 10 | -0.023 | 0.022 |
| 12 | -0.025 | 0.013 |
| 14 | -0.026 | 0.004 |
| 16 | -0.030 | -0.005 |
| 18 | -0.041 | -0.014 |
| 20 | -0.051 | -0.023 |
| 22 | -0.055 | -0.032 |
| 24 | -0.063 | -0.041 |
| 26 | -0.089 | -0.050 |
| 28 | -0.135 | -0.058 |
| 30 | -0.187 | -0.067 |
| 32 | -0.208 | -0.076 |
| 34 | -0.187 | -0.085 |

Table 18 – Night-minus-Day correction ( $\xi_\delta$ ) as a function of declination
Declinations in degrees, $\xi_\delta$ = Dec correction in seconds of arc

| Declination | $\xi_\delta$ |
|---|---|
| -30 | -0.970 |
| -25 | -0.711 |
| -20 | -0.553 |
| -15 | -0.412 |
| -10 | -0.236 |
| -5 | -0.097 |
| 0 | -0.015 |
| 5 | 0.067 |
| 10 | 0.139 |
| 15 | 0.166 |
| 20 | 0.150 |
| 25 | 0.115 |
| 30 | 0.098 |
| 35 | 0.110 |

## Absolute Reduction

During the observing phase of this program, the observations were referenced to an *instrumental frame* through the methods described above to reduce the environmental and instrumental systematics. To aid in monitoring the performance of the equipment, the observed position minus the calculated position (($O-C$)'s) were analyzed. Traditionally, to make the positions *absolute*, in the sense that they do not explicitly rely on any previous observations of these objects, adjustments were determined using the observations of the circumpolar stars and solar system objects. The circumpolar stars were used to determine the position of the celestial pole and the solar system objects used to determine the equinox. One difficulty with this method was that the adjustments thus formed used objects in two different parts of the sky and over a small range of declination and do not appear always to be applicable to objects at all declinations. From the circumpolar stars, corrections to the assumed latitude, instrumental flexure, and the constant of refraction were determined and applied to all observations. From the solar system objects, a correction to the equinox was determined along with a correction to the equator. The *equinox correction* was applied to all the observations, but the *equator correction* conflicts with the adjustments determined by the circumpolar stars. The solution to this difficulty for previous transit circle programs has been to smoothly distribute the offset between the pole and the equator so that positions near the pole were unaffected by the *equator correction* while the full offset was applied to those near the equator. We have tried a modification of the traditional method (Holdenried and Rafferty 1997) in an attempt to remove the need for an *equator correction,* but we were unable to do so.

It has been the practice at the U.S. Naval Observatory to align the *instrumental frame* to a *dynamical frame* through corrections to the equinox point and equatorial plane (the *equinox* and *equator corrections*), derived from observations of the Sun and planets. The *equinox* and *equator corrections* would be solved for simultaneously with corrections to the orbital elements of each planet. With the release of DE200 ephemerides (Standish 1990), it was realized that the orbits of the planets were so well established that it was no longer necessary or desirable to use the observations to correct the orbital elements. The $W1_{J00}$ is the first U.S. Naval Observatory catalog in the series to eliminate the corrections to orbital elements from the alignment solutions.

The alignment of the *instrumental frame* with a *dynamical frame* was accomplished through an adjustment to the equinox point and the equatorial plane. The *equinox correction* assumes that the poles of the two frames were coincident and that a simple rotation is all that was necessary to align the zero point of the right ascensions. As a result of the way in which the right ascensions were determined using ($O$ - $C$)*'s*, the zero point of the *instrumental frame*, before any rotation was applied, was very close to that of the catalog which supplied the data for the apparent places of the clock stars. In the case of the $W1_{J00}$, this catalog was the Fifth Fundamental Catalog (Fricke *et al.* 1988).

Before using the solar system observations, the *instrumental frame* was adjusted to the celestial pole through a correction to the assumed latitude of the transit circle. Using the observations of circumpolar stars, the correction to the assume latitude ($\Delta\phi$), as well as corrections to the refraction constant ($\Delta R$) and the horizontal instrumental flexure ($F$), can be determined using the equation:

$$(O-C)'_\delta - (O-C)_\delta = 2\Delta\phi + \Delta R(\tan z + \tan z') + F(\sin z + \sin z')$$

where:

$(O - C)'_\delta$ = observed minus computed declination of a star at lower culmination
$(O - C)_\delta$ = observed minus computed declination of a star at upper culmination
$z'$ = zenith distance of the lower culmination observations
$z$ = zenith distance of the upper culmination observations
$\Delta R$ correction to the constant of refraction
$\Delta\phi$ = correction to the assumed latitude
$F$ = horizontal instrumental flexure

Since the mean epochs of the upper and lower culmination observations purposely were kept nearly the same, the computed positions canceled out and the solution was, therefore, considered *absolute*. However, a solution including both the flexure ($F$) and the correction to the latitude ($\Delta\phi$) was not feasible because, over the small range of zenith distances involved with the circumpolar stars, $F$ and $\Delta\phi$ were strongly correlated. To alleviate this problem, the value of the flexure directly measured using the two horizontal collimating telescopes was substituted into the equation. Unfortunately this directly measured flexure showed a very large scatter. For the $W1_{J00}$ they had a standard deviation of ±0.″253. Though the flexure measured from collimators was poorly determined, it was necessary to include it to break the previously described correlation when just circumpolar stars were employed. By substituting the mean collimator flexure value for $F$ in the equation above, $\Delta\phi$ and $\Delta R$ can be determined and adjustment to all the declinations can be calculated using:

$$\Delta\delta = \Delta\phi + \Delta R \tan z + F \sin z$$

However, when this procedure was applied to the solar system objects, it frequently resulted in a sizeable systematic offset in the $(O - C)_\delta$'s in the form of the *equator correction*. Since the instrumental pole was regarded as absolutely fixed based on the observations of the circumpolar stars, an additional step was taken in which a solution was made for further corrections to $\Delta\phi$ and $\Delta R$, as well as a correction to $F$, by defining a model that smoothly distributed the offset between the pole and the equator so that positions near the pole were unaffected while the full offset was applied to those near the equator. This method was employed for the $W5_{50}$ (Hughes and Scott 1982), $WL_{50}$ (Hughes *et al.* 1992) and other Washington absolute catalogs, and, because of the location of the stars involved, will be referred to as the *circumpolar solution*. The *circumpolar solution* was very weakly defined because of the sparseness of the set of circumpolar stars contributing $(O - C)_\delta$'s and their limited range of zenith distances. Since the objective was to produce an absolute star catalog that was free of bias, such behavior was a source of concern.

A new approach was developed using stars distributed at all zenith distances and the relations

$$(O-C)_\delta = \Delta\phi + \Delta R \tan z + F \sin z$$

and for lower culmination observations

$$(C-O)'_\delta = \Delta\phi + \Delta R \tan z + F \sin z$$

for the stars observed at all zenith distances, which allows all three unknowns ($\Delta\phi$, $\Delta R$, and $F$) to be solved for by least squares. Because zenith distances are not restricted as they are in the *circumpolar solution*, the strong correlation between $\Delta\phi$ and $F$ is broken. However, the introduction of the star's computed place ($C$), does threaten the *absolute* quality of the calculation. In general, the systematic errors in the *C'*s are not correlated with the zenith distance or, more precisely, with the sine and tangent of the zenith distance. Furthermore, by grouping the $(O - C)_\delta$*'s* into zones arranged symmetrically about the zenith, we can take advantage of a useful feature of odd trigonometric functions. A least squares solution for the coefficients of such functions have the property of being completely independent of the intercept as long as the data points are distributed absolutely symmetrically about the origin. Thus a solution for $\Delta R$ and $F$, from data properly arranged, need not be dependent on $\Delta\phi$. This fact can be seen in Table 19, lines 4 and 5a; whether $\Delta\phi$ was included in the solution, but does not affect the values of $\Delta R$ and $F$. We refer to this method solving for $F$ and $\Delta R$ separately from $\Delta\phi$ as the *all-sky solution.*

The values of $F$ and $\Delta R$ from the *all-sky solution* can be considered independent of the C's, but this was not the case for $\Delta\phi$. The value of $\Delta\phi$ necessarily will link the *instrumental frame* to the system of the C's. As a regression problem, the values for $F$ and $\Delta R$ can be obtained without including $\Delta\phi$ from the *all-sky* method using the model:

$$(O - C)^{*}_{\delta} = \Delta R \tan z + F \sin z$$

Where $(O - C)^{*}_{\delta} = + (O - C)_\delta$ for upper culmination observations and for lower culmination $(O - C)^{*}_{\delta} = - (O - C)_\delta$. These values may then be applied to the observations and a solution for $\Delta\phi$ may be determined from the circumpolar stars using:

$$(O - C)'_{\delta} - (O - C)_{\delta} = 2\Delta\phi$$

By combining the *all-sky* and *circumpolar solutions*, the absolute nature of the *instrumental frame* is retained. We refer to this method as the *combined all-sky solution.*

For the $W1_{J00}$, only the FK5 (Fricke *et al.* 1988) stars were used in the *combined all-sky solution*. Comparisons were made between the *circumpolar* (Table 19, lines 1-3), the *all-sky* (Table 19, line 4), and the *combined all-sky solutions* (Table 19, lines 5a and 5b). For the $W1_{J00}$, the value of the flexure produced by the *all-sky solution* is nearly identical to the value measured with respect to the collimators. This is probably coincidence for we have not seen this happen in past Washington catalogs. The major difference between the two methods was in the value for $\Delta R$. Since the correction to the constant of refraction is a function of tan $z$, even small differences in $\Delta R$ will have a large effect at the extreme zenith distances. To the north the effect was not significant because lower culmination observations, which have the largest zenith distances, were not usually reported in the final catalog; but to the south the effect can be much more of a problem.

Table 19: Solutions for the horizontal instrumental flexure (*F*), the correction to the refraction constant (Δ*R*), and the correction to the assumed latitude (Δø). Solution 1 simultaneously solves for *F*, Δ*R*, and Δø using only the circumpolar stars. This solution was very poor due to the strong correlation between *F* and Δø over the small range of the zenith distances of the circumpolar stars. Solution 2 applies values of the horizontal flexure, measured by means of collimators as a function of the rate of change of temperature (Δt), to the observations and subsequently solves for just Δ*R* and Δø using only the circumpolar stars. Solution 3 applies the mean flexure value determined from the collimators to the observations and subsequently solves for just ΔR and Δø using only the circumpolar stars. Solution 4 simultaneously solves for *F*, Δ*R*, and Δø using observations distributed over the entire sky and averaged into zones 2.5 degrees wide arranged symmetrically about the zenith. When the data were arranged in this manner, the solution for the coefficients *F* and Δ*R* was independent of the intercept, Δø, because of the properties of odd, trigonometric functions. In solution 5 the data were handled in practically the same fashion as for solution 4, the only difference being that, initially, only *F* and Δ*R* were solved for (line 5a). These values were then applied and Δø determined (line 5b). The values for *F* and Δ*R* are identical on lines 4 and 5a because of the just mentioned behavior of odd trigonometric functions. Solutions 5a and 5b used together are referred to as the *combined all-sky solution*.

| | all units=seconds of arc<br>σ=std dev. of mean | Directly Measured Values | σ | | σ | Calculated Values | σ | | σ |
|---|---|---|---|---|---|---|---|---|---|
| no. | solutions | *F* | | *F* | | Δ*R* | | Δø | |
| 1 | Circumpolar | | | -1.629 | 0.524 | -0.081 | 0.084 | 1.637 | 0.161 |
| 2 | Circumpolar | *F*(Δt) | | | | 0.039 | 0.017 | -0.228 | 0.031 |
| 3 | Circumpolar | 0.662 | ±.252 | | | 0.055 | 0.017 | -0.292 | 0.031 |
| 4 | All-Sky | | | 0.687 | 0.083 | -0.024 | 0.022 | -0.186 | 0.007 |
| 5a | Combined All-Sky (all-sky) | | | 0.687 | 0.158 | -0.024 | 0.085 | | |
| 5b | Combined All-Sky (circumpolar) | | | | | | | -0.172 | 0.009 |

An important feature of the *combined all-sky solution* is that it does not utilize observations of solar system objects, thus leaving them free to be used to align the axes of the *instrumental frame* with the *dynamical frame*. We assumed that the *instrumental frame* was a rigid system of orthogonal axes, and that the alignment with the *dynamical frame* should be accomplished by simple orthogonal transformations. If a rectangular coordinate system was employed the transformation can be accomplished by three rotations, one about each of the axes. We chose the axes in the usual manner so that the *z* axis was parallel to the equatorial pole, the *x* axis points to the Vernal Equinox, and rotations were defined as positive in the right-handed sense. A rotation about the *x* axis was described by the angle *i*, about the *y* axis by the angle *j* and about the *z* axis by the angle *k*. To make a solution by linear least squares it was customary to utilize the small angle approximations of sin *a* = *a*, cos *a* =1, and cos *a* cos *b* =1 (where *a* and *b* are small angles) were utilized. In this way the rotation matrix reduces to

$$[R] = \begin{bmatrix} 1 & -k & j \\ k & 1 & -i \\ -j & i & 1 \end{bmatrix}$$

The equation of condition can be formulated as

$$[R] \times \begin{bmatrix} x \\ y \\ z \end{bmatrix} = \begin{bmatrix} x' \\ y' \\ z' \end{bmatrix}$$

where:

$x = \cos\delta \cos\alpha$ ($\alpha$ = right ascension, $\delta$ = declination)
$y = \cos\delta \sin\alpha$
$z = \sin\delta$

and the primed variables are the rectangular coordinates after being rotated. The angles resulting from the solution by using 2062 solar system observations (Sun, Mercury, Venus and Mars) and the ephemeris from DE200 were:

$$i = +0''.097 \pm 0.031$$
$$j = +0''.044 \pm 0.029$$
$$k = +0''.223 \pm 0.023$$

For the purposes of comparison, the same set of data was treated by the traditional method for aligning an *instrumental frame* with the *dynamical*. That was a least squares solution for an *equator* and *equinox correction* employing the same algorithms that in the past were used to solve for corrections to the orbital elements. Forcing the corrections to the orbital elements to be zero resulted in an *equinox correction* of $0''.225 \pm 0.008$. As expected this value was very close to $k$ since the *equinox correction* represents a rotation about the equatorial pole, which was parallel to the $z$-axis. The angles $i$ and $j$, representing an inclination of the equatorial pole, have no analogs in the *traditional solution*.

After the rotations have been applied to the planet observations, a significant offset in declination (i.e. the mean of the planetary $(O - C)_\delta$'s) frequently remains. It could be argued that, to align the *instrumental frame* to the *dynamical frame*, this declination offset must be accounted for in some fashion. However to do so, two problems must be solved. First, as others have demonstrated convincingly (Vityazev 1994), non-rotational differences between coordinate systems can be cross-correlated with the rotational ones making them very difficult to separate. The declination offset was clearly a non-rotational term and as such may confound the determination of rotation angles. Vityazev has shown that non-rotational terms should be removed *before* a solution for rotational terms can be attempted. Unfortunately in the problem we were addressing here there was no *a priori* source of information that would allow us to remove the declination offset from the data.

Our study has shown that if one chooses to solve for the rotation angles and offset simultaneously by least squares using the equations of condition

$$(O - C)_\delta = -i \sin\delta + j \cos\alpha + \lambda$$

$$(O - C)_\alpha = +i \tan\delta \cos\alpha + j \tan\delta \sin\alpha - k$$

where :

$$\lambda = \text{declination offset},$$

that cross correlations of greater than 40% (from the correlation matrix) indicate unstable solutions and poorly distributed data. It was the distribution of the observations in right ascension that was most important. We have found if cross-correlations are greater than 40%, then that data set cannot be used to solve for rotation angles unless the declination offset can be estimated and removed beforehand.

The second problem involving the declination offset is demonstrated in Table 20. This table summarizes solutions, by the standard method, for rotations (between the *instrumental frame* and the *dynamical frame* defined by DE200), and after the rotations have been applied, a determination of λ for data sets comprised of various combinations of the Sun and planets observations (the minor planets were not used since they are not included in DE200). Each line in the table gives the angles of rotation (*i*, *j*, and *k*) as well as λ produced by treating the (*O* - *C*)'s of different combinations of solar system objects. Also, the columns labeled "traditional" give the results obtained by code that solved for *equator* and *equinox corrections* along with corrections to the orbital elements, but, here for comparison purposes only, the corrections to the orbital elements were forced to zero. As the table shows, the angle *k* is equivalent to the traditional *equinox correction* and the λ is equivalent to the traditional *equator correction*. The table is divided into two parts, the upper part contains data sets that were judged to have acceptable distributions; acceptable in the sense, as explained previously, of being suitable for solutions for rotations that are not correlated with λ. The results shown in the lower part of the table and marked by asterisks were judged by this criterion not to be acceptable. Note that λ agrees most closely with the traditional *equator correction* for the acceptable data. It is perhaps more important to note, however, that even among the acceptable data sets λ varies significantly. We have tested other catalogs and noticed the same behavior (Holdenried and Rafferty 1997).

Such behavior might be explained by problems with DE200, but we think this is unlikely. We think it more probable that the observations themselves were the cause. For transit circle observations, a non-zero λ may be introduced through an erroneous correction to the assumed latitude. However, in the case of the $W1_{J00}$, the correction to the assumed latitude was very well determined either from the *all-sky solution* (solving for the flexure, correction to the constant of refraction, and the correction to the assumed latitude, Table 19, lines 4 and 5b) or the *circumpolar solution* (solving for just the correction to the assumed latitude, Table 19, lines 2 and 3). Another source of uncertainty was the effect of making observations in both the daytime and nighttime. The Sun, Mercury, and Venus can only be observed in the daytime while the remainder of the solar system objects was observed only at night. Systematic differences between the daytime and nighttime observations are known to exist and are corrected for by an analysis of the observations of the same stars observed during the day and the night. In the case of the $W1_{J00}$, this analysis produced *night-minus-day corrections* as function of hour angle of the Sun, declination, and temperature. In a study of daytime observations made by another transit circle that could observe fainter day stars, it was found that the clock corrections, fundamental azimuth, clamp difference, corrections to the refraction ($\Delta R$), corrections to the assumed latitude ($\Delta\phi$), and the horizontal flexure ($F$) determined only using the day stars differed from the determinations of those values using only the nighttime observations (Rafferty and Loader 1992). Though *the night-minus-day corrections* could adjusted for such differences if the daytime values behaved similarly for the Six-inch Transit Circle during $W1_{J00}$ that is not a certainty. A further source of uncertainty is the method of observing the solar system objects. For Mercury, the "center of light" is observed with corrections applied for the phase. For Venus, the two illuminated limbs are observed and corrections applied to reduce the observations to the center of the planet. For the $W1_{J00}$ (as well as for the $W5_{50}$, and other Washington catalogs), additional corrections to both Mercury and Venus due to phase were determined empirically and also applied.

Table 20 shows the rotation angles and λ values determined from Mercury and Venus to have the greatest differences from among the various solar system objects. Unfortunately the makeup of DE200 exacerbates these problems. That is, the solar system objects given the highest weight in defining the dynamical coordinate frame of DE200 are daytime and/or extended objects; these are the Sun (i.e. the reflex of the Earth's motion), and (because of radar ranging and spacecraft fly-bys) Mercury, Venus, and Mars. For the $W1_{J00}$, it is surprising how consistent the results are for the outer planets. However for these outer planets there was evidence of systematic errors in DE200 which can be as large as $\pm 0\overset{''}{.}2$, depending on the date. It might be possible to utilize observations of outer planets if an ephemeris was developed based on more recent data than were included in DE200, and if that ephemeris was adopted as the dynamical standard. At the time of these reductions, though, it was the general consensus, that the combination of planets interior to Jupiter represent the best realization of the coordinate frame of DE200. The solar system objects that would be least effected by any of the above described problems are the minor planets. Although five minor planets were used in the perturbation models, there were no ephemerides for them included in DE200.

Table 20. Summary of the solutions by the standard method for the rotation angles (*k*, *j*, and *i*) between the axes of the *instrumental frame* of a $W1_{J00}$ Catalog and DE200 for various combinations of the Sun and planets. A declination offset was also solved for. The columns labeled “traditional” give results obtained from solutions for the equator and equinox corrections along with the corrections to the orbital elements, with the corrections to the orbital elements forced to zero. The planet groups marked with an asterisk are poorly distributed in this catalog, and these solutions for the rotation angles and declination offset are weak.

| all units = seconds of arc<br>σ = standard deviation of mean | | | | | | | | | |
|---|---|---|---|---|---|---|---|---|---|
| | traditional | | | | | | | | |
| Solar system objects | Equinox Corr | σ | *k* | σ | *j* | σ | *i* | σ | n |
| S | -0.130 | 0.023 | -0.134 | 0.033 | 0.059 | 0.042 | -0.033 | 0.044 | 861 |
| M | -0.469 | 0.044 | -0.438 | 0.063 | 0.131 | 0.080 | 0.156 | 0.083 | 328 |
| v | -0.256 | 0.034 | -0.256 | 0.050 | -0.020 | 0.065 | 0.223 | 0.063 | 621 |
| M,v | -0.329 | 0.022 | -0.317 | 0.039 | 0.037 | 0.051 | 0.198 | 0.050 | 949 |
| S,M,v | -0.235 | 0.010 | -0.230 | 0.026 | 0.047 | 0.033 | 0.094 | 0.034 | 1810 |
| S,M,v,m | -0.225 | 0.008 | -0.223 | 0.023 | 0.044 | 0.029 | 0.097 | 0.031 | 2062 |
| S,M,v,m,j | -0.227 | 0.007 | -0.221 | 0.021 | 0.055 | 0.026 | 0.086 | 0.028 | 2389 |
| S,m | | | -0.220 | 0.030 | 0.076 | 0.038 | 0.021 | 0.039 | 1189 |
| S,m,j | | | -0.155 | 0.022 | 0.072 | 0.027 | -0.010 | 0.031 | 1440 |
| S,m,j,s,u,n | | | -0.210 | 0.016 | 0.077 | 0.020 | 0.010 | 0.021 | 2254 |
| S,M,v,m,j,s,u,n | | | -0.241 | 0.016 | 0.068 | 0.020 | 0.070 | 0.021 | 3203 |
| m* | -0.156 | 0.026 | -0.177 | 0.040 | 0.007 | 0.048 | 0.146 | 0.075 | 252 |
| j* | -0.233 | 0.021 | -0.210 | 0.031 | 0.125 | 0.038 | -0.002 | 0.044 | 327 |
| m,j* | | | -0.190 | 0.025 | 0.082 | 0.029 | 0.038 | 0.038 | 579 |
| s,u,n* | | | -0.303 | 0.021 | 0.091 | 0.028 | 0.029 | 0.027 | 814 |
| m,j,s,u,n* | | | -0.255 | 0.016 | 0.091 | 0.020 | 0.039 | 0.021 | 1393 |
| | | | | | | | | | |
| | traditional | | | | | | | | |
| Solar system objects | Equator Corr | σ | *λ* | σ | n | | | | |
| S | 0.093 | 0.024 | 0.106 | 0.022 | 861 | | | | |
| M | -0.222 | 0.044 | -0.150 | 0.040 | 328 | | | | |
| v | -0.265 | 0.035 | -0.220 | 0.033 | 621 | | | | |
| M,v | -0.250 | 0.023 | -0.195 | 0.026 | 949 | | | | |
| S,M,v | -0.090 | 0.011 | -0.056 | 0.018 | 1810 | | | | |
| S,M,v,m | -0.078 | 0.009 | -0.051 | 0.016 | 2062 | | | | |
| S,M,v,m,j | -0.064 | 0.007 | -0.052 | 0.014 | 2389 | | | | |
| S,m | | | 0.031 | 0.020 | 1189 | | | | |
| S,m,j | | | 0.072 | 0.015 | 1440 | | | | |
| S,m,j,s,u,n | | | 0.045 | 0.011 | 2254 | | | | |
| S,M,v,m,j,s,u,n | | | -0.021 | 0.011 | 3203 | | | | |
| m* | 0.010 | 0.027 | 0.003 | 0.026 | 252 | | | | |
| j* | 0.026 | 0.022 | 0.000 | 0.019 | 327 | | | | |
| m,j* | | | 0.001 | 0.016 | 579 | | | | |
| s,u,n* | | | 0.007 | 0.014 | 814 | | | | |
| m,j,s,u,n* | | | 0.004 | 0.010 | 1393 | | | | |

S=Sun, M=Mercury, v=Venus, m=Mars, j=Jupiter, s=Saturn, u=Uranus, and n=Neptune

If we had more confidence in the planetary observations from the transit circle, we would apply λ to complete the alignment of the *instrumental frame* to the *dynamical frame*. At this time we hesitate to do this because of the problems discussed above and because the value of the offset varies so much between the individual and combined groups of solar system objects. As a result, we chose not to apply the declination offsets to the absolute position reported here.

## Differential Reduction to the Hipparcos Catalogue

To make the $W1_{J00}$ more coherent, knowing the presence of unexplained offsets in declination and right ascension, we have differentially processed it using the Hipparcos Catalogue. In 1997, the International Astronomical Union (IAU) passed a resolution designating the Hipparcos Catalogue as the optical realization of the International Celestial Reference Frame (IAU 1998). In 2000, the IAU amended that resolution by excluding those Hipparcos stars that were of questionable reliability, i.e. those stars marked with a C, V, G, O, or X flag, and naming this frame the Hipparcos Celestial Reference Frame (HCRF) (IAU 2001).

A method to adjust star catalogs to a common reference system was developed by USNO staff in the mid-1990s and applied to data used in the formation of the Tycho-2 Catalogue proper motions (Høg *et al.* 2000). This technique was used with the absolute $W1_{J00}$ positions to make them systematically consistent with HCRF.

The procedure started with identifying all stars in common to the $W1_{J00}$ and the Hipparcos Catalogue (the "common stars"). Differences between $W1_{J00}$ and the Hipparcos positions (HIP) at the $W1_{J00}$'s epoch were computed using the Hipparcos proper motions. Only the high quality HIP stars were used; those stars with C, G, V, X, or O codes in the HIP Field H59 were excluded. Next, systematic differences based on right ascension and declination were removed using the following technique. For each star in the $W1_{J00}$ (the "central star"), identification of all the nearby common stars (in RA/Dec space) was made. Stars were declared nearby if they were within a distance of 30 minutes of time in right ascension and 5 degrees in declination to the central star. A minimum of 3 common stars were required else the nearby area was expanded to 45 minutes and 7.5 degrees. Individual common star differences larger than 1″.5 were not included. Weights were computed for each common star according to how close it was to the central star; those closest receiving the most weight. An elliptical paraboloid weighting function was used to avoid the problem of the few, nearby stars getting overwhelmed by the more numerous further ones. The weighted mean residual was then computed and this value was applied to the central star. This process was continued for each $W1_{J00}$ star and each solar system observation.

This above process was used for every object in the $W1_{J00}$, including the solar system observations. Because of the large number of stars in common between the $W1_{J00}$ and Hipparcos and the short difference between their epochs, the star positions of the $W1_{J00}$ are of little value compared with those from Hipparcos, but that is not the case for the solar system positions. Using the stars to remove any systematic and zonal difference between $W1_{J00}$ and Hipparcos, the $W1_{J00}$ solar system positions are directly linked to the International Celestial Reference Frame. Using this method of determining adjustments from the star positions and applying them to the solar system positions is the ideal way of putting the solar systems positions onto the International Celestial Reference Frame.

As stated at the start of this introduction, comparisons of the $W1_{J00}$ absolute star positions with the Hipparcos Star Catalog showed significant offsets for which we have no good explanation; see Fig. 4.

Figure 4: $W1_{J00}$ (absolute) minus Hipparcos

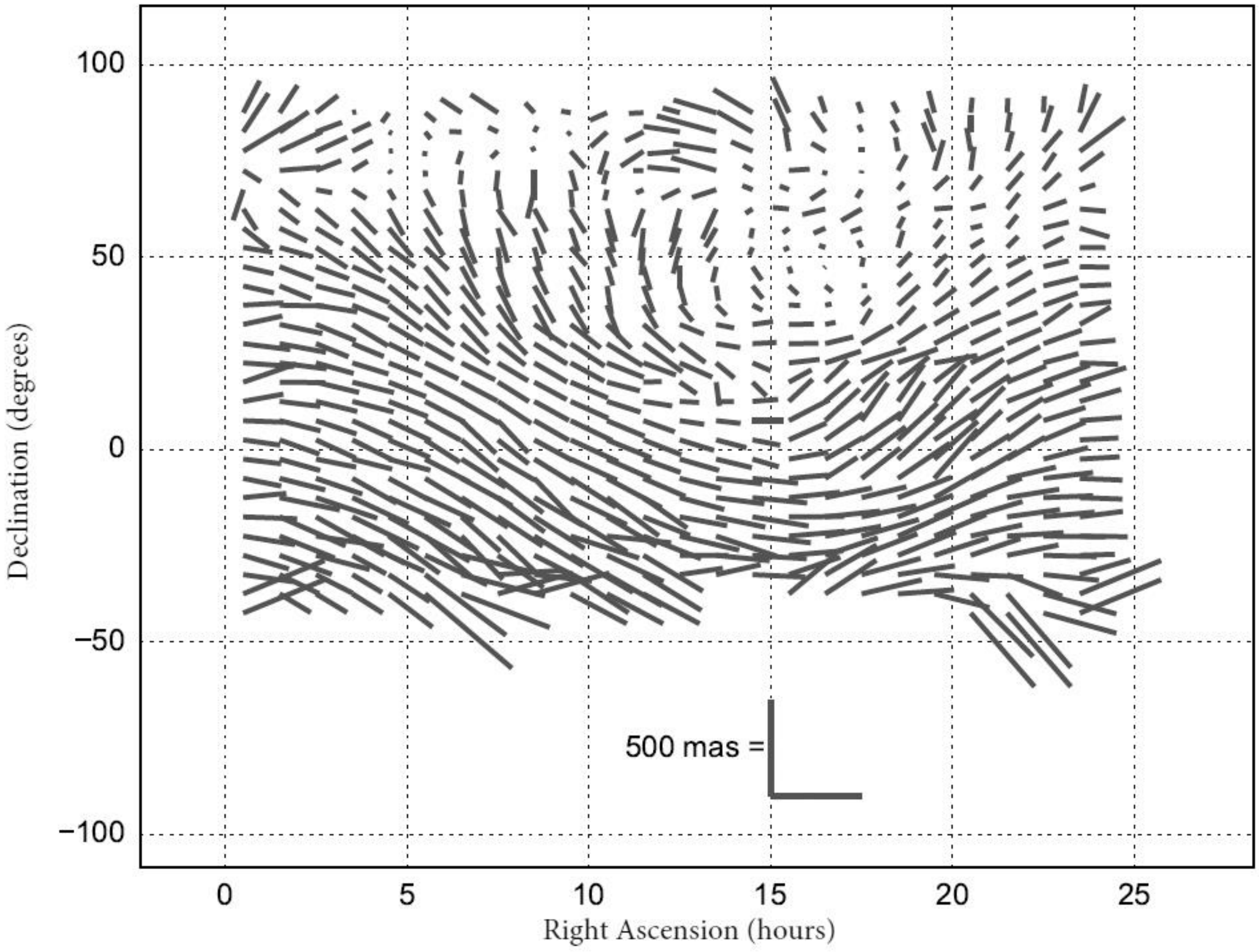

Figure 4: Differences between the $W1_{J00}$ absolute star positions and Hipparcos, by location on the sky. Data are averaged over 1-hour x 5-degree bins. Light smoothing is applied. A 500 milli-arcsecond scale is provided.

The largest offsets appear as a rotation, similar to an incorrectly applied equinox correction or precession value; however, our investigations failed to uncover the actual source of the discrepancies. Therefore presented here are both the absolute positions reduced in the traditional way as well as the positions differentially adjusted to the system of the Hipparcos Star Catalog. Following the Differential Reduction to the Hipparcos Catalogue, all sizable, large-scale offsets are removed, see Figure 5.

Figure 5: $W1_{J00}$ (differential) minus Hipparcos

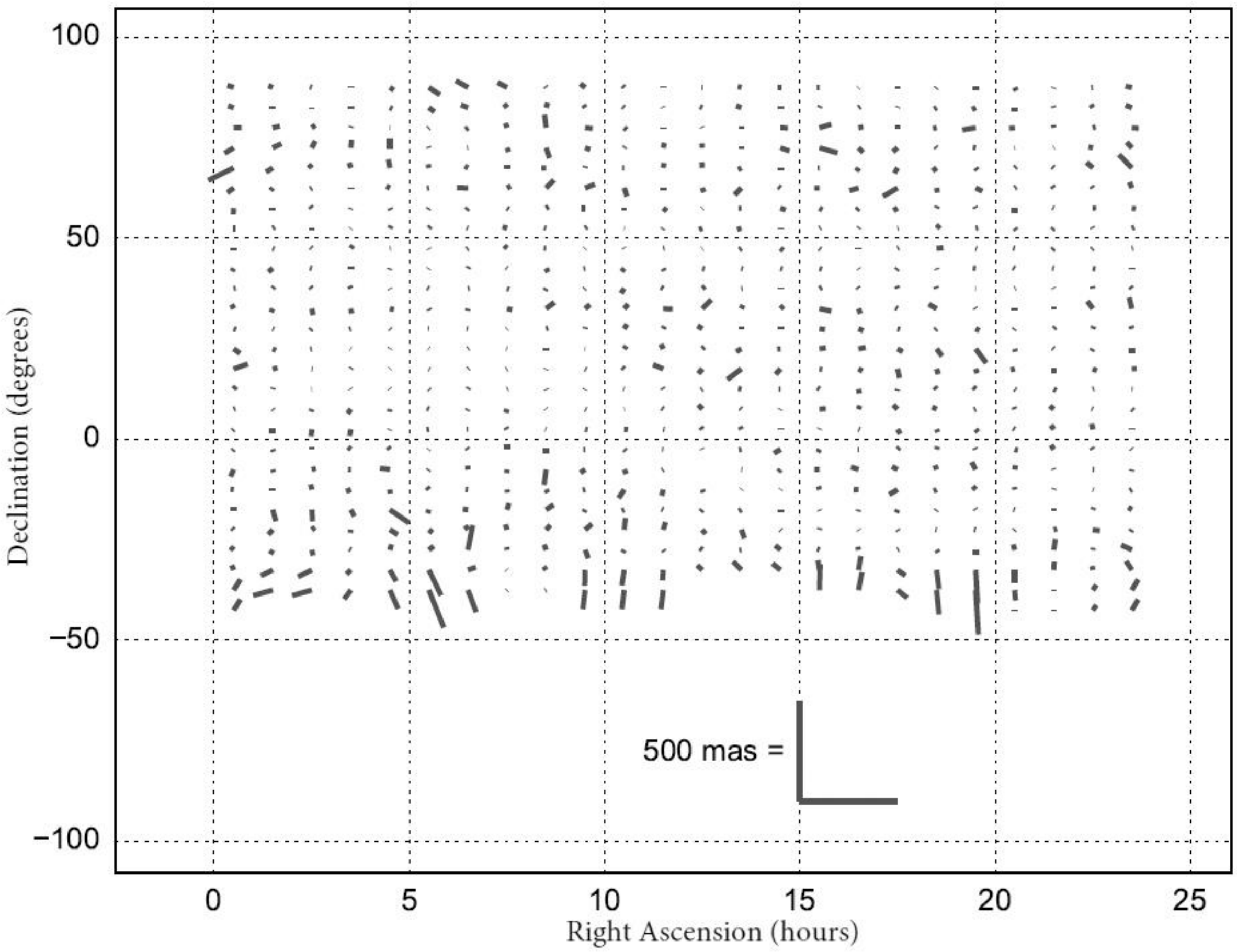

Figure 5: Same as Figure 4, except after the $W1_{J00}$ positions were differentially reduced using the Hipparcos Catalogue

## Errors of Observation and Position

The errors given in the catalog of star positions are standard errors of the mean for each coordinate of each position. The position referred to in this case is the absolute position which is given in the catalog as a difference from the differential Hipparcos position (which is given in full). The differential process, no doubt, has an error of its own but for which we have no estimate except to claim that it is much smaller, probably insignificant and are not included here. Table 21 gives a summary of both the standard deviations and the standard errors of the mean. Errors for the solar system objects are not given in the table of positions because each listed position is the result of a single observation. However, we would estimate that they should be approximately equal to the average of the standard deviations (RA = 268 mas, Dec = 291 mas) listed in Table 21 for the star positions.

Table 21

Summary of the Star Positions Errors in $W1_{J00}$ Catalog

| | Range | Average |
|---|---|---|
| RA standard deviation of a single obs'n | 03 to 920 mas | 268 mas |
| Dec standard deviation of a single obs'n | 02 to 990 mas | 291 mas |
| RA standard error of the mean position | 15 to 460 mas | 98 mas |
| Dec standard error of the mean position | 10 to 400 mas | 107 mas |

## The Final Positions of the Stars and Solar System Objects

As stated at the start of this introduction, the observing program was structured to be absolute, in the sense that the positions were not to explicitly rely on any previous observations. However, comparisons of the absolute star positions with the Hipparcos Star Catalog showed significant offsets for which we have no good explanation. Therefore, it was decided to include data on both the absolute positions reduced similarly to many past Washington transit circle catalogs, as well as the positions differentially adjusted to the system of the Hipparcos Catalog, which represents the ICRS.

The positions presented are differentially adjusted to ICRS via the Hipparcos Star Catalog. To obtain the absolute position reduced in the traditional way, take the ICRS position and add the dRA and dDE values.

For stars, both ICRS and absolute positions are on the equator and equinox J2000.0. (Precisely speaking, the ICRS positions are on the ICRS axes, but the difference between J2000.0 axes and ICRS axes can be ignored for the $W1_{J00}$ data). The epochs of the positions are the mean epochs of observation.

For the solar system objects, both ICRS and absolute positions are apparent places. This follows a tradition laid out by earlier USNO transit circle catalogs. An apparent place is a geocentric direction of an object that takes into account orbital motion, space motion, light-time, light deflection, and annual aberration. This is accomplished by the apparent place routine of the NOVAS astrometric software developed by the Nautical Almanac Office of USNO (Kaplan 1990). The version used here was what was current in 1996. The input to the apparent place routine was provided by the DE200. An apparent place is given with respect to the true equator and equinox "of date"; in this case, it is the date of observation.

## References

Adams, A.N., Bestul, S.M., and Scott, D.K., 1964, Results of observations made with the Six-inch Transit Circle 1949-1956 ($W3_{50}$), Publications of the United States Naval Observatory, Second Series, XIX(1)

Adams, A.N. and Scott, D.K., 1968, Results of observations made with the Six-inch Transit Circle 1956-1962 ($W4_{50}$), Publications of the United States Naval Observatory, Second Series, XIX(2)

Blaauw, A., 1955, Co-ordination of Galactic Research, IAU Symposium No.1 (Cambridge University Press)

Chauvenet, W., 1960, A Manual of Spherical and Practical Astronomy, Dover Publications, Inc., New York

Corbin T.E., 1978, The Proper Motions of the AGK3R and SRS Stars, Proc. IAU Coll. 48 Modern Astrometry, Prochazka F.V. & Tucker R.H. eds., 505-514

Eichelberger, W.S.,1904, Reduction Tables for Transit Circle Observations, Publications of the U.S. Naval Observatory, Second Series, Vol. IV, appendix II, pp 26-27

European Space Agency, 1997, The Hipparcos and Tycho Catalogues, ESA Publications Division, Vols 1-17

Fricke, W., Schwan, H., and Lederle, T., 1988, Fifth Fundamental Catalog, Part I. The Basic Fundamental Stars, Astronomisches Rechen-Institut, Heidelberg

Gylden, H., 1956, Refractions Tables of the Pulkovo Observatory, 4th Edition, Academy of Sciences Press, Moscow

Hemenway, P.D., 1966, The Washington 6-inch Transit Circle, Sky and Telescope, Vol XXXI, No. 2

Høg E. and Miller R.J., 1986, Internal Refraction in a USNO Meridian Circle, Astronomical Journal 92, 495-502

Høg, E., Fabricius, C., Makarov, V.V., Bastian, U., Schwekendiek, P., Wicenec, A., Urban, S., Corbin, T., and Wycoff, G., 2000, Construction and verification of the Tycho-2 Catalogue, Astronomy and Astrophysics 2000 357 367-386

Høg, E., 1961, Determination of Division Corrections, Astronomische. Nachrichten, Vol 286, p 65

Holdenried, E.R. and Rafferty, T.J., 1997 New methods of forming and aligning the instrumental frames of absolute transit circle catalogs, Astronomy and Astrophysics Supplement Series, Vol 125, pp 595-603

Hughes, J.A. and Scott, D.K., 1982, Results of observations made with the Six-inch Transit Circle 1963-1971 ($W5_{50}$), Publications of the United States Naval Observatory, Second Series, XXIII(3)

Hughes, J.A., Smith, C.A., and Branham, R.L., 1992, Results of observations made with the Seven-inch Transit Circle 1967-1973 ($WL_{50}$), Publications of the United States Naval Observatory, Second Series, XXVI(2)

Hughes, J.A., Espenscheid, P., and McCarthy D.D., 1975, Coordinates of U.S. Naval Observatory Installations, United States Naval Observatory Circular, No 153

IAU, 1998, Transitions of the IAU Vol. XXIIIB, Processings of the 23rd General Assembly, Kyoto, Japan, August 18-20, 1997, Ed. J. Andeser

IAU, 2001, Transactions of the IAU Vol XXIVB, Processings of the. 24th General Assembly, Manchester, 2000, Ed. H. Rickman

Kaplan, G., (1990), Bulletin of the American Astronomical Society, Vol. 22, pp. 930- 931

Kristian, J. and Sandage, A., 1970, Precise Positions of Radio Sources, II. Optical Measurements, Astrophysical Journal 162, 390-403

McCarthy, D. D., 1984, provided the data from the Bureau International de l'Heure, Paris

Nautical Almanac Office (2000), The Astronomical Almanac, U.S. Government Printing Office, Washington, DC

Podobed, V.V., 1962, Fundamental Astrometry, The University of Chicago Press

Rafferty, T.J. and Klock, B.L., 1982, Experiences with the U.S. Naval Observatory Glass Circles, Astronomy and Astrophysics, Vol 114, pp 95-101

Rafferty, T.J. and Klock, B.L.,1986, Circle scanning systems of the U.S. Naval Observatory, Astronomy and Astrophysics, Vol 164, pp 428-432

Rafferty, T.J. and Loader, B.R., 1993, Improvements in the use of daytime star observations from a transit circle, Astronomy and Astrophysics, Vol 271, pp 727-733

Rafferty, T.J., 1980, Effects of Different Sources of Variation of Latitude Data on Meridian Circle Catalogues, Astronomy and Astrophysics Supplement Series, 50, pp 27-47

Robertson, J., 1940, Catalog Of 3539 Zodiacal Stars For The Equinox 1950.0, Astronomical Papers Amer. Eph. X, II, U.S. Government Printing Office, Washington, DC

Standish, E.M. Jr., 1990, The Observational Basis for JPL's DE200, the planetary Ephemerides of the Astronomical Almanac, Astronomy and Astrophysics, 233, pp 252-271

Sandage, A., and Kristian, J., 1970, Precise Positions of Radio Sources, III. Comparison of Optical and Radio Measurements, Astrophysical Journal, Vol 162, pp 399-403

Seidelman, P. K., ed., 1992, Explanatory Supplement to the Astronomical Almanac, University Science Books

Vityazev, V.V., 1994, The Rotor: A New Method to Derive Rotation Between Two Reference Frames, Astronomical and Astrophysical Transactions, 4, pp 195-218

Wade, C. M., 1970, Precise Positions of Radio Sources, I. Radio Measurements, Astrophysical Journal 162, 381-390

Watts, C.B., 1950, Description of the Six-inch Transit Circle, Publication of the United States Naval Observatory, Second Series, Vol XVI, Part 2

## Format of the $W1_{J00}$ Stars and Solar System Objects Data

As stated at the start of this introduction, comparisons of the $W1_{J00}$ absolute star positions with the Hipparcos Star Catalog showed the presence of unexplained offsets in declination and right ascension. Presented here are the positions differentially adjusted to the system of the Hipparcos Star Catalog and their differences from the absolute positions.

The $W1_{J00}$ position files ***W1J00_stars.dat*** and ***W1J00_solsys.dat***, along with the ReadMe file ***W1J00 ReadMe***, can be found in the at the Centre de Domnées astronomiques de Strasbourg (Strasbourg Astronomical Data Center): http://cdsweb.u-strasbg.fr/

| FileName | Lrecl | Records | Explanations |
|---|---|---|---|
| ***W1J00 ReadMe*** | 80 | | $W1_{J00}$ ReadMe file |
| ***W1J00_stars.dat*** | 113 | 7267 | $W1_{J00}$ star positions (means) |
| ***W1J00_solsys.dat*** | 79 | 4383 | $W1_{J00}$ solar system object positions |

## Format of the $W1_{J00}$ Star Data File: ***W1J00_stars.dat***

For the stars, the positions are from the Differential Reductions to the Hipparcos Catalogue. Also given are the differences from the positions using the Absolute Method, which used some of the solar system observations to rotate them to J2000.0 via DE200. The positions of the stars are on the system of J2000.0 which means that although epoch of the observations is that of the mean observation, the orientation of the celestial reference system is fixed by the epoch of 2000 January 1.5.

The $W1_{J00}$ star data can be found in the file ***W1J00_stars.dat*** at the Centre de Domnées astronomiques de Strasbourg (Strasbourg Astronomical Data Center): http://cdsweb.u-strasbg.fr/

Byte-by-byte Description of file: ***W1J00_stars.dat***

| FileName | Lrecl | Records | Explanations |
|---|---|---|---|
| W1J00_stars.dat | 113 | 7267 | W1J00 star positions (means) |

```
Bytes Format Units  Label Explanations

1-4     I4   ---    W1J00 W1J00 identifier
6-7     I2   h      RAh   Right Ascension ICRS, at Ep_RA (hours) (1)
9-10    I2   min    RAm   Right Ascension ICRS, at Ep_RA (minutes) (1)
12-17   F6.3 s      RAs   Right Ascension ICRS, at Ep_RA (seconds) (1)
19-21   I3   mas    e_RAs Standard error of the mean, RA
23-25   I3   ---    NumRA Number of observations used for RA
27-32   F6.3 yr     Ep_RA Mean epoch of RA, minus 1900
34-39   F6.3 s      dRA   *RA (Absolute) - RA (ICRS) (2)
41      A1   ---    DE-   Declination (ICRS) at Ep_DE (sign) (1)
42-43   I2   deg    DEd   Declination (ICRS) at Ep_DE (degrees) (1)
45-46   I2   arcmin DEm   Declination (ICRS) at Ep_DE (minutes) (1)
48-52   F5.2 arcsec DEs   Declination (ICRS) at Ep_DE (seconds) (1)
54-56   I3   mas    e_DEs Standard error of the mean, DE
58-60   I3   ---    NumDE Number of observations used for DE
```

```
62-67   F6.3 yr     Ep_DE Mean epoch of DE, minus 1900
69-73   F5.2 arcsec dDE   *DE (Absolute) - DE (ICRS)  (2)
75-80   F6.3 mag    Vmag  Visual magnitude (3)
82-87   I6   ---    HIC   Hipparcos Input Catalog number (4)
88-89   A2   ---    Comp  Hipparcos Input Catalog component (4)
91-98   I8   ---    DM    Durchmusterung number (4)
100-103 I4   ---    FK5   FK5 number (4)
105-113 A9   ---    WDS   Washington Double Star Catalog identifier (4)
```

```
Note (1): Mean positions are on the International Celestial Reference System (ICRS),
differentially reduced using the Hipparcos catalog. The epochs for the right ascension
and declination coordinates are found in the Ep_RA and Ep_DE fields respectively.
Note (2): The observing program was designed to be an "absolute" catalog, independent
of other catalogs.  Data reductions were carried out according to previous absolute
transit circle catalogs.  Following the reductions, comparisons of Stars with the
Hipparcos Catalogue revealed remaining, unaccounted for systematic deviations in the
absolute positions.  The authors decided to differentially reduce the absolute positions
using the Hipparcos data; those are the position presented in the RAh, RAm, RAs, DE-,
DEd, DEm and DEs fields.  For users who may be interested in the absolute positions,
values of the absolute position minus the differential position are provided in the dRA
and dDE fields.
Note (3): Visual magnitude for identification purposes.  This was drawn from a variety
of sources, but largely from the Hipparcos Input Catalogue.
Note (4): Cross references to several widely known catalogs are provided for convenience
of the user.
```

## Format of the W1$_{J00}$ Solar System Data File: *W1J00_solsys.dat*

For the solar system objects, the positions are from the Differential Reductions to the Hipparcos Catalogue. Also given are the differences from the positions using the Absolute Method, where the same rotations that were applied to the stars were also applied to the solar system object observations. Unlike the stars, for the solar system objects the celestial reference frame was moved by precession and nutation to the time of the observation. This is accomplished by the apparent place routine of the NOVAS astrometric software developed by the Nautical Almanac Office of USNO (Kaplan 1990). The version used here was what was current in 1996. The input to the apparent place routine was provided by the DE200. Furthermore the positions for the solar system objects are geocentric. This follows a tradition laid out by earlier USNO transit circle star catalogs.

The W1$_{J00}$ solar system data can be found in the file ***W1J00_solsys.dat*** at the Centre de Domnées astronomiques de Strasbourg (Strasbourg Astronomical Data Center): http://cdsweb.u-strasbg.fr/

Byte-by-byte Description of file: ***W1J00_solsys.dat***

| FileName | Lrecl | Records | Explanations |
|---|---|---|---|
| W1J00_solsys.dat | 79 | 4383 | W1J00 solar system object positions |

```
Bytes Format Units Label Explanations
1-7   A7     ---    Obj    Solar system object identifier
9-10  I2     h      RAh    Right Ascen. (apparent) at Epoch (hours) (1, 2)
12-13 I2     min    RAm    Right Ascen. (apparent) at Epoch (minutes) (1, 2)
15-20 F6.3   s      RAs    Right Ascen. (apparent) at Epoch (seconds) (1, 2)
22-27 F6.3   s      dRA    *RA (Absolute) - RA (HIP) (3)
29    A1     ---    DE-    Declination (apparent) at Epoch (sign) (1, 2)
30-31 I2     deg    DEd    Declination (apparent) at Epoch (degrees) (1, 2)
33-34 I2     arcmin DEm    Declination (apparent) at Epoch (minutes) (1, 2)
```

| | | | | |
|---|---|---|---|---|
| 36-40 | F5.2 | arcsec | DEs | Declination apparent place at Epoch (seconds) (1, 2) |
| 42-46 | F5.2 | arcsec | dDE | *DE (Absolute) - DE (HIP) (3) |
| 48-60 | F13.5 | day | Epoch | Julian date (UT1) of observation |
| 62-63 | A2 | --- | Obs | Observer code (4) |
| 65 | A1 | --- | Clamp | Clamp orientation of instrument (5) |
| 67 | A1 | --- | RALimb | Limbs or center of light measured in RA (6) |
| 69 | A1 | --- | DELimb | Limbs or center of light measured in DE (6) |

Note (1): Positions are apparent places, reduced to be systematically consistent with the Hipparcos Catalogue. An apparent place is a geocentric direction of an object that takes into account orbital motion, space motion, light-time, light deflection, and annual aberration. Apparent place is given with respect to the true equator and equinox "of date"; in this case, it is the date of observation found in the Epoch field.
Note (2): In some cases, an observation is made in only right ascension or declination. If only the right ascension coordinate was observed, then the declination seconds field (DEs) will be blank, as will dDE and DELimb. If only the declination coordinate was observed, then the right ascension seconds field (RAs) will be blank, as will dRA and RALimb.
Note (3): The observing program was designed to be an "absolute" catalog, independent of other catalogs. Data reductions were carried out according to previous absolute transit circle catalogs. Following the reductions, comparisons of stars with the Hipparcos Catalogue revealed remaining, unaccounted for systematic deviations in the absolute positions. The authors decided to differentially reduce the absolute positions using the Hipparcos data; those are the position presented in the RAh, RAm, RAs, DE-, DEd, DEm and DEs fields. For users who may be interested in the absolute positions, values of the absolute position minus the differential position are provided in the dRA and dDE fields.
Note (4): The Six-inch Transit Circle measurements were manual, in the sense that a person measured the position while looking through the instrument. Table 3 Rafferty, T.J, Holdenried, E.R., and Urban, S.E. (2016) provides names for the observer code in the Obs field.
Note (5): The Six-inch was reversed (rotated 180° in azimuth) interchanging the east and west pivots approximately every 30 days. The orientation of the instrument was referenced to the location of the clamping device, which fixed the altitude of the instrument after it was pointed to that of a star, and thus referred to either as Clamp West or Clamp East. The flag as the following meaning:

W = Clamp West
E = Clamp East

Note (6): The Sun and planets subtend sizable disks in the instrument. Primarily depending on the object, either the limbs or the center of light were observed. The flag has the following meaning:

L = limb was measured
C = center of light was measured
R = ring was measured (Saturn only)
P = preceding limb was measured (Venus only)
F = following limb was measured (Venus only)
N = North limb was measured (Venus only)
S = South limb was measured (Venus only)

Despite how an object was observed, the W1J00 positions refer to the center of the object.

# Part II

# $W2_{J00}$

## RESULTS OF THE USNO POLE-TO-POLE OBSERVATIONS

**MADE WITH THE**

## SIX-INCH AND SEVEN-INCH TRANSIT CIRCLES

## 1985-1996

**OBSERVATIONS OF THE PLANETS AND MINOR PLANETS**
**CATALOG OF 44,395 STARS ON THE**
**INTERNATIONAL CELESTIAL REFERENCE FRAME (ICRF)**

**By**

**E.R. HOLDENRIED AND T.J. RAFFERTY**

## Introduction

Plans for an observing program concurrently covering both hemispheres (Pole-to-Pole) using the USNO Six-inch Transit Circle and Seven-inch Transit Circle date back to the 1970's (Hughes 1978). The result was to be an absolute catalog tied to the dynamical reference frame. This required both transit circles to be able to observe in the daytime and located at latitudes such that a fundamental determination could be made of the azimuth using circumpolar stars. It was also planned that the bulk of the stars, termed the program stars, would be observed in zones of 15 degrees declination along with a suitable distribution of reference stars to allow differential reduction on a semi-nightly basis.

At this same time, during the 1970's, the European Space Agency (ESA) was studying the feasibility of an astrometric satellite called Hipparcos (Høg 1978). Though the estimated accuracy of Hipparcos was a significant improvement over that of a transit circle, the plans were to tie the Hipparcos positions to Fifth Fundamental Catalog (FK5) (Fricke *et al.* 1988) reference frame. However the FK5 exhibits known systematic errors, therefore the USNO undertook the Pole-to-Pole project to provide an improved global stellar reference frame. Renovation and testing of the Seven-inch Transit Circle delayed the start of the observing program until 1985. The launch of the Hipparcos satellite took place in August, 1989. Even with a revised mission made necessary by the failure of the apogee booster, the satellite was able to operate until August, 1993. The Hipparcos Catalogue (ESA 1997) was released in mid-1997. In the end, the Hipparcos Catalogue was referenced to the International Celestial Reference Frame (ICRF) (Ma *et al.* 1998) and not to FK5. Observations for the Pole-to-Pole project were completed in April, 1995 by the Six-inch Transit Circle and in February, 1996 by the Seven-inch Transit Circle. The Pole-to-Pole project is the latest and largest of a long series of transit circle catalogs produced by the U.S. Naval Observatory. It is also, because of advancing technologies, certainly the last.

It should be pointed out that, unlike most previous USNO transit circle catalogs, in the end the Pole-to-Pole project is not an absolute catalog,that is absolute in the sense that the positions do not explicitly rely on any other stellar observations. With the availability of Hipparcos observational data, it was decided to differentially adjust the Pole-to-Pole to the ICRF using the Hipparcos star positions. A catalog on the ICRF was judged be more useful than one tied to the dynamical reference frame as was the tradition.

**Observing Program -** This Pole-to-Pole catalog contains the combined results of observations made with the Six-inch Transit Circle in Washington, D.C. USA (see Figure 1) and the Seven-inch Transit Circle at the Black Birch station near Blenheim, New Zealand (see Figure 2), between April 1985 and February 1996. This is the second Washington catalog to be referred to the Equinox of J2000.0 and will be referred to from this point on as the $W2_{J00}$.

**Figure 1: Six-inch Transit Circle, Washington DC**

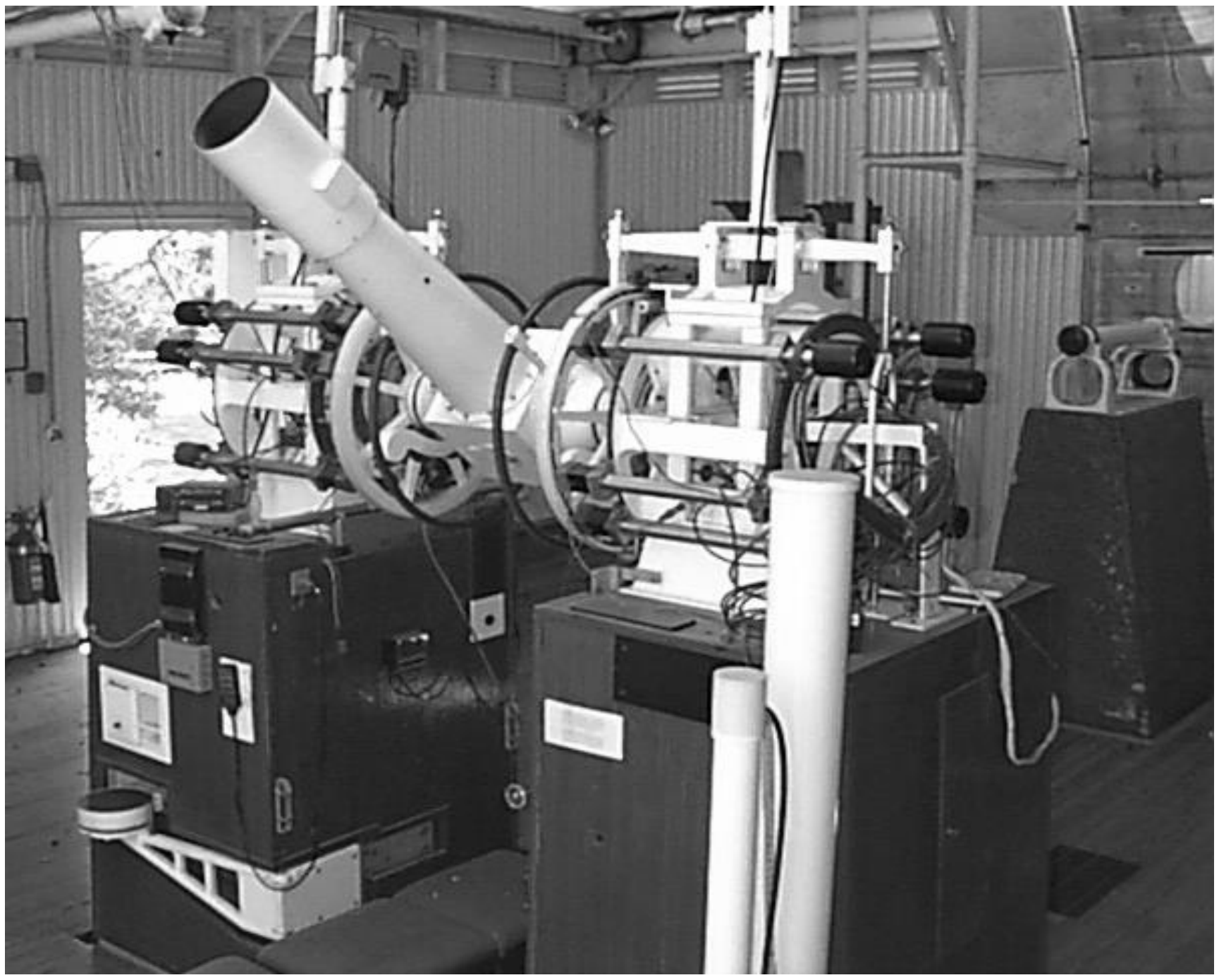

**Figure 2: Seven-inch Transit Circle, Black Birch, New Zealand**

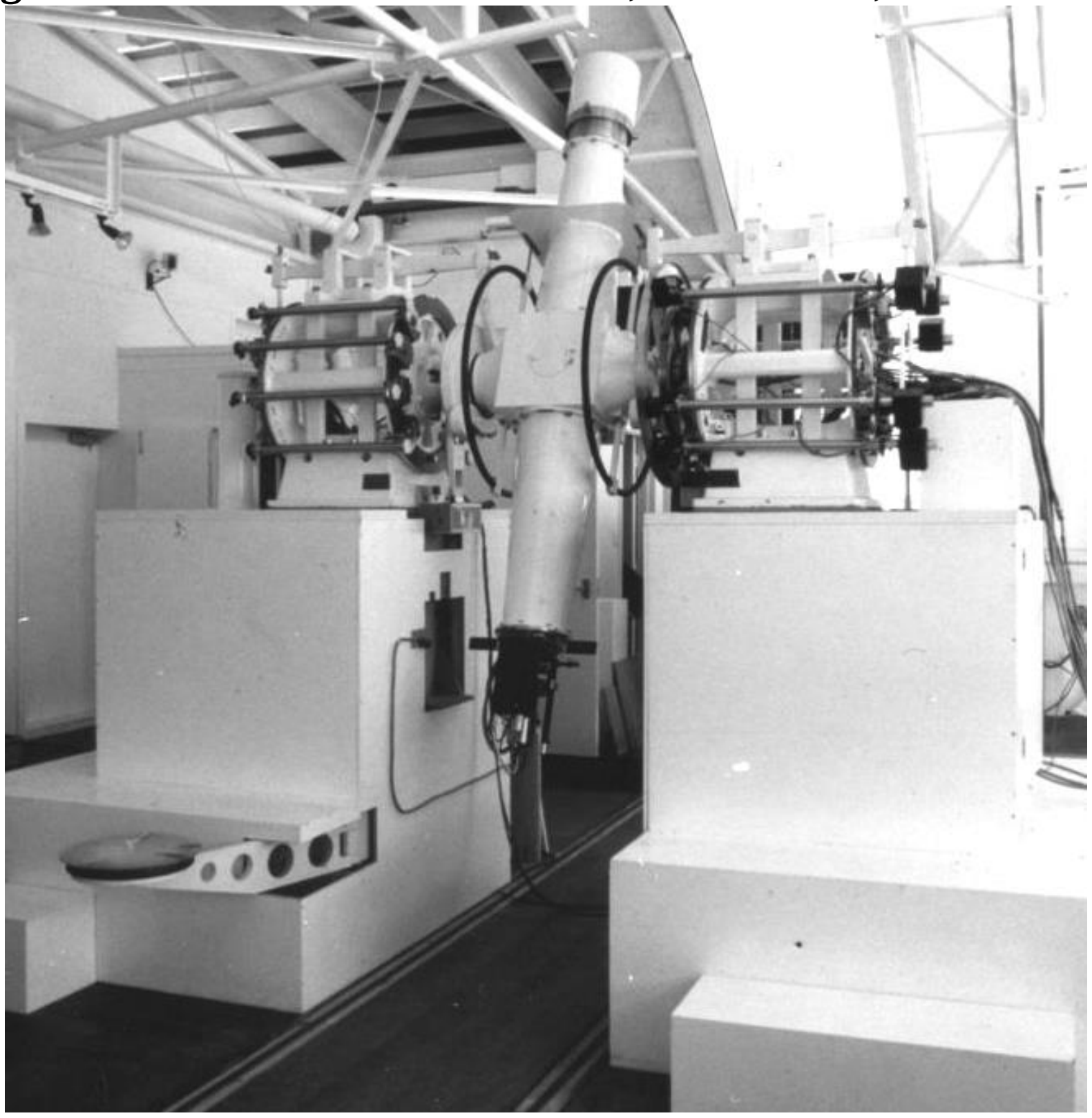

The majority of the celestial objects observed in this program fall into three categories; FK Stars (Fricke *et al.* 1988 and 1991), the program stars and solar system objects. Tables 1 and 2 give the number of stars in each category and the number of observations made of the stars and solar system objects. In the case of the planet observations made with the Seven-inch Transit Circle, a large number of the observations of Mars, Jupiter, and Saturn had to be rejected due to problems with the centering algorithm and were not included in the Table 2. The FK Stars include the FK5, FK5 Supplemental, and Faint Fundamental Stars. The largest number of the program stars were International Reference Stars (IRS) (Corbin 1991), but also included in this class were AGK3R and SRS stars (Corbin 1978) that were not in the IRS. For brevity's sake we shall denote this entire class as IRS.

Table 1: Observations made by the Six-inch Transit Circle located in Washington DC, USA, during 1985-1995

| **Stellar Objects** | | |
|---|---|---|
| **Star Class** | **Number of Obn's** | **Number in Class** |
| IRS | 143310 | 20720 |
| Clocks | 37520 | 230 |
| FK Stars | 95021 | 3288 |
| Refraction (LC) | 5748 | 121 |
| Azimuth (UC) | 7607 | 23 |
| Azimuth (LC) | 7210 | 23 |
| *Day | *9241 | *84 |
| Radio | 3187 | 106 |
| | | |
| Totals | 308844 | 24695 |

| **Sun and Planets** | |
|---|---|
| **Object** | **Number of Obn's** |
| *Sun | *1863 |
| *Mercury | *596 |
| *Venus | *1426 |
| *Mars (day) | *139 |
| Mars (night) | 574 |
| Jupiter | 706 |
| Saturn | 710 |
| Uranus | 711 |
| Neptune | 604 |
| Totals | 7329 |

| **Minor Planets** | |
|---|---|
| **Object** | **Number of Obn's** |
| Ceres | 474 |
| Pallas | 385 |
| Juno | 310 |
| Vesta | 504 |
| Hebe | 286 |
| Iris | 272 |
| Flora | 255 |
| Metis | 216 |
| Eunnomia | 261 |
| Totals | 2963 |

* Observations not reduced

UC = Upper culmination

LC = Lower culmination

Because of the inherent high scatter of the daytime observations and the difficulty of putting them on the same system as the nighttime observations, these observations were not reduced.

Every effort was made to obtain six observations of each star distributed equally between clamps and circles (the terms clamp and circle will be defined later). However, some stars were added to the program too late to obtain that ideal distribution. For the late arrivals it was decided that a minimum of three good observations in right ascension and declination would be required. Of course, the FK stars that served as clock, azimuth, refraction, IRS reference, and day stars accrued many more than the minimum number of observations. In some cases, not enough good observations were made of a star and it does not appear in the final results. Also some solar system observations were found to be discordant and rejected.

Table 2: Observations made by the Seven-inch Transit Circle located at Black Birch, New Zealand, during 1987-1996

| **Stellar Objects** | | |
|---|---|---|
| **Star Class** | **Number of Obn's** | **Number in Class** |
| IRS | 174561 | 21912 |
| Clocks | 29728 | 223 |
| FK Stars | 101464 | 3440 |
| Refraction (LC) | 8086 | 98 |
| Azimuth (UC) | 12295 | 49 |
| Azimuth (LC) | 11812 | 49 |
| *Day | *36343 | *348 |
| Radio | 3690 | 117 |
| | | |
| Totals | 377979 | 26234 |

* Observations not reduced

UC = Upper culmination

LC = Lower culmination

| **Sun and Planets** | |
|---|---|
| **Object** | **Number of Obn's** |
| *Sun | *1166 |
| *Mercury | *461 |
| *Venus | *867 |
| *Mars (day) | *521 |
| Mars (night) | 172 |
| Jupiter | 246 |
| Saturn | 188 |
| Uranus | 579 |
| Neptune | 558 |
| Totals | 4757 |

| **Minor Planets** | |
|---|---|
| **Object** | **Number of Obn's** |
| Ceres | 328 |
| Pallas | 359 |
| Juno | 333 |
| Vesta | 381 |
| Hebe | 275 |
| Iris | 321 |
| Flora | 236 |
| Metis | 188 |
| Eunnomia | 292 |
| Hygiea | 281 |
| Melphomene | 242 |
| Nemausa | 68 |
| Amphitrite | 251 |
| Totals | 3555 |

Table 3 shows the ranges of the magnitude of the stars, the standard error of the mean positons, number of observation per star, and the sky coverage.

Table 3: Summary of the Star Positions in $W2_{J00}$ Catalog

| | Range | Average |
|---|---|---|
| Magnitudes | -1.6 to 9.91 mag | 6.84 |
| RA standard error of the mean | 3 to 441 mas | 68 mas |
| Dec standard error of the mean | 1 to 448 mas | 76 mas |
| RA Number of observations/star | 3 to 411 | 14 |
| Dec Number of observations/star | 2 to 418 | 14 |
| Declination Coverage | -89° 08′ 15″ to +89° 02′ 16″ | |

**Personnel and Acknowledgments -** The program was carried out under the leadership of J.A. Hughes (1985-1992), C.A. Smith (1993), and F.S. Gauss (1993-1998), directors of the Astrometry Department. The observing program was overseen by T.E. Corbin (1985-1993) and T.J. Rafferty (1993-1996). The operations of the Black Birch station in New Zealand were directed by M.D.

Robinson (1985-1989), E.R. Holdenried (1989-1990), T.J. Rafferty (1990-1992), and C.S. Cole (1992-1996). Many Department members served as observers and are listed in Tables 4 and 5.

The preliminary daily reductions and editing of the data were carried out by a team composed of various observers and included at one time or another Holdenried, Miller, Rafferty, Urban, Wycoff, Hall, Crull, Dick, Loader, and Jordan. The final reductions were carried out by Holdenried and Rafferty with close and frequent consultation with C.A. Smith and T.E. Corbin.

Table 4: Six-inch Transit Circle Observers (1985-1995)

| **Name** | **Observer Code** | **Tenure** | | |
|---|---|---|---|---|
| C.S. Cole | CSC | Feb-87 | to | Mar-91 |
| T.E. Corbin | TEC | Apr-85 | to | Apr-95 |
| H.E. Crull | HEC | Dec-88 | to | Apr-95 |
| S.J. Dick | SJD | Apr-87 | to | Jan-88 |
| J.C. Doty | JCD | Feb-87 | to | Apr-95 |
| R. Etheridge | RE | Nov-85 | to | Aug-86 |
| F.S. Gauss | FSG | Apr-85 | to | Apr-89 |
| M.E. Germain | MEG | Oct-93 | to | Apr-94 |
| D.M. Hall | DMH | Sep-86 | to | Apr-95 |
| G.S. Hennessy | GSH | Oct-93 | to | Nov-94 |
| J.L. Hershey | JH | Apr-85 | to | Mar-88 |
| R.B. Hindsley | RBH | Dec-86 | to | Jun-94 |
| E.R. Holdenried | ERH | Apr-85 | to | Apr-95 |
| E.S. Jackson | ESJ | Apr-85 | to | Apr-95 |
| I. Jordan | IAN | Jun-88 | to | Mar-90 |
| V. Kallarakal | KAL | Apr-85 | to | Apr-95 |
| J.C. Martin | JCM | May-90 | to | Jan-95 |
| J.M. Muse | JMM | Nov-90 | to | Oct-91 |
| R.J. Miller | RJM | Apr-85 | to | Apr-95 |
| M.D. Robinson | MDR | May-91 | to | Apr-95 |
| T.J. Rafferty | TJR | Apr-85 | to | Apr-95 |
| C.B. Sande | CBS | Feb-90 | to | Jan-93 |
| D.K. Scott | DS | May-85 | to | Nov-86 |
| C.A. Smith | CAS | Apr-85 | to | Mar-91 |
| S.E. Urban | SEU | Sep-85 | to | Apr-95 |
| G.L. Wycoff | GLW | Sep-85 | to | Apr-95 |
| Z.G. Yao | ZGY | Apr-85 | to | Apr-95 |

Instrumentation support was provided by Gauss, Hughes, Robinson, Rafferty, Urban, Clinton G. Hollins, Russell Millington, Paul A. Heermans, Charles J. Carpenter, and William L, Dunn, Sr., as well as the personnel of the USNO Instrument Shop: John W. Pohlman, Jr., Gary Wieder, Edward C. Matthews, John W. Bowles, Dave Smith, Stephen J. Boretos, and Boyd Simpson.

James Hughes died before the completion of this project. His thoughtful planning and leadership of this undertaking played an important role in its success. It is our wish to present this catalog as a memorial to him.

Table 5: Seven-inch Transit Circle Observers (1987-1996)

| **Name** | **Tenure** | | |
|---|---|---|---|
| C.S. Cole | May 1991 | to | February 1996 |
| W.B. Dunn | June 1987 | to | November 1987 |
| E. Durham | December 1988 | to | September 1991 |
| E.R. Holdenried | July 1987 | to | June 1990 |
| R. Hudson | April 1992 | to | February 1996 |
| I. Jordan | May 1990 | to | September 1995 |
| T. Love | January 1994 | to | February 1994 |
| B.R. Loader | June 1987 | to | February 1996 |
| R. Millington | June 1987 | to | February 1996 |
| J. Priestley | September 1993 | to | September 1993 |
| M.D. Robinson | June 1987 | to | December 1995 |
| T.J. Rafferty | June 1987 | to | March 1992 |
| R.C. Stone | June 1987 | to | December 1988 |
| A. Wadsworth | November 1991 | to | September 1995 |

## Instrumentation, Accessories, and Procedures

The Six-inch Transit Circle (Six-inch) was built by the Warner and Swasey Company and has operated from the U.S. Naval Observatory in Washington, DC since 1897. The visual, two axis micrometer was the same as used during the previous programs, the $W5_{50}$ (Hughes and Scott 1982) and the $W1_{J00}$ (see Part I of this paper). For more information on transit circles than is provided here or in the $W1_{J00}$ introduction, see Podobed (1962), Watts (1950) and Hemenway (1966).

The Seven-inch Transit Circle (Seven-inch) was built by the USNO Instrument Shop in 1948. Its previous program, the $WL_{50}$ (Hughes *et al.* 1992), was a visual catalog made from El Leoncito, Argentina. For the $W2_{J00}$, the telescope was located on the Black Birch ridge at an elevation of 1350m, 20 km southwest of the city of Blenheim, New Zealand. The station, referred to as the Black Birch Astrometric Observatory, was at latitude of $-41^\circ\,44'\,41''\!.4$ and a longitude of $173^\circ\,48'\,11''\!.99$ East.

**New Equipment -** For this catalog, because of its long and continuous series of excellent visual catalogs, the Six-inch remained a visual instrument. However major changes to the Six-inch made prior to the start of the $W2_{J00}$ observing program included a second glass circle, two additional magnitude screens, and an upgrade to the photoelectric circle scanning system. At the beginning of Circle Two, the photoelectric scanners were replaced with CCD devices (Rafferty and Klock, 1986).

Major changes to the Seven-inch made after its completion of the $WL_{50}$ and prior to the beginning of the $W2_{J00}$ observing program included the replacement of the visual micrometer with one using an image dissector as the detector (Hughes *et al.* 1986), the installation of a temperature compensating objective built by the Farrand Optical Corporation of New York, the installation of two graduated glass circles mounted on steel wheels fabricated by Heidenhain Corporation of Germany, and the installation of a new photoelectric system for scanning the graduated circles. As in the case of the Six-inch, at the beginning of Circle Two, the circle scanning system was upgraded to use CCD's.

## Procedures

As was stated previously, the plan for the $W2_{J00}$ was to create an absolute catalog following the same observational and reduction procedures, with some minor modifications, as were used in forming the $W1_{J00}$. Even though ultimately the $W1_{J00}$ was not realized as an absolute catalog the reduction procedures for a differential catalog are, for the most part, the same. Thus the introduction to the $W1_{J00}$ provides most of the details of the methods used to form the $W2_{J00}$.

**Instrument Reversal -** Both transit circles were equipped with clamping devices that prevented any motion of the telescope in altitude during an observation. These devices were located near one of the pivots of the instrument and provide a convenient way of referencing the orientation of the telescope; that is the telescope could be in either a “Clamp East” or “Clamp West” orientation. Both transit circles were reversed (rotated 180 degrees about the vertical), thus changing clamp, approximately every 30 days. This was done to mitigate any clamp-dependent systematic errors.

**Circle Rotation -** Midway through the observing program the wheels supporting the graduated circles of each of the transit circles were turned so that after this rotation the same altitude of the telescope was measured with different circle graduations (divisions). The observations taken before the circle rotation were referred to as from “Circle One” and after from “Circle Two”.

**Observing Tours -** Observations were grouped into “tours”. Usually two tours were taken per night, dividing the night in half between two observers. Each tour contained determinations of the collimation, level, nadir, azimuth, and flexure taken at two to three hour intervals for the nighttime tours. For the Seven-inch, azimuth determinations were made hourly due to concern over instabilities of the piers. For each tour, observations were made of selected groups of stars to determine corrections to the clock, azimuth, and refraction. In addition during a tour a subset of FK5 stars, following a concept developed by Küstner (Küstner 1900) and hence referred to as Küstner stars, distributed over the entire sky was observed to check for nightly variations of the instrument or atmosphere over large angles. The IRS, grouped in zones of 15 degrees of declination, were observed with FK5 reference stars to allow nightly differential reductions. Only one zone of IRS stars were observed during a tour. The differential reductions could be produced practically instantaneously and thus provided a real-time check on the quality of the observations. Differential observations also can be used to reduce the random and systematic errors in the data.

**Star Selector -** The requirement imposed by the even distribution in time and zenith distance of the clock, azimuth, refraction, and Küstner stars as well as the need to choose IRS and their reference stars while maintaining a balance of all observations over the Clamps and Circles necessitated the development of an automatic method of selecting the stars to be observed for each tour. The logical criteria for this Star Selector software were constructed by T. Corbin, while the actual coding was done by F.S. Gauss. The software also benefited the observing efficiency, allowing the observer to choose the interval between stars (usually one and a half to two minutes), to identify holes in the star list as the program progressed and to more easily add new lists of stars and balance the types of stars observed during a tour.

## Right Ascension

**Micrometer Screw Corrections -** The Six-inch traveling micrometer was driven by a screw to which was attached a circular encoder for measuring the micrometer's position; screw errors were applied to the encoder output. The position of the Seven-inch traveling micrometer, although driven by a screw, was directly measured by a linear encoder. These measures did not have to be corrected. The Six-inch micrometer screw errors were measured three times; prior to the program in 1984, at the change between Circles in 1989, and finally at the end of the program in 1995. The errors determined from the measures in 1984 and 1989 agreed closely enough that they were combined. Progressive and periodic errors were found. Both the progressive and the periodic errors derived from the 1995 measures differed significantly from the 1989 results. Therefore time interpolated values of the screw errors (progressive and periodic) were calculated for observations taken between the 1989 and 1995 sets (i.e. all observations on Circle Two), and applied to each micrometer encoder reading.

**Inclination Corrections -** Normally the micrometer of a transit circle is adjusted so that the scale by which right ascension is measured exhibits no inclination with respect to the sky. That is, the wires used to measure an object’s position in right ascension must be parallel to the meridian. In the case of the Six-inch, during the change from Circle One to Circle Two, these wires were incorrectly adjusted leaving them inclined to the meridian and resulting in a significant correction that must be applied to the right ascension measures and is dependent on the vertical place in the field of the object observed. The micrometer was correctly adjusted after the third clamp on Circle Two. In the case of the Seven-inch, it appears from an analysis of the observations that an inclination was also introduced when changing from Circle One to Circle Two. In this case the initial adjustment of the wires was correct, but some kind of relaxation phenomenon caused them to shift later. The inclination was derived from star observations and found to vary with clamp. Thus, for Circle Two, a clamp dependent correction for the inclination is applied to the right ascension measures.

**Corrections for Pivot Irregularities -** The telescope rests on two cylinders, called pivots, attached to opposite sides of the cube to which the telescope tubes are fixed and it is on these pivots that the instrument rotates to point to designated altitudes. Although the pivots are carefully machined to be round, the deviations from roundness must be measured and applied as corrections to the right ascension data. In the case of the Six-inch, pivot irregularities were measured on four different occasions after the pivots were lapped in 1963, the last measures being taken in 1989 at the break between circles. No significant change was detected so all measures were averaged and applied as a function of the altitude of the telescope. The pivot irregularities of the Seven-inch were measured in 1973, and at the break between circles in this program in 1992. No significant change was evident so a mean was formed and applied as described above.

**Collimation -** The optical collimation of each telescope was determined at intervals of between two and three hours by means of horizontal collimating telescopes mounted to the north and south of the main telescope but within the telescope pavilion. A mean collimation was formed for a tour and applied to each observation. For further information see the description in the $W1_{J00}$ introduction.

**Level -** The level of each telescope was measured by auto-collimating on the reflection of the micrometer wires in a basin of mercury placed beneath the instrument in the direction of the nadir. The amount of displacement from the collimation point was the value taken as the level. It was measured at intervals of between two and three hours, a linear rate computed, and the value interpolated to the time of transit. For further information see the description in the $W1_{J00}$ introduction.

**Azimuth -** The azimuth of the mires, also called marks, (pin-hole light sources mounted several hundred meters to the north and south of each transit circle) with respect to each telescope, was measured at intervals of between one and three hours depending on the telescope. Because of indications of an unstable azimuth, the azimuths of the mires of the Seven-inch were measured much more frequently (as often as once an hour) than were the azimuths of the Six-inch, which were observed about every two to three hours. The azimuth of the mires with respect to the celestial pole was determined by observations of a special set of circumpolar stars (azimuth stars). The azimuth of the mires with respect to the telescope was added to the azimuth of the mires with respect to the celestial pole to form the azimuth of the telescope. A linear rate was computed from the difference between consecutive observations of the azimuths of the mires and the value interpolated to the time of transit. For further information see the description in the $W1_{J00}$ introduction.

**Clock Corrections and Sidereal Time -** Cesium frequency standards were used for both transit circles. A once per second pulse from the clocks was used to trigger an interrupt-driven routine in the data acquisition computer that maintained the time in a common area accessible to all programs and was accurate to approximately 30 microseconds. The clock time was corrected for the variation of longitude (provided by IERS), the Equation of the Equinoxes, and a correction derived from observations of a special set of stars (clock stars). The clock stars for both telescopes consisted of the same set of 203 FK5, Part I, stars evenly distributed in right ascension and declination between declinations -30 and +30 degrees. A clock correction was derived from the mean observed minus calculated position (*O-C*) of clock stars observed during a tour. On the average between five and nine, and never less than four, clock stars were observed per tour.

**Clamp Differences -** Differences between the right ascension (*O-C*)'s of the same stars observed on Clamp East and Clamp West were grouped in 5-degree zones of zenith distance and averaged. Half of the average zonal difference was added to or subtracted from all observations in that zone depending on the clamp on which the observations were made. For further information see the description in the $W1_{J00}$ introduction.

**Circle Differences -** Differences between the right ascension (*O-C*)'s of the same stars observed on Circle One and Circle Two were grouped in 5-degree zones of zenith distance and averaged. This was done after the clamp differences were applied. Half of the average zonal difference was added to or subtracted from all observations in that zone depending on the circle on which the observations were made. For further information see the description in the $W1_{J00}$ introduction.

**Tour Adjustments -** Differential adjustments were applied to each tour from a least squares fit to a set of Hipparcos reference stars. This set of stars was distributed over the entire sky and consisted of those Küstner stars in the tour that were observed by Hipparcos and were not found to exhibit multiplicity. A number of numerical models incorporating coefficients that depended on various combinations of the zenith distance, the sine and cosine of the zenith distance, and arguments of time and powers of time, were tested for each tour. The one providing the best fit was used. After the tour corrections were applied, the (*O-C*)'s of the Hipparcos stars, when averaged by star and collected as a function of declination, exhibited slight systematic deviations about zero. A cubic spline was fit to these deviations and applied to all observations as a function of declination. Table 6 presents information about the estimated standard deviation of the models for the tour corrections.

Table 6: Right ascension tour corrections

| **Estimated standard deviation of the models<br>Right Ascensions (units = arc seconds)** | | | | | |
|---|---|---|---|---|---|
| **Telescope** | **Circle** | **Median** | **Minimum** | **Maximum** | **No. of Tours** |
| Six-inch | Circle One | 0.186 | 0.022 | 0.469 | 1969 |
| Six-inch | Circle Two | 0.182 | 0.010 | 0.430 | 2719 |
| Seven-inch | Circle One | 0.194 | 0.052 | 0.383 | 1452 |
| Seven-inch | Circle Two | 0.201 | 0.046 | 0.421 | 1853 |

**Image Dissector -** It was discovered after the program had been under way that an error in the Seven-inch's image dissector processing software was causing a systematic offset in the observations that depended on observed magnitude. The amount of this offset was one pixel, and only occurred under certain conditions. This problem was corrected after the circle rotation but affected all of the data from Circle One. A correction for the effect was developed from a statistical analysis of the Circle One observations and applied to the data. However, because it is a statistical correction, individual observations may be biased, although at a reduced level, by some residual error.

## Declination

**Micrometer Screw Corrections -** For the Six-inch, the progressive error of the declination screw was measured before the observing program began and after it ended, and was judged not to have changed significantly between the two determinations; therefore the two sets were averaged. This averaged screw error was applied to the mean declination screw value of each observation. No periodic screw error was detected. Corrections for the inclination of the micrometer wires and reduction to the meridian were also applied to the mean screw value. The micrometer of the Seven-inch had no movable declination slide so no corrections were necessary.

**Nadir -** The nadir point, in conjunction with the level, was determined by auto-collimating over a mercury basin. This was done at intervals of between two and three hours, a linear rate computed, and the value interpolated, as a function of time, to the time of transit.

**Circle Diameter Corrections -** The Høg method (Høg 1961), that used three microscope diameters, for determining the circle diameter corrections was used for both transit circles. Changes in the circle diameter corrections for both transit circles were noticed after the rotation of the circles midway through their observing programs and different sets of circle diameter corrections were used for each Circle.

**Assumed Latitude and the Variation of Latitude Corrections -** An assumed latitude of +38° 55′ 14.″257 was used in the reductions for the Six-inch (Hughes *et al*. 1975) and -41° 44′ 41.″387 was used for the Seven-inch furnished by New Zealand Department of Land and Survey. Corrections for the variation of latitude were provided by the IERS.

**Refraction Corrections -** Air temperature was measured to 0.1°C using Hy-cal platinum resistance probes. Air pressure was measured to 0.1mm of mercury using Setra barometers. Dew point was measured to 5.0°F using Honeywell probes treated with lithium chloride activation solution and dried. Corrections for refraction came from the Fourth Edition of Pulkovo Refraction Tables. See the $W1_{J00}$ introduction for details.

The latitude, height above sea level, and gravity constant for the Washington Six-inch transit circle are the same values used in the $W1_{J00}$. For the Seven-inch transit circle in New Zealand, the latitude is -41.°744829722, the height above sea level 1366m, and the gravity constant 0.99814549.

**Clamp Differences -** Differences between the declination (*O-C*)'s of the same stars observed on Clamp East and Clamp West were grouped in zones of 5 degrees of zenith distance and averaged. Half of the average zonal difference was added to or subtracted from all observations in that zone depending on the clamp on which the observations were made. For further details see the description in the $W1_{J00}$ introduction.

**Circle Differences -** Differences between the declination (*O-C*)'s of the same stars observed on Circle One and Circle Two were grouped in zones of 5 degrees of zenith distance and averaged.

This was done after the clamp differences were applied. Half of the average zonal difference was added or subtracted to all observations in that zone depending on the circle on which the observations were made. For further details on see the description in the $W1_{J00}$ introduction.

**Flexure -** Measurements of the instrumental flexure determined from the horizontal collimators were made for each transit circle but these exhibited very large variations. Since a more consistent determination of the flexure can be determined from the star observations (Holdenried and Rafferty 1997), the flexure determined from the horizontal collimators was not applied.

**Latitude, Constant of Refraction, and Flexure Corrections -** The new method, developed for the $W1_{J00}$, to determine the corrections to the constant of refraction and flexure using all FK5 stars, and the correction to the assumed latitude from the circumpolar observations (Holdenried and Rafferty 1997) was not used for the $W2_{J00}$. Although Six-inch observations gave excellent results, when used with the Seven-inch observations systematic differences as a function of zenith distance were still apparent. The source of these differences was not discovered. Instead a cubic spline was used to adjust the observed positions to the Hipparcos system. For consistency the same method was used on the Six-inch observations.

**Tour Adjustments -** Differential adjustments were applied to each tour from a least squares fit to a set of Hipparcos reference stars. A number of models were tested for each tour and the one providing the best fit was used. The models incorporated coefficients that depended on zenith distance, the tangent and sine of the zenith distance, and time. Table 7 presents information about the estimated standard deviation of the models for the tour corrections.

Table 7: Declination tour corrections

| **Estimated standard deviation of the models<br>Declination (units = arc seconds)** | | | | | |
|---|---|---|---|---|---|
| **Telescope** | **Circle** | **Median** | **Minimum** | **Maximum** | **No. of Tours** |
| Six-inch | Circle One | 0.221 | 0.017 | 0.596 | 1947 |
| Six-inch | Circle Two | 0.216 | 0.010 | 0.644 | 2714 |
| Seven-inch | Circle One | 0.274 | 0.060 | 0.757 | 1447 |
| Seven-inch | Circle Two | 0.293 | 0.020 | 0.648 | 1854 |

**Image Dissector -** As with the right ascensions, it was discovered after the program had been under way for a considerable time that an error in the image dissector processing software was causing a systematic offset in the observations of the Seven-inch, that depended on observed magnitude. This problem was corrected after the circle rotation but affected all of the data from Circle One. A correction for the effect was developed from a statistical analysis of the Circle One observations and applied to the data. However, because it is a statistical correction, individual observations may be biased, although at a reduced level, by some residual error.

## Combined Observations

The locations of the two transit circles allowed for a nearly 70 degree overlap in the declinations accessible to each telescope. For those stars in this overlap region, the observations were combined in a weighted mean using:

$$\boldsymbol{\bar{o}} = \sum_{i=1}^{n} \boldsymbol{\omega_i\, o_i} \,/ \sum_{i=1}^{n} \boldsymbol{\omega_i}$$

where:

$\boldsymbol{o_i}$ = single observation
$\boldsymbol{\omega_i}$ = weight for $\boldsymbol{o_i}$ based on its zenith distance
$\boldsymbol{\bar{o}}$ = mean observed position
$n =$ number of observations

The weights (given in Table 8) were based on the mean standard deviation of a single observation as a function of zenith distance and were an attempt to account for the degradation suffered by observations made through large air masses.

Table 8: Weights applied when combining observations from the two transit circles to form a single position. Zenith distance in units of degrees.

| **Weights for combined observations** | | |
|---|---|---|
| **Zenith Distance** | **RA obs** | **Dec obs** |
| 80.0 | 0.00 | 0.00 |
| 75.0 | 0.16 | 0.03 |
| 70.0 | 0.29 | 0.14 |
| 65.0 | 0.42 | 0.24 |
| 60.0 | 0.56 | 0.34 |
| 55.0 | 0.68 | 0.43 |
| 50.0 | 0.78 | 0.53 |
| 45.0 | 0.87 | 0.62 |
| 40.0 | 0.93 | 0.72 |
| 35.0 | 0.97 | 0.82 |
| 30.0 | 0.98 | 0.89 |
| 25.0 | 0.99 | 0.94 |
| 20.0 | 1.00 | 0.97 |
| 15.0 | 1.00 | 0.99 |
| 10.0 | 1.00 | 1.00 |
| 5.0 | 1.00 | 1.00 |
| 0.0 | 1.00 | 1.00 |

## Errors of Observation and Position

The weighted standard deviation of a single observation was determined using:

$$\sigma_o = \sqrt{\sum_{i=1}^{n} \omega_i (o_i - \bar{o})^2 / (n-1)}$$

and the weighted standard deviation of the mean using:

$$\boldsymbol{\bar{\sigma}_{\bar{o}}} = \sqrt{\sum_{i=1}^{n} \boldsymbol{\omega_i (o_i - \bar{o})^2 / (n-1)\, n}}$$

where:

$\boldsymbol{o_i}$ = single observed position
$\boldsymbol{\omega_i}$ = weight for $\boldsymbol{o_i}$ based on its zenith distance
$\boldsymbol{\bar{o}}$ = mean observed position
$n$ = number of observations

The standard deviation of the mean is given with the position of each star. For the stars observed with both transit circles, the mean position and standard deviation of the mean as determined by each instrument are given as well as the weighted mean and weighted standard deviation of the mean of the combined data.

Tables 9 and 10 group into five degree zones of declination: the average standard deviations of a single observation, the average standard error of the mean, and the number of stars. The average standard deviation of a single observation was close to 200 mas in right ascension and 215 mas in declination. The average standard deviation of the mean position for a star varied by the number of observations. Since the majority of stars in each zone were IRS, which averaged six observations each (two on Circle One and four on Circle Two), the average standard deviation of the mean was close to 70 mas in right ascension and 77 mas in declination. In the declination zone -5 to +5 degrees, both the Six-inch and Seven-inch observed the same IRS stars doubling the number of observations each received, and this manifests itself in a sharp drop in the average standard deviation of the mean.

Table 9: Weighted standard deviations and weighted standard errors of the mean in right ascension for each transit circle as well as these statistics for the final positions averaged over zones of five degrees.

| **Right ascension errors** | | | | | | | | | | | |
|---|---|---|---|---|---|---|---|---|---|---|---|
| | | | **Six-inch** | | | **Seven-inch** | | | **Total** | | |
| **Declination Range** | | | **s.d. mas** | **s.e.m mas** | **no. stars** | **s.d. mas** | **s.e.m mas** | **no. stars** | **s.d. mas** | **s.e.m. mas** | **no. stars** |
| 90 | to | 85 | 234 | 68 | 94 | | | | 234 | 68 | 94 |
| 85 | to | 80 | 218 | 71 | 263 | | | | 218 | 71 | 263 |
| 80 | to | 75 | 203 | 70 | 431 | | | | 203 | 70 | 431 |
| 75 | to | 70 | 208 | 74 | 567 | | | | 208 | 74 | 567 |
| 70 | to | 65 | 203 | 72 | 728 | | | | 203 | 72 | 728 |
| 65 | to | 60 | 203 | 74 | 863 | | | | 203 | 74 | 863 |
| 60 | to | 55 | 197 | 72 | 1006 | | | | 197 | 72 | 1006 |
| 55 | to | 50 | 197 | 72 | 1140 | | | | 197 | 72 | 1140 |
| 50 | to | 45 | 193 | 71 | 1272 | | | | 201 | 74 | 1272 |
| 45 | to | 40 | 192 | 71 | 1396 | | | | 192 | 71 | 1396 |
| 40 | to | 35 | 188 | 70 | 1512 | | | | 188 | 70 | 1512 |
| 35 | to | 30 | 188 | 70 | 1571 | 128 | 94 | 14 | 188 | 70 | 1571 |
| 30 | to | 25 | 190 | 72 | 1743 | 165 | 63 | 196 | 189 | 71 | 1743 |
| 25 | to | 20 | 191 | 70 | 1727 | 181 | 55 | 233 | 191 | 69 | 1727 |
| 20 | to | 15 | 199 | 74 | 1766 | 187 | 48 | 230 | 199 | 73 | 1766 |
| 15 | to | 10 | 192 | 72 | 1781 | 191 | 45 | 206 | 193 | 71 | 1781 |
| 10 | to | 5 | 191 | 71 | 1800 | 193 | 44 | 251 | 192 | 70 | 1800 |
| 5 | to | 0 | 193 | 73 | 1809 | 209 | 76 | 1774 | 208 | 55 | 1811 |
| 0 | to | -5 | 194 | 74 | 1700 | 208 | 74 | 1769 | 208 | 55 | 1773 |
| -5 | to | -10 | 189 | 52 | 289 | 206 | 72 | 1801 | 205 | 70 | 1801 |
| -10 | to | -15 | 185 | 48 | 206 | 200 | 70 | 1800 | 200 | 69 | 1800 |
| -15 | to | -20 | 183 | 52 | 203 | 200 | 70 | 1808 | 200 | 69 | 1808 |
| -20 | to | -25 | 163 | 51 | 198 | 199 | 69 | 1660 | 199 | 69 | 1660 |
| -25 | to | -30 | 149 | 56 | 188 | 198 | 68 | 1573 | 196 | 68 | 1573 |
| -30 | to | -35 | 123 | 98 | 66 | 196 | 69 | 1747 | 196 | 69 | 1747 |
| -35 | to | -40 | | | | 199 | 69 | 1799 | 199 | 69 | 1799 |
| -40 | to | -45 | | | | 200 | 69 | 1618 | 200 | 69 | 1618 |
| -45 | to | -50 | | | | 198 | 70 | 1609 | 198 | 70 | 1609 |
| -50 | to | -55 | | | | 201 | 70 | 1284 | 201 | 70 | 1284 |
| -55 | to | -60 | | | | 204 | 71 | 1150 | 204 | 71 | 1150 |
| -60 | to | -65 | | | | 201 | 70 | 994 | 201 | 70 | 994 |
| -65 | to | -70 | | | | 205 | 70 | 783 | 205 | 70 | 783 |
| -70 | to | -75 | | | | 208 | 70 | 635 | 208 | 70 | 635 |
| -75 | to | -80 | | | | 222 | 73 | 485 | 222 | 73 | 485 |
| -80 | to | -85 | | | | 229 | 70 | 298 | 229 | 70 | 298 |
| -85 | to | -90 | | | | 237 | 65 | 107 | 237 | 65 | 107 |

Table 10: Weighted standard deviations and weighted standard errors of the mean in declination for each transit circle as well as these statistics for the final positions averaged over zones of five degrees.

| **Declination errors** | | | | | | | | | | | | |
|---|---|---|---|---|---|---|---|---|---|---|---|---|
| | | | **Six-inch** | | | **Seven-inch** | | | **Total** | | | |
| **Declination Range** | | | **s.d. mas** | **s.e.m. mas** | **no. stars** | **s.d. mas** | **s.e.m. mas** | **no. stars** | **s.d. mas** | **s.e.m. mas** | **no. stars** |
| 90 | to | 85 | 251 | 72 | 94 | | | | 251 | 72 | 94 |
| 85 | to | 80 | 226 | 73 | 263 | | | | 226 | 73 | 263 |
| 80 | to | 75 | 216 | 74 | 431 | | | | 216 | 74 | 431 |
| 75 | to | 70 | 212 | 75 | 567 | | | | 212 | 75 | 567 |
| 70 | to | 65 | 215 | 76 | 728 | | | | 215 | 76 | 728 |
| 65 | to | 60 | 213 | 78 | 863 | | | | 213 | 78 | 863 |
| 60 | to | 55 | 206 | 74 | 1006 | | | | 206 | 74 | 1006 |
| 55 | to | 50 | 209 | 76 | 1140 | | | | 209 | 76 | 1140 |
| 50 | to | 45 | 201 | 74 | 1272 | | | | 201 | 74 | 1272 |
| 45 | to | 40 | 194 | 71 | 1396 | | | | 194 | 71 | 1396 |
| 40 | to | 35 | 200 | 73 | 1512 | | | | 200 | 73 | 1512 |
| 35 | to | 30 | 201 | 75 | 1571 | 142 | 189 | 14 | 201 | 75 | 1571 |
| 30 | to | 25 | 201 | 75 | 1743 | 210 | 114 | 196 | 202 | 75 | 1743 |
| 25 | to | 20 | 207 | 76 | 1727 | 227 | 91 | 233 | 209 | 76 | 1727 |
| 20 | to | 15 | 206 | 77 | 1766 | 219 | 72 | 230 | 208 | 77 | 1766 |
| 15 | to | 10 | 209 | 81 | 1781 | 218 | 65 | 206 | 211 | 80 | 1781 |
| 10 | to | 5 | 201 | 79 | 1800 | 216 | 58 | 251 | 203 | 78 | 1800 |
| 5 | to | 0 | 194 | 80 | 1809 | 217 | 93 | 1774 | 215 | 64 | 1811 |
| 0 | to | -5 | 186 | 80 | 1700 | 218 | 87 | 1769 | 211 | 63 | 1773 |
| -5 | to | -10 | 183 | 59 | 289 | 227 | 86 | 1801 | 225 | 84 | 1801 |
| -10 | to | -15 | 179 | 57 | 206 | 220 | 80 | 1800 | 218 | 80 | 1800 |
| -15 | to | -20 | 173 | 62 | 203 | 223 | 80 | 1808 | 221 | 79 | 1808 |
| -20 | to | -25 | 173 | 70 | 198 | 221 | 78 | 1660 | 219 | 77 | 1660 |
| -25 | to | -30 | 165 | 86 | 188 | 221 | 76 | 1573 | 220 | 76 | 1573 |
| -30 | to | -35 | 124 | 148 | 66 | 217 | 77 | 1747 | 217 | 77 | 1747 |
| -35 | to | -40 | | | | 220 | 77 | 1799 | 220 | 77 | 1799 |
| -40 | to | -45 | | | | 225 | 78 | 1618 | 225 | 78 | 1618 |
| -45 | to | -50 | | | | 222 | 78 | 1609 | 222 | 78 | 1609 |
| -50 | to | -55 | | | | 222 | 77 | 1284 | 222 | 77 | 1284 |
| -55 | to | -60 | | | | 227 | 79 | 1150 | 227 | 79 | 1150 |
| -60 | to | -65 | | | | 228 | 79 | 994 | 228 | 79 | 994 |
| -65 | to | -70 | | | | 235 | 81 | 783 | 235 | 81 | 783 |
| -70 | to | -75 | | | | 247 | 83 | 635 | 247 | 83 | 635 |
| -75 | to | -80 | | | | 266 | 87 | 485 | 266 | 87 | 485 |
| -80 | to | -85 | | | | 275 | 84 | 298 | 275 | 84 | 298 |
| -85 | to | -90 | | | | 298 | 79 | 107 | 298 | 79 | 107 |

## Epochs

The average epoch of a position in right ascension is 1991.53 and in declination 1991.52. However, because the Seven-inch started observing about a year after the Six-inch, there is a pronounced dependence on declination of the epochs of individual stars. This is shown in Table 11 where the epochs have been averaged over the same 5-degree zones in declination that were used in the preceding tables for the errors.

Table 11: Mean epochs for the final positions.

| Declination Range in degrees | | | RA | Dec |
|---|---|---|---|---|
| 90 | to | 85 | 1990.80 | 1990.76 |
| 85 | to | 80 | 1990.73 | 1990.72 |
| 80 | to | 75 | 1990.70 | 1990.68 |
| 75 | to | 70 | 1990.73 | 1990.71 |
| 70 | to | 65 | 1990.76 | 1990.73 |
| 65 | to | 60 | 1990.80 | 1990.76 |
| 60 | to | 55 | 1990.79 | 1990.74 |
| 55 | to | 50 | 1990.86 | 1990.83 |
| 50 | to | 45 | 1990.83 | 1990.80 |
| 45 | to | 40 | 1990.83 | 1990.80 |
| 40 | to | 35 | 1990.86 | 1990.83 |
| 35 | to | 30 | 1990.79 | 1990.76 |
| 30 | to | 25 | 1990.84 | 1990.80 |
| 25 | to | 20 | 1990.87 | 1990.82 |
| 20 | to | 15 | 1990.84 | 1990.80 |
| 15 | to | 10 | 1990.85 | 1990.82 |
| 10 | to | 5 | 1990.90 | 1990.85 |
| 5 | to | 0 | 1991.51 | 1991.45 |
| 0 | to | -5 | 1991.58 | 1991.58 |
| -5 | to | -10 | 1992.14 | 1992.15 |
| -10 | to | -15 | 1992.13 | 1992.16 |
| -15 | to | -20 | 1992.14 | 1992.16 |
| -20 | to | -25 | 1992.13 | 1992.14 |
| -25 | to | -30 | 1992.12 | 1992.14 |
| -30 | to | -35 | 1992.16 | 1992.18 |
| -35 | to | -40 | 1992.16 | 1992.17 |
| -40 | to | -45 | 1992.23 | 1992.25 |
| -45 | to | -50 | 1992.21 | 1992.22 |
| -50 | to | -55 | 1992.15 | 1992.16 |
| -55 | to | -60 | 1992.27 | 1992.27 |
| -60 | to | -65 | 1992.21 | 1992.22 |
| -65 | to | -70 | 1992.24 | 1992.25 |
| -70 | to | -75 | 1992.22 | 1992.21 |
| -75 | to | -80 | 1992.25 | 1992.26 |
| -80 | to | -85 | 1992.30 | 1992.30 |
| -85 | to | -90 | 1992.26 | 1992.23 |

## Double Stars

A few double stars observed by both transit circles showed significant differences. For example, in some cases the image dissector on the Seven-inch could not split doubles that the observers on the Six-inch were able to resolve. In those situations where it was clear that each telescope observed a particular double differently, the observations by one instrument or the other were dropped. Double stars outside the overlap zone for the two telescopes, of course, cannot be compared in this way and may have undetected errors in their positions.

## Solar System Observations

The purpose of the daytime observations of the Sun, Mercury, Venus, Mars, and bright stars was to create an absolute catalog tied to the dynamical reference frame. Because these observations were not necessary for the link to the ICRF and the quality of these observations makes it difficult to adjust them to the nighttime system, the daytime observations were dropped.

The same corrections that were developed for the observations of the stars also were applied to the nighttime planetary observations. It is necessary to apply additional corrections to the observations of most of the planets due to their orbital motions, appearances, and distances. These additional corrections must be calculated using data from an ephemeris. For the planets, ephemerides data from DE405 (Standish and Williams 2012) were used, and for the minor planets, James Hilton (1999) of USNO provided the ephemerides.

**Orbital Motion Corrections -** Corrections for orbital motion were applied to bring the mean, measured position into coincidence with the meridian. The same equation for the orbital motion correction used for the $W1_{J00}$ was also used for the $W2_{J00}$ .

**Visual Appearance Corrections -** Corrections for the visual appearance of each solar system object were based on their appearance in the transit circle and the method of measurement used.

The software algorithm (Stone 1990) used by the Seven-inch system to determine the centers of the planets from digital pixel data produced by the image dissector was found to be faulty after five years of observations had been acquired. In 1991 the software was changed to address the faults; unfortunately it was impossible to recover data taken before this time resulting in the loss of observations of Mars, Jupiter and Saturn from 1987 to the time of the software change in 1991. Problems developed with the algorithm used by the Seven-inch for Saturn as the rings tilted edge on during the last year of the program and these observations were also dropped.

Uranus, Neptune, and the minor planets were observed center of light.

The Six-inch, observing visually, dealt with the planetary objects as follows:

Mars, Jupiter, and Saturn - The four limbs were observed for all the nighttime observations, except for three observations of Mars when the center of light was taken. In the case of Saturn, the four limbs were observed about 65% of the time, otherwise the edges of the rings were taken. The equations for the corrections for phase were applied to the limb observations made by the Six-inch using the same equations used for the $W1_{J00}$ observations of Mars and Jupiter.

Minor Planets - No visual appearance corrections were applied as all presented point source images.

Uranus and Neptune - Center of light was observed and no corrections for phase were applied.

Plots of the (*O*-*C*)'s as functions of the phase corrections show systematic offsets symmetrical around opposition (see Figure 3). Equations used for the Six-inch data for the phase corrections, as well as the algorithms developed for the Seven-inch data, are based on the geometric changes in the appearances of these planets. The failures to account for all the phase effects are likely a result of limb darkening or other illumination effects. The use of (*O*-*C*)'s caused some concern that the residual effect was in the ephemeris rather than in the observations themselves. The Six-inch results for Saturn clarified the situation, the observations of Saturn's limbs showed the systematic offsets whereas the observations of the rings did not (no such phase corrections could be determined for the Seven-inch Saturn observations because the algorithm used was fitted to both the limbs and rings). The empirically determined additional phase corrections are in Table 12.

Table 12: Empirically determined additional phase corrections for Mars, Jupiter, and Saturn
*v* = visual appearance correction

**Additional phase corrections**

| | | **Six-inch** | **Seven-inch** |
|---|---|---|---|
| Mars | RA | ±0.15 + *v* x 0.282 | ±0.88 |
| | Dec | ±0.10 + *v* x 1.290 | *v* x 3.450 |
| | | | |
| Jupiter | RA | *v* x 1.901 | ±0.25 + *v* x 4.250 |
| | Dec | *v* x 6.426 | ±0.15 + *v* x 4.445 |
| | | | |
| Saturn | RA | *v* x 7.000 | none |
| limb obs | Dec | none | none |
| Saturn | RA | none | none |
| ring obs | Dec | none | none |

Figure 3: Example of the plots used to determine the additional phase correction based on the applied phase correction and the (O-C)'s in RA and Dec that are symmetrical around opposition. The example plots are observations of Jupiter with the Seven-inch transit circle.

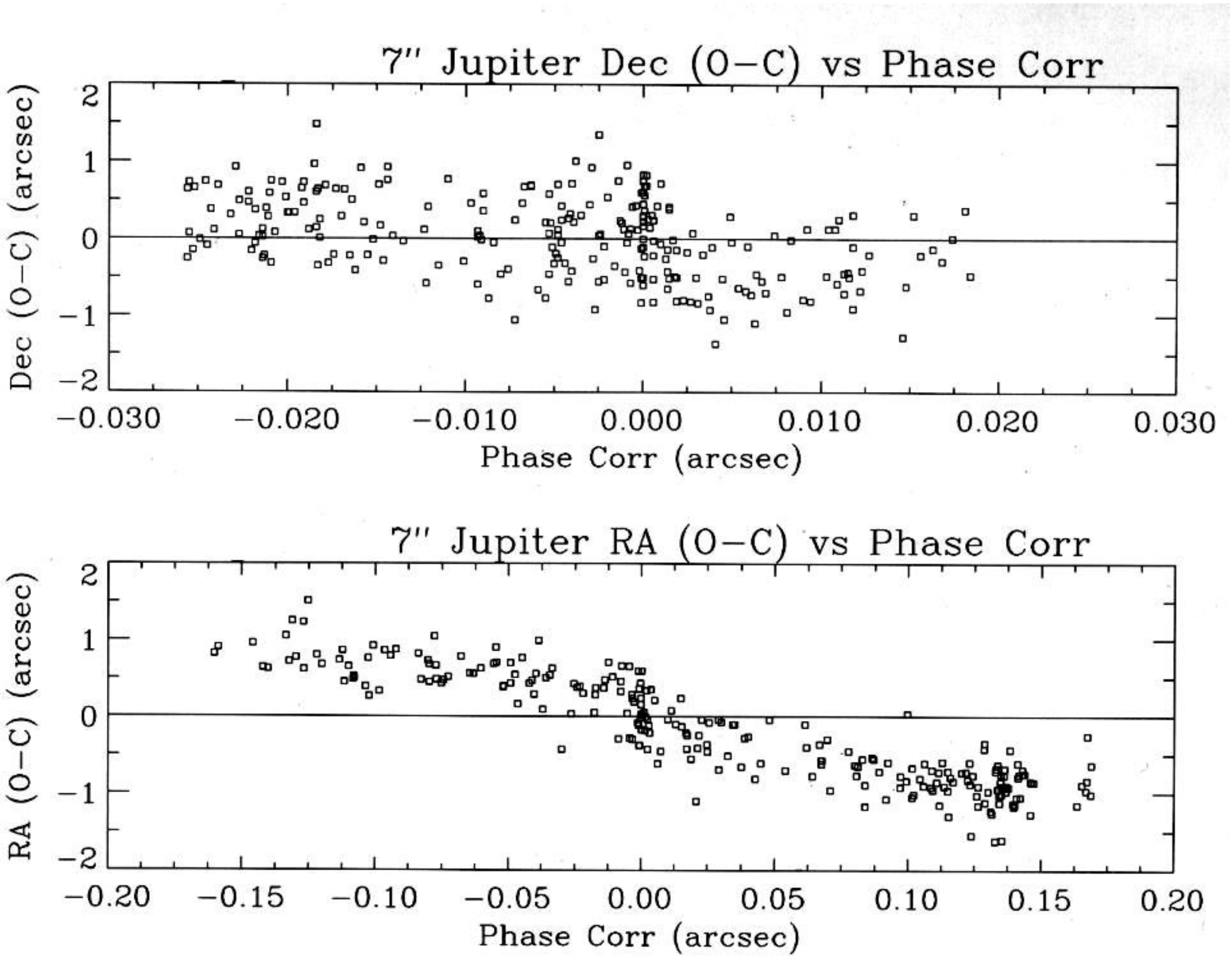

**Horizontal Parallax Correction -** In declination, a correction based on the horizontal parallax at the time of the observation, the Earth's radius vector, and the difference between the geocentric latitude and observed declination was applied. The same equation for the horizontal parallax correction that was used for the $W1_{J00}$ was also used for the $W2_{J00}$. For the Six-inch, 0.998691 was used for the Earth's radius vector and $38^{\circ}\!.920626945$ for the geocentric latitude. For the Seven-inch, 0.998735 was used for the Earth's radius vector and $-41^{\circ}\!.744829722$ for the geocentric latitude.

**Delta-T (*ΔT*) -** Unlike the (*O-C*) positions for the stars, the (*O-C*)'s for the solar system objects must involve Delta-T (*ΔT*), which is the difference between dynamical time (now referred to as Terrestrial Time or TT) and Universal Time (UT1). The necessary corrections are to the ephemeris places of the solar system objects for the motion of the object during the interval between its transit over the ephemeris meridian and the instrumental meridian. The correction is found by multiplying the daily motion of the object by *ΔT* and adding it to the calculated position determined from ephemeris, which in the case of the $W2_{J00}$ was DE405. The values of Delta-T used were interpolated from the yearly values taken from page K9 of the 2000 Astronomical Almanac (Nautical Almanac Office 2000), which are given in the table below.

| Year | $\Delta T$ (sec of time) |
|---|---|
| 1985.0 | 54.34 |
| 1986.0 | 54.87 |
| 1987.0 | 55.32 |
| 1988.0 | 55.82 |
| 1989.0 | 56.30 |
| 1990.0 | 56.86 |
| 1991.0 | 57.57 |
| 1992.0 | 58.31 |
| 1993.0 | 59.12 |
| 1994.0 | 59.98 |
| 1995.0 | 60.78 |
| 1996.0 | 61.63 |
| 1997.0 | 62.29 |

## References

Corbin, T.E., 1991, NASA, NSSDC 91-11

Corbin, T.E., 1978, The Proper Motions of the AGK3R and SRS Stars, Proc. IAU Coll. 48 Modern Astrometry, Prochazka, F.V. & Tucker, R.H. eds., 505-514

European Space Agency, 1997, The Hipparcos and Tycho Catalogues, ESA Publications Dvision, Vols 1-17

Explanatory Supplement to the Astronomical Ephemeris and the American Ephemeris, 1992, Her Majesty's Stationery Office

Fricke, W., Schwan, H., and Lederle, T., 1988, Fifth Fundamental Catalogue (FK5), Part I, Veroff. Astro. Rechen-Inst., No. 32

Fricke, W., Schwan, H., and Corbin, T.E., 1991, Fifth Fundamental Catalogue (FK5), Part II, Veroff. Astro. Rechen-Inst., No. 33

Hemenway, P.D., 1966, The Washington 6-inch Transit Circle, Sky and Telescope, Vol XXXI, No. 2

Holdenried, E.R. and Rafferty, T.J., 1997, New methods of forming and aligning the instrumental frame of absolute transit circle catalogs, Astronomy and Astrophysics Suppl. Ser. 125, pp 595-603
Høg, E., 1961, Astron. Nachr, vol 286, p 65

Høg, E., 1978, The ESA Astrometry Satellite, Modern Astrometry, IAU Colloquium No. 48, pp 557-559

Hilton, J., 1999, US Naval Observatory Ephemerides of the Largest Asteroids, The Astronomical Journal, Vol. 117, Issue 2, pp. 1077-1086

Hughes, J.A., 1978, SRS Observations: Future U.S. Naval Observatory Meridian Programs, Modern Astrometry, IAU Colloquium No. 48, pp 497-501

Hughes, J.A., Espenscheid, P., and McCarthy D.D., 1975, Coordinates of U.S. Naval Observatory Installations, United States Naval Observatory Circular, No 153

Hughes, J.A. and Scott, D.K., 1982, Results of observations made with the Six-inch Transit Circle 1963-1971 ($W5_{50}$), Publications of the United States Naval Observatory, Second Series, XXIII(3)

Hughes, J.A., Smith, C.A., and Branham, R.L., 1992, Results of observations made with the Seven-inch Transit Circle 1967-1973 ($WL_{50}$), Publications of the United States Naval Observatory, Second Series, XXVI(2)

Hughes, J.A., Robinson, M.D., Gauss, F.S., and Stone, R.C., 1986, The Seven-inch Transit Circle and its New Zealand Program, Astrometric Techniques, IAU Symposium No. 109, pp 483-496

Küstner, F., 1900, Beobachtungen von 4070 Steren... Veröff. d. Kgl. Stw. zu Bonn, No. 4

Ma, C., Arias, E.F., Eubanks, T.M., Fey, A.L., Gontier, A.M., Jacobs, C.S., Sovers, O.J., Archinal, B.A., and Charlot, P., 1998, The International Celestial Reference Frame as Realized by Very Long Baseline Interferometry, Astronomical Journal, 116(1), pp 516-546

Nautical Almanac Office, 2000, The Astronomical Almanac, U.S. Government Printing Office, Washington, DC

Podobed, V.V., 1962, Fundamental Astrometry, The University of Chicago Press

Rafferty, T.J. and Klock, B.L., 1986, Circle scanning systems of the U.S. Naval Observatory, Astronomy and Astrophysics, vol 164, pp 428-432

Standish, E.M. Jr. and Williams, J.G., 2012, Orbital Ephemerides of the Sun, Moon, and Planets, Explanatory Supplement to the Astronomical Almanac, Chapter 8

Stone, R.S., 1990, Digital Centering Algorithms for the Sun, Moon, and Planets, Astronomical Journal, 99(1), pp 424-430

Watts, C.B., 1950, Description of the Six-inch Transit Circle, Publication of the United States Naval Observatory, Second Series, Vol XVI, Part 2

## Format of the $W2_{J00}$ Stars and Solar System Objects Data

As stated in the introduction, the $W2_{J00}$ observing program was structured to be absolute, in the sense that the positions were not explicitly relying on any previous observations. However, with the availability of Hipparcos observational data, it was decided to differentially adjust the observations to the ICRF using the Hipparcos star positions (ESA, 1997). The $W2_{J00}$ observing program used both the Six-inch Transit Circle and Seven-inch Transit Circle. For stars that were observed by both telescopes, positions based on the combination of observations from both telescopes are given as well as the positions from each of the telescopes.

The $W2_{J00}$ position files ***W2J00_stars.dat*** and ***W2J00_solsys.dat***, along with the ReadMe file ***W2J00 ReadMe***, can be found in the at the Centre de Domnées astronomiques de Strasbourg (Strasbourg Astronomical Data Center): http://cdsweb.u-strasbg.fr/

| FileName | Lrecl | Records | Explanations |
|---|---|---|---|
| ***W2J00 ReadMe*** | 80 | | $W2_{J00}$ ReadMe file |
| ***W2J00_stars.dat*** | 197 | 44395 | $W2_{J00}$ star positions (means) |
| ***W2J00_solsys.dat*** | 63 | 11566 | $W2_{J00}$ solar system object positions |

## Format of the $W2_{J00}$ Star Data File: ***W2J00_stars.dat***

The positions of the stars are on the system of J2000.0 which means that although epoch of the observations is that of the mean observation, the orientation of the celestial reference system is fixed by the epoch of 2000 January 1.5.

The $W2_{J00}$ star data can be found in the file ***W2J00_stars.dat*** at the Centre de Domnées astronomiques de Strasbourg (Strasbourg Astronomical Data Center): http://cdsweb.u-strasbg.fr/

Byte-by-byte Description of file: ***W2J00_stars.dat***

| Bytes | Format | Units | Label | Explanations |
|---|---|---|---|---|
| 1-5 | I5 | --- | W2J00 | W2J00 identifier |
| 7-8 | I2 | h | RAh | Right Ascension ICRS, at Ep_RA (hours) (1) |
| 10-11 | I2 | min | RAm | Right Ascension ICRS, at Ep_RA (minutes) (1) |
| 13-18 | F6.3 | s | RAs | Right Ascension ICRS, at Ep_RA (seconds) (1) |
| 20-22 | I3 | mas | e_RAs | Weighted standard error of mean, RA |
| 24-26 | I3 | --- | NumRA | Number of observations used for RA |
| 28-35 | F8.3 | yr | Ep_RA | Mean epoch of RA |
| 37 | A1 | --- | DE- | Declination (ICRS) at Ep_DE (sign) (1) |
| 38-39 | I2 | deg | DEd | Declination (ICRS) at Ep_DE (degrees) (1) |
| 41-42 | I2 | arcmin | DEm | Declination (ICRS) at Ep_DE (minutes) (1) |
| 44-48 | F5.2 | arcsec | DEs | Declination (ICRS) at Ep_DE (seconds) (1) |
| 50-52 | I3 | mas | e_DEs | Weighted standard error of the mean, DE |
| 54-56 | I3 | --- | NumDE | Number of observations used for DE |
| 58-65 | F8.3 | yr | Ep_DE | Mean epoch of DE |

```
67-70   F4.1 mag    Vmag   Visual magnitude (2)
72-77   I6   ---    HIP    Hipparcos Catalogue number (3)
79-80   I2   h      RA6h   RA ICRS from 6-inch, at Ep_RA6 (hours) (4)
82-83   I2   min    RA6m   RA ICRS from 6-inch, at Ep_RA6 (minutes) (4)
85-90   F6.3 s      RA6s   RA ICRS from 6-inch, at Ep_RA6 (seconds) (4)
92-94   I3   mas    e_RA6s Weighted standard error of the mean, RA6
96-98   I3   ---    NumRA6 Number of observations used for RA6
100-107 F8.3 yr     Ep_RA6 Mean epoch of RA6
109     A1   ---    DE6-   Dec (ICRS) from 6-inch, at Ep_DE6 (sign) (4)
110-111 I2   deg    DE6d   Dec (ICRS) from 6-inch, at Ep_DE6 (degrees) (4)
113-114 I2   arcmin DE6m   Dec (ICRS) from 6-inch, at Ep_DE6 (minutes) (4)
116-120 F5.2 arcsec DE6s   Dec (ICRS) from 6-inch, at Ep_DE6 (seconds) (4)
122-124 I3   mas    e_DE6s Weighted standard error the mean, DE6
126-128 I3   ---    NumDE6 Number of observations used for DE6
130-137 F8.3 yr     Ep_DE6 Mean epoch of DE6
139-140 I2   h      RA7h   RA ICRS from 7-inch, at Ep_RA7 (hours) (5)
142-143 I2   min    RA7m   RA ICRS from 7-inch, at Ep_RA7 (minutes) (5)
145-150 F6.3 s      RA7s   RA ICRS from 7-inch, at Ep_RA7 (seconds) (5)
152-154 I3   mas    e_RA7s Weighted standard error of the mean, RA7
156-158 I3   ---    NumRA7 Number of observations used for RA7
160-167 F8.3 yr     Ep_RA7 Mean epoch of RA7
169     A1   ---    DE7-   Dec (ICRS) from 7-inch, at Ep_DE7 (sign) (5)
170-171 I2   deg    DE7d   Dec (ICRS) from 7-inch, at Ep_DE7 (degrees) (5)
173-174 I2   arcmin DE7m   Dec (ICRS) from 7-inch, at Ep_DE7 (minutes) (5)
176-180 F5.2 arcsec DE7s   Dec (ICRS) from 7-inch, at Ep_DE7 (seconds) (5)
182-184 I3   mas    e_DE7s Weighted standard error of the mean, DE7
186-188 I3   ---    NumDE7 Number of observations used for DE7
190-197 F8.3 yr     Ep_DE7 Mean epoch of DE7

Note (1): Mean positions are on the International Celestial Reference System (ICRS),
differentially reduced using the Hipparcos catalog. The positions are a weighted mean
of the Six-inch Transit Circle and Seven-inch Transit Circle observations, found in
columns 79-197. The epochs for the right ascension and declination coordinates are
found in the Ep_RA and Ep_DE fields respectively.
Note (2): Visual magnitude for identification purposes.
Note (3): Cross references to the Hipparcos Catalogue are provided for convenience of
the user.
Note (4): Columns 1-65 give positions that are a weighted mean of the Six-inch Transit
Circle and Seven-inch Transit Circle observations.  Columns 79-137 give the positions
from the Six-inch Transit Circle only.
Note (5): Columns 1-65 give positions that are a weighted mean of the Six-inch Transit
Circle and Seven-inch Transit Circle observations.  Columns 139-197 give the positions
from the Seven-inch Transit Circle only.
```

## Format of the W2$_{J00}$ Planet and Minor Planet Data File: *W2J00_solsys.dat*

The solar system objects include Mars, Jupiter, Saturn, Uranus, Neptune, twelve minor planets (Amphitrite, Eunomia, Flora, Hebe, Hygiea, Iris, Juno, Melphomene, Metis, Nemausa, Pallas, and Vesta), and the dwarf planet Ceres. Daytime observations of the Sun, Mercury, Venus, and Mars were made but not included in the final catalog due to the problems inherent in reducing observations made in the daylight. Unlike the stars, for the solar system objects the celestial reference frame was moved by precession and nutation to the time of the observation. This is accomplished by the apparent place routine of the NOVAS astrometric software developed by the Nautical Almanac Office of USNO (Kaplan 1990). The version used here was what was current in 1996. The input to the apparent place routine was provided by the DE200. Furthermore the positions for the solar system objects are geocentric. This follows a tradition laid out by earlier USNO transit circle star catalogs.

The $W2_{J00}$ solar system data can be found in the file ***W2J00_solsys.dat*** at the Centre de Domnées astronomiques de Strasbourg (Strasbourg Astronomical Data Center): http://cdsweb.u-strasbg.fr/

Byte-by-byte Description of file: ***W2J00_solsys.dat***

```
Bytes Format Units   Label    Explanations

1-10   A10   ---     Obj      Solar system object identifier
12-13  I2    h       RAh      Right Ascen. (apparent) at Epoch (hours) (1, 2)
15-16  I2    min     RAm      Right Ascen. (apparent) at Epoch (minutes) (1, 2)
18-23  F6.3  s       RAs      Right Ascen. (apparent) at Epoch (seconds) (1, 2)
25     A1    ---     DE-      Declination (apparent) at Epoch (sign) (1, 2)
26-27  I2    deg     DEd      Declination (apparent) at Epoch (degrees) (1, 2)
29-30  I2    arcmin  DEm      Declination (apparent) at Epoch (minutes) (1, 2)
32-36  F5.2  arcsec  DEs      Declination (apparent) at Epoch (seconds) (1, 2)
38-50  F13.5 day     Epoch    Julian date (UT1) of observation
52-54  A3    ---     Obs      Observer (3)
56     A1    ---     Clamp    Clamp orientation of instrument (4)
58     A1    ---     RALimb   Limbs or center of light measured in RA (5)
60     A1    ---     DELimb   Limbs or center of light measured in DE (5)
62-63  A2    ---     Tel      Telescope used (6)

Note (1): Positions are apparent places, reduced to be systematically consistent with
the Hipparcos Catalogue.  An apparent place is a geocentric direction of an object
that takes into account orbital motion, space motion, light-time, light deflection,
and annual aberration.  Apparent place is given with respect to the true equator and
equinox "of date"; in this case, it is the date of observation found in the Epoch
field.
Note (2): In some cases, an observation is made in only right ascension or
declination.  If only the right ascension coordinate was observed, then the
declination seconds field (DEs) will be blank, as will the DELimb.  If only the
declination coordinate was observed, then the right ascension seconds field (RAs) will
be blank, as will the RALimb.
Note (3): The Six-inch Transit Circle measurements were manual, in the sense that a
person measured the position while looking through the instrument.  The Seven-inch
Transit Circle's observations were automated, so the observer is not listed.  The
observer code has the following meaning:
    CSC = C.S. Cole
    TEC = T.E. Corbin
    HEC = H.E. Crull
    SD  = S.J. Dick
    JCD = J.C. Doty
    RE  = R.   Etheridge
    FSG = F.S. Gauss
    MEG = M.E. Germain
    DMH = D.M. Hall
    GSH = G.S. Hennessy
    JH  = J.L. Hershey
    RBH = R.B. Hindsely
    ERH = E.R. Holdenried
    ESJ = E.S. Jackson
    IAN = I.   Jordan
    KAL = V.   Kallarakal
    JCM = J.C. Martin
    JMM = J.M. Muse
    RJM = R.J. Miller
    MDR = M.D. Robinson
```

TJR = T.J. Rafferty
CBS = C.B. Sande
DS = D.K. Scott
CAS = C.A. Smith
SEU = S.E. Urban
GLW = G.L. Wycoff
ZGY = Z.G. Yao

Note (4): Both transit circle telescopes were reversed (rotated 180 degrees in azimuth) interchanging the east and west pivots approximately every 30 days. The orientation of the instrument was referenced to the location of the clamping device, which fixed the altitude of the instrument after it was pointed to that of a star, and thus referred to either as Clamp West or Clamp East. The flag has the following meaning:

W = Clamp West
E = Clamp East

Note (5): Some planets subtend sizable disks in the instrument. Primarily depending on the object, either the limbs or the center-of-light were observed. The flag has the following meaning:

L = limb was measured
C = center of light was measured
R = ring was measured (Saturn only)

Despite how an object was observed, the W2J00 positions refer to the center of the object.

Note (6): The observing program used two telescopes, the Six-inch Transit Circle in Washington DC, and the Seven-inch Transit Circle at Black Birch New Zealand. The "Tel" flag has the following meaning:

6" = Six-inch Transit Circle, Washington D.C., USA
7" = Seven-inch Transit Circle, Black Birch, New Zealand